\documentclass[a4paper,11pt]{article}
\pdfoutput=1 % if you are submitting a pdflatex (i.e. if you have
\usepackage{jheppub} % for details on the use of the package, please
\usepackage[utf8]{inputenc}
\usepackage{setspace}

\definecolor{refkey}{gray}{0.45}
\definecolor{labelkey}{RGB}{155,48,48}
\usepackage{tikz}
\usetikzlibrary{arrows,snakes,backgrounds}
\usetikzlibrary{decorations.markings}
\usepackage{setspace}

\usepackage{physics}
\usepackage[obeyFinal]{todonotes}

\def\ba{\begin{align}}\def\ea{\end{align}}
\def\beq{\begin{eqnarray}}\def\eeq{\end{eqnarray}}
\def\be{\begin{equation}}\def\ee{\end{equation}}

\def\mes[#1]{d^{3}{#1}}

\def\del{\partial}

\newcommand{\half}{\frac{1}{2}}

\def\del{\partial}

\def\order{\ensuremath{\mathcal{O}}}

\reversemarginpar
\tikzset{middlearrow/.style={
		decoration={markings,
			mark= at position 0.5 with {\arrow[scale=2]{#1}} ,
		},
		postaction={decorate}
	}
}

\tikzset{2multiarrow/.style={
		decoration={markings,
			mark= at position 0.35 with {\arrow[scale=2]{#1}} ,
			mark= at position 0.75 with {\arrow[scale=2]{#1}} ,
		},
		postaction={decorate}
	}
}

\tikzset{3multiarrow/.style={
		decoration={markings,
			mark= at position 0.25 with {\arrow[scale=2]{#1}} ,
			mark= at position 0.5 with {\arrow[scale=2]{#1}} ,
			mark= at position 0.75 with {\arrow[scale=2]{#1}} ,
		},
		postaction={decorate}
	}
}

\tikzset{4multiarrow/.style={
		decoration={markings,
			mark= at position 0.2 with {\arrow[scale=2]{#1}} ,
			mark= at position 0.4 with {\arrow[scale=2]{#1}} ,
			mark= at position 0.6 with {\arrow[scale=2]{#1}} ,
			mark= at position 0.8 with {\arrow[scale=2]{#1}} ,
		},
		postaction={decorate}
	}
}

\author[a]{Upamanyu Moitra,}
\author[a]{Sunil Kumar Sake,}
\author[a]{Sandip P. Trivedi,}
\affiliation[a]{\it Department of Theoretical Physics,
	Tata Institute of Fundamental Research,\\  Colaba, Mumbai, India, 400005\\}

\vspace{1cm}

\abstract{We formulate the path integral for Jackiw-Teitelboim gravity in the second order formalism working directly with the metric and the dilaton.  We consider the theory  both in Anti-de Sitter(AdS) and de Sitter space(dS) and   analyze the path integral for the disk topology and the ``double trumpet"  topology with two boundaries.
	We also consider its behavior in   the presence of conformal matter. 
	In the dS case the path integral   evaluates the wavefunction of the universe  which arises in the no-boundary proposal. In the asymptotic AdS or dS  limit  without matter we get agreement with the first order formalism. More generally, away from this limit, the path integral is more complicated due to the presence of  modes from the gravity- dilaton sector and also matter sector with short wavelengths along the boundary that are  smaller than the AdS or dS scales. In the double trumpet case, for both AdS and dS,  we find that bosonic matter gives rise to a diverging contribution in  the moduli space integral rendering the path integral ill-defined. The divergence occurs when the size of the wormhole neck vanishes and is related to the Casimir effect.  For fermions this divergence can be avoided by imposing  suitable  boundary conditions. In this case, in dS space the resulting path integral  gives a finite contribution for  two disconnected universes to be produced by quantum tunneling.     }

\title{ Jackiw-Teitelboim Gravity in the Second Order Formalism }

\preprint{\parbox{3cm}{TIFR/TH/20-53}}

\begin{document}
	\maketitle
	\flushbottom
	\newpage
	\section{Introduction}
	\label{intro}
 Jackiw-Teitelboim (JT) gravity is a theory of two-dimensional gravity which  has received considerable attention recently \cite{JACKIW1985343,Teitelboim:1983ux,Saad:2019lba,Saad:2018bqo,Almheiri:2014cka,Jensen:2016pah,Maldacena:2016upp,Engelsoy:2016xyb,Harlow:2018tqv,Blommaert:2018iqz,Yang:2018gdb,Blommaert:2019hjr,Stanford:2017thb,Mertens:2017mtv,Kitaev:2017awl,Nayak_2018,Moitra:2019bub,Moitra:2019xoj,Moitra:2018jqs,Lin:2018xkj,Stanford:2019vob,Stanford:2020qhm,Mertens:2019tcm,Maldacena:2019cbz,Iliesiu:2019xuh,Mertens:2019bvy,Almheiri:2019psf,Penington:2019npb,Penington:2019kki,Blommaert:2020seb,Lin:2019qwu,Maldacena:2018lmt,Almheiri:2019qdq,Almheiri:2019hni,Penington:2019kki,Goel:2020yxl,Balasubramanian:2020xqf,Hartman:2020khs,Iliesiu:2020zld,Goto:2020wnk,Mefford:2020vde,Fitkevich:2020okl,Johnson:2020exp,Maxfield:2020ale,Suh:2019uec,Saad:2019pqd,Lala:2019inz,Gross:2019ach,Cotler:2019nbi,Brown:2018bms,Li:2018omr,Chen:2019uhq,Xian:2019qmt,Grumiller:2020elf,Numasawa:2020sty,Bhattacharjee:2020nul,Johnson:2020mwi,Jian:2020qpp,Hartman:2020khs,Chen:2020tes,Chen:2020jvn,Engelhardt:2020qpv,Okumura:2020dzb,Wieland:2020ogk,Verlinde:2020upt,Pollack:2020gfa,Cotler:2019dcj,Alishahiha:2020jko,Grumiller:2020fbb,Johnson:2020heh,Iliesiu:2020qvm,Ishii:2019uwk,Joshi:2019wgi,Hadar:2018izi,Goto:2018iay,Kim:2018nir,Okumura:2018xbh,Grumiller:2017qao,Callebaut:2018nlq,Lemos:1993qn,Lemos:1996bq,Alishahiha:2018swh,Blommaert:2018oro,Gonzalez:2018enk,Alkalaev:2013fsa,Dubovsky:2017cnj,Haehl:2017pak,PhysRevResearch.2.043310,PhysRevLett.54.959,Hollowood:2020kvk,Li:2020ceg,Hollowood:2020cou,Blommaert:2019wfy,Jafferis:2019wkd,Balasubramanian:2020coy,Almheiri:2019yqk,Narayan:2020pyj,Manu:2020tty,Narayan:2020nsc,Kolekar:2018sba,Kolekar:2018chf,Kolekar:2018sba,Sachdev:2019bjn,Sachdev_1993,Davison:2016ngz,Larsen:2018cts,Larsen:2018iou,Castro:2018ffi,Hong:2019tsx,Dhar:2018pii,Gaikwad:2018dfc,Mandal:2017thl,Kimura:2020zke}.
In this paper we carry out a path integral quantization of the theory  in the second order formalism which involves the metric and the dilaton. This is to be contrasted with the first order formalism, used in much of the recent discussion\cite{Saad:2019lba,Iliesiu:2019xuh,Constantinidis:2008ty,Isler:1989hq,Witten:1991we}, which involves  the spin connection and Vierbein along with the dilaton. 
Some of the motivation for our work comes from wanting to compare the results we get from the second order formalism with those obtained in the first order formalism. 	The second order formalism  allows matter to be added in a direct way, and this  permits  us to generalize our  study   of JT gravity   to also include  matter.  	Finally, one might hope to glean some lessons about quantizing gravity in higher dimensions   from the second order formalism. 

We study both JT gravity in Anti-de Sitter (AdS) space and in de Sitter (dS)  space here. Our analysis includes     the path integral for   spacetimes with the  topology of a disk which have Euler character $\chi=1$, with one boundary,  and also  spacetimes of the ``double trumpet" kind with  $\chi=0$ and    two boundaries.

For the disk topology we find  in the asymptotic AdS or dS limit, obtained by taking the dilaton and length of the boundary to diverge while keeping their ratio fixed, that  the results of the second order path integral quantization agree with those obtained from the first order formalism. In particular, we find like in the first order formalism   that the dynamics of the gravity-dilaton system  is given  by the  reparametrization modes of the boundary (called time reparametrization modes) which are governed by an  action involving the Schwarzian derivative. However away from the this limit, for example, even when   working with a boundary of large but fixed length, we find that the path integral is more complicated to evaluate. 
This happens  because the measure for diffeomorphisms is more complicated in general, due to the  mixing of   small and large diffeomorphisms with each other, and also because   the determinants which  now arise have a complicated dependence  on the large diffeomorphisms. The underlying reason for all this is that   away from the asymptotic limit there are modes with wavelengths shorter than the AdS or dS scale which need to be included in the path integral and their dynamics  is not simple. 
In particular one needs  to include an arbitrarily large number of higher derivative terms, beyond the Schwarzian, valued in $\text{Diff}(S^1)/SL(2,R)$ in order to incorporate their coupling with the large diffeomorphisms. 

The matter we include is conformally invariant - mostly free bosons or fermions, although some of our results are more general. 	
In the presence of matter, for the disk,  we find again that in the asymptotic AdS or dS limit the path integral can be carried out but  away from this  limit the matter determinant from quantum fluctuations  has  a dependence  on the large diffeomorphisms   which requires us to go beyond the Schwarzian action and include the higher derivative terms mentioned above. It is worth mentioning that simple dimensional counting shows that the  quantum effects of matter only arise away from the asymptotic limit when one is working at finite boundary length, and including them in a systematic manner along with the quantum effects from the gravity-dilaton sector is quite non-trivial. Such an analysis would need to be carried out to go beyond the semi-classical limit which has been analyzed in considerable detail recently where  the number of matter fields $N\rightarrow \infty$, and the gravity-dilaton is  treated as  being classical.

In AdS space the path integral for a single boundary has the interpretation of computing the partition function of the boundary theory at finite temperature. 
In dS space the path integral we carry out evaluates the wavefunction of the universe obtained from the no-boundary proposal first put forward by Hartle and Hawking \cite{PhysRevD.28.2960}. The contour in this case is more complicated and goes over spacetime regions with different signatures. We consider both the Hartle-Hawking (HH) contour, which involves initially a Euclidean region with signature $(2,0)$ and then de Sitter space of signature $(1,1)$, and also an alternate contour suggested by Maldacena \cite{Maldacena:2019cbz} which involves a spacetime with signature $(0,2)$, with metric $-AdS_2$, followed by a region of dS space\footnote{In this paper we will mostly consider the branch of the wavefunction which  in the classically allowed region  corresponds to a universe expanding as time increases, as discussed further in section \ref{dsjtpi}.}. We find agreement between both contours and discuss how to carry out the path integral  by analytically continuing fluctuations across regions of different signature. The short distance modes we were referring to above, which render the calculations more complicated away from the asymptotic limit, in the dS case refer to modes which are  still dynamical and have not yet exited the horizon. It is important to mention that especially in the dS case one would like to obtain the wavefunction when the universe has finite size (and the dilaton takes a finite value) and it is therefore important to try and go beyond the asymptotic limit after  including these modes, although we will not be able to make much progress in this direction here.

The contribution of the double trumpet geometry is suppressed compared to the disk topology, since it has higher Euler character, by a factor of $e^{\Phi_0/4G}$ where $\Phi_0$ can be thought of as the horizon value of the volume of the   internal space  which gives rise to the JT AdS or dS  theory after dimensional reduction.  The double trumpet can be thought of as a wormhole  joining the two boundaries. Most of the comments above about agreement with the first order formalism in the absence of matter and in the asymptotic AdS or dS limits 
apply to the double trumpet geometry as well. The path integral now  involves two sets of large diffeomorphisms which act independently at the two boundaries and also moduli, including one which corresponds to the size of the ``neck" of the  wormhole (called $b$ below).  We show how the correct measure for summing over these diffeomorphisms and moduli arise in the second order formalism. 

Once matter is added we find that its  quantum effects  give rise to a contribution in the path integral which diverges when the  neck goes to zero size. The  quantum effects can be thought of as giving rise to a Casimir effect which diverges when the neck becomes vanishingly small. We show that this happens both for bosons and also for fermions. The bosons have  periodic while  the fermions have   anti-periodic boundary conditions along the time or temperature direction. 
This divergence is analogous to the tachyon divergence which arises on the world sheet for Bosonic string theory. 
Alternatively,  we can also consider periodic boundary conditions along this  direction for the fermions, as  would be appropriate for example if one is evaluating an index 
$\text{Tr}((-1)^Fe^{-\beta H})$ instead of the  partition function\footnote{We are grateful to Shiraz Minwalla for emphasizing these alternate boundary conditions.}. In this case we find that the quantum effects from fermions do not diverge when $b \rightarrow 0$ and the path integral is well behaved.   

In dS space the double trumpet gives rise to an  amplitude for two  universes to arise after quantum tunnelling from ``nothing". The divergence in the $b \rightarrow 0$ limit mentioned above arises in this case as well and can also be avoided by taking the fermions to have periodic boundary conditions along the spatial directions of the two universes. In fact,  the periodic boundary conditions for the fermions cannot be satisfied for  the disk topology,    so  with these boundary conditions the the leading contribution in the wavefunction, in an  expansion in Euler character,  arises from the amplitude  to pair produce  the two universes from the double trumpet wormhole.	 

This paper is organized as follows. In section \ref{purejtads2} we elaborate on the quantization of JT gravity in second order formalism for the case of Euclidean AdS spacetime with disk topology. In section \ref{jtwithmatterads2}, we extend this analysis to include additional matter fields for the disk topology.  Following this, in section \ref{dbads} we repeat the analysis for the case of Euclidean AdS spacetime with ``double trumpet" topology, both with and without additional matter. In section \ref{dsjtpi} and \ref{desitdobt} we redo the analysis of sections \ref{purejtads2},\ref{jtwithmatterads2} and \ref{dbads}, now in the $dS$ spacetime.  Finally we end with conclusions in section \ref{conclusion}. Appendices \ref{coordtrsfs}-\ref{dsdbdets} contain important additional details.

%	\newpage
	\section{Pure JT theory path integral in  AdS} 
	\label{purejtads2}
	In this section we will consider   JT gravity in  Euclidean AdS space. 
	The path integral for the system is given by 
	\begin{equation}
		Z_{JT}=\int  \frac{{D}[\phi]{D}[g_{\mu\nu}]}{\text{Vol}(\Omega)} \exp{- S_{JT}}\label{piexp}
	\end{equation}
	where $S_{JT}$,  the action for  Jackiw-Teitelboim gravity, involves the metric and a scalar, the dilaton. In  Euclidean signature the action is given by 
	\begin{align}
	S_{JT}=-\frac{1}{16\pi G}\pqty{\int d^2x\sqrt{g}\phi (R+2)+2\int dx \sqrt{\gamma}\phi (K-1)}\label{eadsjtact}, 
	\end{align}
	where  we have set the $AdS_2$ length $R_{AdS}=1$. 
	
	Note that the sum in eq.(\ref{piexp}) is over all metric and dilaton configurations and our main task here  will be to  make this precise. 
	This problem has invited considerable attention recently, \cite{Saad:2019lba,Stanford:2017thb}.  In general, one must sum over all topologies subject to the boundary conditions  that are imposed. 
	Our approach will be to  work directly in the second order formalism which involves a sum of metric configurations and not with the first order formalism which has been used in much of the previous literature. This will  also allow us to    include matter  easily, as we will see later. 
	We will restrict ourselves in this section  to the relatively simple case of the disk topology with one boundary.
	
	Before proceeding let us note that the action of JT gravity actually includes one additional term which is topological, 
	\be
	\label{Stop}
	S_{top}=-{\Phi_0 \over 16 \pi G} \pqty{ \int \sqrt{g} R +2 \int \sqrt{\gamma} K}= -{\Phi_0\over 4G} \chi
	\ee
	where $\chi$ is the Euler characteristic of the manifold, related to the number of handles, $H$ and boundaries, $B$ by  
	\be
	\label{defchi}
	\chi= 2- 2 H - B
	\ee
	Such a topological term arises for example when one constructs the JT action by dimensionally reducing from higher dimensions in the near horizon region of  a near extremal black hole and in that case it accounts for the ground state entropy of the extremal black hole. 
	We will mostly ignore $S_{top}$ for now, since we will be working on the disk topology with fixed $\chi=1$,  and work with the action eq. (\ref{eadsjtact}).
	
	We will formulate  the path integral for  a  boundary  of fixed length  $l$ with  the dilaton taking a fixed value $\phi_B$ at this boundary. 
	An important limit in which the path integral eq.(\ref{eadsjtact})  has been studied is the asymptotic AdS limit. In this limit we introduce a cut-off $\epsilon$ to regulate the theory and take the limit $\epsilon \rightarrow 0$, with the dilaton and length of the boundary scaling  like 
	\be
	\label{scaled1}
	\phi\rightarrow {1\over J \epsilon}
	\ee
	\be
	\label{scalelength1}
	l \rightarrow { \beta\over \epsilon}
	\ee
	with $J, \beta$ fixed. {By rescaling $\epsilon$ we can set $\beta =1$ so there is actually only one  dimensionless parameter specifying the limit given by $\beta J$.} So, we take the boundary conditions for $\phi$ and $l$ in this section as follows. 

	\begin{align}
	\phi&\rightarrow \frac{2\pi}{J\beta \epsilon}\label{scaled}\\
	l&\rightarrow \frac{2\pi}{\epsilon}\label{scalelength}
	\end{align}
	  We will see below when dealing with determinants that there is also the issue of taking the cut-offs, introduced to regulate the determinants, to infinity;  the asymptotic AdS limit then needs to be defined more precisely   keeping track of the correct order of limits. 
	
	In this asymptotic AdS limit we will find complete agreement between the path integral in the second order and first order formalisms. In particular, we will show below how the sum over large diffeomorphisms, which correspond to fluctuations of the boundary, arises,  with the  correct measure, in the second order formalism as well. 
	In the more general case where the dilaton takes a fixed value $\phi_B$ at the boundary of length $l$ 
	we will show how the path integral can be defined quite precisely, but will not be able to carry out the evaluation till the very end. 
	Interestingly, we will find that the more general case differs from the asymptotic AdS one in important  ways, even when $\phi_B, l \gg 1$. This more general case will also be of interest when we turn to de Sitter space later in the paper.

	We have not specified  yet what  the $\text{Vol}(\Omega)$  factor in eq.(\ref{piexp}) refers to. In defining the path integral for any gauge theory one would only like to sum over physically distinct configurations. 
	This can be achieved by summing over all configurations and then dividing by the volume of the gauge group. For our case we would therefore divide by the volume of all diffeomorphisms which leave  the geometry - along with the boundary - unchanged. These diffeomorphisms which will be defined more precisely below be  will referred to  as ``small diffeomorphisms"   and $\text{Vol}(\Omega)$, in eq.(\ref{piexp}),  then    refers to the volume of these diffeomorphisms, .  In contrast, there will also be a set of ``large diffeomorphisms",  these are   physically distinct configurations corresponding to different boundaries and we will   sum over them without treating them as gauge transformations.

	\subsection{The path integral defined more precisely}
	\label{pjtadsstrategy}
	
	We first consider in our discussions below the general case of a disk  with a boundary of  length $l$  where the  dilaton takes value $\phi_B$.
	As a limiting case we will then turn to  the asymptotic AdS boundary conditions, eq.\eqref{scaled} and \eqref{scalelength}.

	Let us begin by specifying the measure for the  sum over metrics more carefully. The starting point  is as follows. We   consider the space of metrics satisfying the required boundary conditions itself to be a Riemannian manifold and  denote this space as ${\cal R}$. A point in this space is a metric $g_{ab}$ on a manifold with disk topology and boundary of length $l$. 
	The tangent space of all metric deformations at any particular point in $\mathcal{R}$,  $\mathcal{T}_g\mathcal{R}$, corresponds to small deformations $\delta g_{ab}$. This space is endowed with an ultra local inner product which takes the form, 
	\begin{equation}
	\langle\delta_1 g,\delta_2 g\rangle=\int d^2x \sqrt{g}g^{ac}g^{bd}\delta_1 g_{ab}\delta_2 g_{cd}\label{metnrm}
	\end{equation}
	for two deformations, $\delta_1g, \delta_2g$. The inner product then  defines a metric on $\mathcal{R}$ and the measure for summing over different metrics is defined using  the volume element which follows from this metric. 
	
	In two dimensions things become especially simple  because  a general metric, $g_{ab}$, after a coordinate transformation  can always  be locally written  in terms of a conformal factor $\sigma$ as 
	\be
	\label{confg}
	g_{ab}=e^{2\sigma} {\hat g}_{ab}
	\ee
	where ${\hat g}_{ab}$ is a fiducial metric. For the disk topology, the manifold can be covered by a single coordinate chart and  we can take ${\hat g}_{a b }$  to be a constant negative curvature metric  with curvature ${\hat R}=-2$ in this chart. 
	
	Furthermore, a small metric deformation in general in two dimensions  can be decomposed in the following manner, \cite{DHoker:1985een}:
	\begin{equation}
	\delta g_{ab}=\delta \sigma g_{ab}\oplus \text{range} P  \oplus  \text{Ker} P^\dagger\label{metpersplit}.
	\end{equation}
	Note that this is an orthogonal decomposition with respect to the inner product, eq.\eqref{metnrm}
	In eq.(\ref{metpersplit})  $\delta \sigma$ is a perturbation in the conformal factor, $P$ is an operator acting on vector fields, $V$, as
	\begin{equation}
	(PV)_{ab}=\nabla_a V_b+\nabla_b V_a-\nabla\cdot V g_{ab}\label{popdef}
	\end{equation}
	and $P^\dagger$, which is the adjoint of $P$, acts on traceless metric perturbations as
	\begin{equation}
	(P_1^\dagger\delta g)_b=-2\nabla^a\delta g_{ab}\label{P1daggerdef}
	\end{equation} 
	It is well known that the kernel of $P^\dagger$ in general  corresponds to moduli, which together with the conformal factor then determine the metric, upto coordinate transformations. 
	In fact Eq.\eqref{metpersplit} is  the  statement that any perturbation around a given metric can be written as a combination of an infinitesimal conformal transformation, an infinitesimal diffeomorphism and an infinitesimal change in the moduli. The Kernel of $P^\dagger$ vanishes for the disk since it has no moduli.

	The orthogonal decomposition in eq.(\ref{metpersplit} ) means that the measure for summing over metrics  can then be written as 
	\be
	\label{measurement}
	Dg_{ab}=  D [\sigma]    D[PV]
	\ee
	Here $D[\sigma]$ involves the volume element in the space of conformal deformations  which arises from the inner product,  eq.(\ref{metnrm}). 
	For  deformations, $\delta_1 \sigma, \delta_2 \sigma$  eq.(\ref{metnrm}), this takes the form,   
	\be
	\label{inner sigma}
	(\delta_1 \sigma, \delta_2 \sigma)=\int d^2x \sqrt{g} \delta_1 \sigma \delta_2 \sigma
	\ee
	Similarly the measure $D[PV]$ includes the volume element which arises from eq.(\ref{metnrm}) for the metric perturbations of the form $\delta g_{ab}= (PV)_{ab}$.
	
	We now come to the main new element in this problem. The set of diffeomorphisms we sum over, whose measure we schematically denoted as $D[PV]$ above, includes both ``small" and ``large" diffeomorphisms as mentioned above. 
	Small diffeomorphisms leave the boundary unchanged and roughly speaking ``fall off fast enough" towards the boundary. 
	Large diffeomorphisms in contrast do not leave the boundary unchanged, in fact they can be thought of as modes which describe the  fluctuations of the boundary. 
	
	Before proceeding, let us note that in general the space of vector fields on the disk also has a natural inner product given by 
	\begin{equation}
	\langle \delta_1 V, \delta_2 V\rangle=\int d^2 x \sqrt{g} g_{ab}\delta_1 V^a \delta_2 V^b\label{vectinp}
	\end{equation}
	For $P^\dagger$ to be the adjoint of $P$ it is easy to see that a boundary term 
	\be
	\label{bdtermpd}
	BT=2\int d^2 x\nabla_c(\sqrt{g}g^{ac}g^{bd}\delta g_{ab} V_d)
	\ee
	which arises during the manipulation
	\be
	\label{bat}
	\langle \delta g_{ab},P_1V\rangle= 2\int d^2 x\nabla_c(\sqrt{g}g^{ac}g^{bd}\delta g_{ab} V_d) + \langle (P_1^\dagger \delta g)_a,V_b\rangle
	\ee
	must vanish. 
	
	For the boundary term eq.(\ref{bdtermpd})  to vanish,  the vector field $V_a$ must satisfy appropriate boundary conditions. 
	We choose the small diffeomorphisms to correspond to vector fields which satisfy the following two boundary conditions, 
	\begin{align}
	n_a V^a&=0,\nonumber\\ t^a n^b PV_{ab}&=0\label{veclapdetbc}
	\end{align}
	where $t^a, n^b$ are the tangent and normal vector to the boundary respectively. It is easy to see that the first condition ensures that the boundary remains unchanged and together  the two  boundary  conditions ensure that the 
	the boundary term  vanishes for $\delta g=PV$ in eq.\eqref{bdtermpd}.  Acting on the space of all such small diffeomorphisms $P^\dagger P$ is therefore an adjoint operator.
		
	As was mentioned above,  the small diffeomorphisms, which we have now defined precisely, correspond to the gauge transformations and therefore    $\text{Vol}(\Omega)$, in 
	eq.(\ref{piexp})
	is given by 
	\be
	\label{cola}
	\text{Vol}(\Omega)= \text{Vol}(\text{sDiffeo})
	\ee
	where $\text{Vol}(\text{sDiffeo})$  denotes the volume of the group generated by the small diffeomorphisms.

	The additional large diffeomorphisms we would like to include arise from zero modes of 
	$P^\dagger P$. We turn to describing them next. 
	For now then putting together all the information we have acquired so far  the partition function for the  disk topology is given by 
	\be
	\label{part}
	Z_{JT}=\int   D[\phi] D[\sigma] { { D[PV] }  \over \text{Vol}(\text{sDiffeo}) }  e^{-S_{JT}}
	\ee
	where  $D[PV]$ together refer to the measure for the sum  over the small and  large diffeomorphisms.

	\subsection{ Large diffeomorphisms}
	\label{ldiffs}
	Physically, as has already been noted \cite{Maldacena:2016upp},  one can think of the large diffeomorphisms as follows. Consider carrying out the path integral by first  fixing a metric, summing over all configurations of  the dilaton for this metric, and then summing over all metrics. As the dilaton varies, due to the boundary condition that $\phi=\phi_B$ on the boundary,  the boundary must also fluctuate. 
	The diffeomorphisms we are including correspond to these fluctuations of the boundary and they can be thought of as different ways of cutting out a single connected component, meeting our boundary conditions,  from a given disk geometry. In particular, we will  consider such diffeomorphisms which preserve the boundary length to be $l$.

	We will see in the next subsection that on carrying out the path integral for the dilaton first, along the contour we choose, we obtain a delta function constraint that localizes
	the metric path integral to geometries with constant curvature $R=-2$. We restrict ourselves to describing the large diffeomorphisms for such a geometry here.  
	
	In general, any vector field on the disk can be written in terms of two scalar fields $\xi, \psi$ as 
	\be
	\label{veca}
	V= d \xi + * d \psi
	\ee
	
	For a constant curvature metric with $R=-2$ it is easy to see that zero modes of $P$, and therefore  of $P^\dagger P$,  arise from  scalars $\psi, \phi$ which satisfy the equation
	\begin{align}
	\label{laps}
	\nabla^2 \psi  & = 2 \psi,\nonumber\\
	\nabla^2 \xi  & =  2 \xi,
	\end{align}
	
	The large diffeomorphisms arise from modes where $\xi=0$, with $\psi$  satisfying eq.(\ref{laps}). To be more explicit, take the   metric for $AdS_2$ in ``polar coordinates" given by, 
	\be
	\label{metads2}
	ds^2={dr^2 \over (r^2-1)} + (r^2-1) d\theta^2 
	\ee
	The $\theta$ coordinate is periodic $\theta\in [0,2\pi]$ and can be thought of as the Euclidean time direction. 
	Solutions to eq.(\ref{laps}) in this coordinate system, with $\psi \sim e^{i m \theta}$, which are regular at the origin, $r=1$  take the form   
	\be
	\label{psiform}
	\psi_m={\hat c}_m e^{im\theta}  (r+|m|)\left({r-1\over r+1}\right)^{|m| \over 2}
	\ee
	The modes with $m=0,1,-1$ give rise to killing vectors, corresponding to the $SL(2,R)$ isometries of $AdS_2$.

	For other values $\abs{m}>1$ we get zero modes of $P^\dagger P$ which correspond to the large diffeomorphisms of interest. 
	The corresponding vector field for  $\psi$ in eq.(\ref{psiform})  is  given by, for $V_{L,m}^a=(V_{L,m}^r, V_{L,m}^\theta)$ by  
	\begin{align}
	V_{L,m}^a
	=&\hat{c}_{m} e^{i m \theta  } \left(\frac{r-1}{r+1}\right)^{\frac{\abs{m} }{2}}\pqty{i  m  (\abs{m}+r),- \frac{\left(\abs{m}  (\abs{m} +r)+r^2-1\right)}{r^2-1}}\label{vecask}
	\end{align}
	where the subscript $L,m$ denotes that it is a large diffeomorphism with mode number $m$. The resulting metric perturbations are 
	\begin{align}
	&\delta \hat{g}_{rr}=\frac{2 i \hat{c}_{m} m \left(m^2-1\right) e^{i m \theta  } }{\left(r^2-1\right)^2}\left(\frac{r-1}{r+1}\right)^{\frac{\abs{m} }{2}}\nonumber\\
	&\delta \hat{g}_{r\theta}=-\frac{2 \hat{c}_{m} \left(m^2-1\right) \abs{m}  e^{i m \theta  } }{r^2-1}\left(\frac{r-1}{r+1}\right)^{\frac{\abs{m} }{2}}\nonumber\\
	&\delta \hat{g}_{\theta\theta}=-2 i \hat{c}_{m} m \left(m^2-1\right) e^{i m \theta  } \left(\frac{r-1}{r+1}\right)^{\frac{\abs{m} }{2}}\label{crdgvc}
	\end{align}
	Note that these large diffeomorphisms do not satisfy the boundary conditions eq.\eqref{veclapdetbc} in general. 
	
	For $r\rightarrow \infty$ the vector field takes the form,
	
	\begin{equation}
	V^{a}_{L,m}=\pqty{i \hat{c}_{m} m r e^{i m \theta  }+O\left(\frac{1}{r}\right),-\hat{c}_{m} e^{i m \theta  }+O\left(\frac{1}{r^2}\right)}\label{vcuasy}
	\end{equation}
	In particular, since  the $\theta$ coordinate transforms  as $\theta\rightarrow \theta + V^\theta$; 
	the vector field  generates reparametrizations of the $\theta$ coordinate  (Euclidean time) in this limit. 
	Also note that the metric in the large $r$ limit is given by 
	\begin{align}
	\delta \hat{g}_{rr}&=\frac{2 i \hat{c}_{m} m \left(m^2-1\right) e^{i m \theta  }}{r^4}+O\left(\frac{1}{r^5}\right)\nonumber\\
	\delta  \hat{g}_{r\theta}&=-\frac{2 \hat{c}_{m} \left(m^2-1\right) \abs{m}  e^{i m \theta  }}{r^2}+O\left(\frac{1}{r^3}\right)\nonumber\\
	\delta \hat{g}_{\theta\theta}&=-2 i \hat{c}_{m} m \left(m^2-1\right) e^{i m \theta  }+\order\left({1\over r}\right)\label{larrdg}
	\end{align}
	and we see that $ \delta{\hat g}_{\theta r}$ and the fractional change in the components ${\delta {\hat{ g}}_{rr}\over {\hat{ g}}_{rr}}, {\delta{\hat{ g}}_{\theta\theta} \over {\hat{ g}}_{\theta\theta}}$ vanish in this limit. As a result  these diffeomorphisms give rise to asymptotic isometries, in the limit $r\rightarrow \infty$. Note that the requirement that the vector field and the associated metric perturbations be real gives the condition 
	\begin{align}
	\hat{c}_{-m}=\hat{c}_m^*\label{ldfcoefreal}
	\end{align}
	
	The action for the metric perturbations generated by the large diffeomorphisms stays finite even in the asymptotic limit when the 
	 dilaton and total length scaling like eq.(\ref{scalelength}),  as is well known and as we will  also see below in section \ref{asymadslim}. The physical reason for this is the low-dimension of spacetime  we are working in here and the fact that the diffeomorphisms are asymptotic isometries in this limit. 
	Due to their finite action these large diffeomorphisms need to be included in the path integral. 
	
	To be very explicit, for the metric, eq.(\ref{metads2}),  we note that  the boundary of length $l$ is located at $r=r_B$, {for the on-shell configuration for which the dilaton is proportional to $r$,\eqref{valphio}}, where
	\be
	\label{caleb}
	\sqrt{r_B^2-1}  = {l\over 2 \pi}
	\ee
	when $l\gg1$ we have $r_B \simeq {l \over 2 \pi} \gg1$. Once a  diffeomorphism is turned on we go to new coordinates 
	${\tilde r}= r+V^r, {\tilde \theta}=\theta+ V^\theta$, where $(V^r, V^\theta)$ is the vector field leading to the diffeomorphism. The boundary will now be located at ${\tilde r}=r_B$ and so
	\begin{align}
	r=r_B-V^r \label{rudisk}
	\end{align}
	at the boundary.
	For large diffeomorphisms where $V\cdot n$ does not vanish, unlike for small diffeomorphism, eq.(\ref{veclapdetbc}), the boundary will change. Let us also note that 
	for the diffeomorphisms eq.(\ref{vcuasy}) with $|m|>1$ the boundary length does not change to linear order in ${\hat c}_m$ showing that these give rise to length preserving diffeomorphisms at the boundary. 	In the asymptotic AdS limit, eq.\eqref{scaled} and \eqref{scalelength} we have that 
	\be	\label{relive}
	r_B \simeq {1\over \epsilon}
	\ee
	and we see that $r_B \rightarrow \infty$. 
	
	More generally, away from the asymptotic AdS limit, when we consider a boundary of finite  length $l$ and finite  boundary value of the dilaton $\phi_B$, the large diffeomorphisms continue to be  give rise to physically distinct geometries   and we need to    include these  modes in the path integral in the general case as well. 
	
	The resulting measure in the space of small and large diffeomorphisms is actually quite complicated in general.
	This is because the inner product which follows from eq.\eqref{metnrm} is not orthogonal between the small and large diffeomorphisms, and as a result the metric in the space of diffeomorphisms 
	has off-diagonal components between the large and small diffeomorphisms. 
	In the asymptotic AdS limit though  these  off-diagonal elements vanish, in a precise manner which we estimate below. 
	As a result the measure simplifies allowing the path integral to be explicitly carried out. More generally, for fixed $\phi_B,l$ carrying out the path integral is more challenging. 
	
	To estimate how the off-diagonal components in the space of diffeomorphisms vanishes in the limit when $r_B \rightarrow  \infty$  let us first consider the diagonal components of the metric. 
	Starting from eq.\eqref{popdef} for two small diffeomorphisms $V_{s_1}, V_{s_2}$ meeting boundary conditions eq.\eqref{veclapdetbc} we get that 
	\be
	\label{innersmalld}
	\langle PV_{s_1}, P V_{s_2}\rangle = \langle V_{s_1}, P^\dagger P V_{s_2}\rangle .
	\ee
	The subscript $s$ in the diffeomorphisms is to indicate that it is a small diffeomorphism. 
	Note that the inner product on the left is between two metric deformations, eq.\eqref{metnrm}, and on the right between two vector fields, eq.\eqref{vectinp}. The subscripts $s_i$ is to indicate that the vector fields correspond to small diffeomorphisms.
	On the other hand, the inner product between two large diffeomorphisms  can be written as a boundary term, since they are zero modes of $P^\dagger P$. With metric, eq.(\ref{metads2}) and boundary at $r= r_B$, this  takes the form, 
	\be
	\label{innerlarged}
	\langle PV_{L,m_1} PV_{L,m_2}\rangle  =  \delta_{m_1, -m_2} \frac{8 \pi  \hat{c}_{m_1} \hat{c}_{-m_1} \abs{m_1}\left(m_1^2-1\right)  \left(  \left(2 m_1^2+r_B^2-1\right)+2 \abs{m_1} r_B\right)}{r_B^2-1}\left(\frac{r_B-1}{r_B+1}\right)^{\abs{ m_1} }
	\ee
	where $V_{L,m}$ is given in eq.\eqref{vecask}
	and $\delta_{m_1,-m_2}$ is the Kronecker delta symbol. 
	
	A general small diffeomorphism can be decomposed in a basis of vector fields going like $\sim e^{i m \theta}$ which are eigenmodes of $-i \partial_\theta$ and also eigenmodes of $P^\dagger P$ with eigenvalues, $\lambda$. Denoting such a basis element as $V_{s,m,\lambda}$, the inner product between a large diffeomorphism $V_{L,m}$ and $V_{s,\lambda, -m}$ can also be expressed as a boundary term and is given by 
	\be
	\label{inns}
	\langle PV_{s,\lambda,-m}, PV_{L,m}\rangle= \sqrt{\gamma} \, n^r  V_{s,\lambda, -m}^\theta  (PV_{L,m})_{\theta r}\big\vert_\del
	\ee
	where $\sqrt{\gamma}= { \sqrt{r_B^2-1}}$ is the boundary volume element,   $n^r$ is the unit normal and $\vert_\del$ denotes the boundary values at $r=r_B$. 
	It is easy to see that the expression above   does not vanish in general. 
	
	As discussed in the appendix \ref{offdiagme} in the asymptotic AdS limit, when $r_B \rightarrow \infty$, one gets that the  ratio  
	\be
	\label{normalisedjef}
	{\langle PV_{s,\lambda,-m}, PV_{L,m}\rangle \over \sqrt{\langle PV_{s,\lambda,-m}, PV_{s,\lambda,m}\rangle \langle PV_{L,m}, PV_{L,-m}\rangle}} \sim {1\over r_B^{3/2}}
	\ee
	It is this ratio which determines the importance of the off-diagonal terms compared to the diagonal ones in the volume element for the sum over all diffeomorphisms. Since it vanishes we learn that the off-diagonal terms can be neglected when $r_B \rightarrow \infty$ and the measure for the small and large diffeomorphisms decouple. 
	
	We learn that 
	\be
	\label{measurediff}
	\int {D[PV]\over \text{Vol}(\text{sDiffeo})} = \int{ {D[PV_s] D[PV_L]} \over \text{Vol}(\text{sDiffeo})} = \int\sqrt{\text{det}'(P^\dagger P)} D[PV_L].
	\ee
	where in the last equality we have carried out the integral over the small diffeomorphisms since the action is independent of them. The prime in $\text{det}'(P^\dagger P)$ indicates that the zero modes have been removed. In fact, for the disk, with the boundary conditions, eq.\eqref{veclapdetbc}, there are no zero modes, which we we explain in appendix \ref{ckvdisk} . 
	
	The path integral in the asymptotic AdS limit then becomes, 
	\be
	\label{pathia}
	Z_{JT}=\int D[\phi] D [\sigma] D[PV_L]{\sqrt{\text{det}'(P^\dagger P)}} e^{-S_{JT}}
	\ee
	
	From eq.(\ref{innerlarged}) we also learn that when $r_B \rightarrow \infty$ 
	\be
	\label{inlaw}
\langle PV_{L,m}, PV_{L,-m}\rangle= 8 \pi  \hat{c}_{m} \hat{c}_{-m} \abs{m}\left(m^2-1\right)
	\ee
	leading to the measure 
	\be
	\label{mlarged}
	D[PV_L] = \prod_{m\geq 2} d{\hat c}_m d{\hat c}_{m}^* 8 \pi \abs{m} \left(m^2-1\right).
	\ee
	where we used eq.\eqref{ldfcoefreal} in obtaining the above form of the measure. {Note that in writing the above measure we have excluded the modes $m=0,\pm 1$. This is because these modes correspond to the exact isometries of $AdS_2$ and so do not lead to physically distinct configuration. This is the same reason that the integral over time reparametrization modes is taken to be $Diff(S_1)\over SL(2,R)$ with the modding out by $SL(2,R)$ corresponding to the elimination of the three zero modes $m=0,\pm 1$.}
	
	Before closing this subsection let us introduce  a   variable $u$  which is the rescaled proper length along the boundary.{ Using a small diffeomorphism, we can bring the boundary line element to the form,}
	\be
	\label{elements}
	ds^2\bigg\vert_{\del}={du^2\over \epsilon^2},\qquad u\in [0,2\pi]
	\ee
	For the   metric eq.(\ref{metads2}), when the   boundary length $l\rightarrow \infty$, we have near the boundary, where $r\gg 1$, 
	\be
	\label{element}
	ds^2={du^2\over \epsilon^2}\simeq  {dr^2\over r^2}+r^2  d\theta^2
	\ee

	 Using the form  of  the vectors fields specifying a large diffeomorphism $V^r_L, V^\theta_L$ in eq.(\ref{vcuasy}) it is easy to then show that $u={\tilde \theta}$. 
	For infinitesimal transformations we can therefore write 
		\be
	\label{largedi}
	V_L^r= - r\delta \theta'(u), V_L^\theta= \delta \theta(u)
	\ee
	where 
	\be
	\label{default}
	\delta \theta(u)=-\sum_{|m|>1} {\hat c}_m e^{i m \theta(u)}
	\ee
	
	On comparing we  find  that the measure obtained in eq.\eqref{mlarged}  agrees with that which arises in the first order formalism as discussed in \cite{Saad:2019lba} and {\cite{Stanford:2017thb}}. 	Let us also note, as was mentioned above,  that the $\theta$ direction can be thought of as the Euclidean time direction. From eq.(\ref{largedi}) we also see that large diffeomorphisms act as 
	reparametrizations of $\theta$, when $l \rightarrow \infty$. For this reason we will sometimes refer to the large diffeomorphisms as time 
	reparametrizations below. 
	
Indeed, we can generalize the above arguments to an arbitrary intial boundary specified by $\theta=\theta_0(u)$\footnote{We thank the referee for raising a query in this context.}. In this case, we take the vector fields eq.\eqref{largedi} with 
	\begin{align}
	\delta\theta(u)=\theta_0'(u)\epsilon(u)\label{theprelarb}
	\end{align}
	where $\epsilon(u)$ corresponds to the infinitesimal reparametrization of the boundary proper time $u$ as $u\rightarrow u+\epsilon(u)$. In this case boundary is now ``wiggly" and to leading order given by  $r(u)=\frac{1}{\epsilon \theta_0'(u)}$. However the inner product between two large diffeomorphisms specified as given in eq. \eqref{default} remains the same, in turn giving rise to the same measure for the modes $\hat{c}_m$ as in eq.\eqref{mlarged}. It is easy to see that this matches with the result obtained using the symplectic 2-form eq.(2.10) of \cite{Stanford:2017thb} which gives for the inner product of two vector fields as 
\begin{align}
\omega(\delta_1\theta,\delta_2\theta)&=\int_0^{2\pi} du \pqty{ \frac{\del_u\delta_1\theta}{\theta_0'(u)}\del_u\pqty{\frac{\del_u\delta_2\theta(u)}{\theta_0'(u)}}-\delta_{1}\theta(u)\del_u\delta_2\theta(u)}\nonumber\\
&=-\int_0^{2\pi} d\theta\,\, \delta_1\theta\,\pqty{ {{\del_\theta^3\delta_2\theta}}+\del_\theta\delta_2\theta}\label{wittensympmes}
\end{align}
where in obtaining the second line from the first, we have used $\del_u=\theta_0'(u)\del_\theta$ and did an integration by parts.

In this way, we see that the measure obtained in the second order formalism for summing over the boundary time reparametrizations exactly matches with the measure considered in the analysis of the Schwarzian theory \cite{Stanford:2017thb}. Since the action which is given by the Schwarzian term also agrees one can use the result in \cite{Stanford:2017thb},  to argue that the path integral in the second order formalism is also one loop exact.
 
 The fact that the measure for the time reparametrizations in the second order formalism agrees with the symplectic measure of \cite{Stanford:2017thb} can be understood more abstractly. The metric on the space of large diffeomorphisms locally given by eq.\eqref{mlarged} in terms of the holomorphic coordinates $\hat{ c}_m$ is a Hermitian metric. The K\"{a}hler 2-form constructed from this metric matches with the symplectic 2-form constructed in \cite{Stanford:2017thb} and is therefore closed. Thus, the manifold of time reparametrizations is a K\"{a}hler manifold with the Riemannian structure and the symplectic structure being  mutually compatible.

	\subsection{Integral over the Dilaton}
	\label{dilint}
	Having described the diffeomorphisms, small and large, in some detail, and the measure for summing over them we now come back to a systematic evaluation of the path integral. 
	Our starting point is eq.(\ref{part}). We will first carry out the integral over the dilaton, then over the conformal factor $\sigma$ and finally  turn to the sum over the diffeomorphisms. 
	Also, to begin here we will consider the case of a general boundary of length $l$  with the dilaton taking a value $\phi=\phi_B$ on the boundary, and consider the asymptotic AdS limit as a special case in subsection \ref{asymadslim}.

	We  carry out the dilaton path integral in the background of a general metric $g_{ab}$ given by eq.(\ref{confg}). To begin we  write a general dilaton field as 
	\be
	\label{gendi}
	\phi=\phi_0(r) + \delta \phi(r, \theta)
	\ee
	where $\phi_0(r) $ is a fixed function of $r$  given by 
	\be
	\label{valphio}
	\phi_0(r)= A r
	\ee
	The constant $A$ is  fixed to take the value 
	\be
	\label{valA}
	A=\phi_B/r_B
	\ee
	so that for the  metric in eq.\eqref{metads2}   the locus  $r=r_B $  also has the required boundary value of the dilaton $\phi=\phi_B$ with $\delta\phi=0$. 
	
	Let us note in passing that $\phi_0(r)$ satisfies the classical equations of motion in JT gravity.
	
	The path integral for the dilaton  requires us to sum over various values for $\delta \phi$. To carry out this sum we adopt the prescription, now being commonly adopted, \cite{Saad:2019lba},  of rotating the contour for $\delta \phi$ to lie along the  imaginary axis $[-i \infty, i \infty]$ so that after taking $\delta \phi \rightarrow i \delta \phi$, $\delta \phi$ has the range $ [-\infty, \infty]$. 
	Next we expand $\delta \phi$ into eigenmodes of the operator $({\hat \nabla}^2 -2)$ where ${\hat \nabla^2}$ is the scalar laplacian obtained from a metric ${\hat g}_{ab}$ of constant negative curvature ${\hat R}=-2$. 
	We require  that  $\delta \phi$ vanishes at the boundary (at $r=r_B$) so that the dilaton, eq.\eqref{gendi} takes the value $\phi_B$ at the boundary as required. 
	
	With this prescription  the path integral  over the dilaton gives 
	\be
	\label{pathdil}
	\int D[ \phi  ] e^{-S_{JT}}= \delta (R[\sigma]+2) e^{-S_{JT, \partial}}
	\ee
	where  
	\be
	\label{defRsig}
	R[\sigma]=- e^{-2\sigma}(2+ 2 {\hat \nabla}^2\sigma ) 
	\ee
	is the Ricci scalar which arises from the metric eq.(\ref{confg}),  $\delta(R[\sigma] + 2)$,  denotes a delta function which has non-trivial support only when the curvature $R[\sigma]=-2$, and  $S_{JT, \partial}$ is the boundary part of the $JT$ action, eq.\eqref{eadsjtact},  which with $\phi=\phi_B$ takes the form, 
	\be
	\label{bjtact}
	S_{JT,\partial}=-\frac{1}{8\pi G}\int_{\del} dx \sqrt{\gamma}\phi_B (K-1)
	\ee
	After carrying out the dilaton path integral the partition function therefore takes the form, 
	\be
	\label{formpart}
		Z_{JT}=\int {D[\sigma] D[PV]\over \text{Vol}( \rm{sdiffeo} )}\delta (-e^{-2\sigma}(2+ 2 {\hat \nabla}^2\sigma )+2) e^{-S_{JT,\partial}}
	\ee
	
	Let us note before proceeding that one could have considered another  contour for doing the dilaton integral. In fact from the higher dimensional point of view it is perhaps more natural 
	to consider a contour where the dilaton is real with a range $[-\Phi_0,\infty]$, where $\Phi_0$ is the prefactor of the topological term, eq.(\ref{Stop}), since this ensures that the volume of the internal space does not become negative. We will not have anything further to say about such a contour here and leave it for   future consideration. 
	
	We have glossed over one subtlety above. The path integral as we mentioned at the beginning is being done for  a general metric of the form eq.(\ref{confg}) and the measure  for the dilaton integral will therefore depend on $\sigma$ the conformal factor. 
	This measure arises from an inner product which, to begin  with for two dilaton perturbations $\delta_1 \phi, \delta_2 \phi$, takes the form 
	\be
	\label{measured}
	\langle \delta_1 \phi, \delta_2\phi\rangle = \int d^2x \sqrt{g} \delta_1\phi \delta_2 \phi
	\ee
	and  $g_{ab}$ is the metric including the conformal factor. The dependence of the measure on the conformal factor is the same as for a scalar field satisfying Dirichlet boundary conditions, and can be obtained from the conformal anomaly, as is discussed in appendix \ref{deterests}. Since, as we will see shortly below,  satisfying the delta function in eq.\eqref{formpart} results in setting $\sigma=0$,
	this  dependence results  at best in a  constant multiplying the partition function. We can  therefore ignore this subtlety, since we are not   keeping track of the overall multiplicative constant in the partition function. 
	
	\subsection{Integral over the Liouville mode}
	\label{louimodeads}
	Next, we turn to the path integral over the Liouville mode. The  delta function we obtained in eq.(\ref{formpart}) makes this easy. 
		The argument of the delta function manifestly vanishes when $\sigma=0$. Linearizing around it we get 
	\be
	\label{deluxe}
	\delta(e^{-2\sigma}(-2-2 {\hat \nabla^2\sigma}) + 2 )=\delta((-2) ({\hat \nabla}^2-2) \delta \sigma)
	\ee
	
	It is easy to see that in conformal gauge the action, eq.\eqref{eadsjtact}, gives rise to a well-defined variational principle with $\delta \sigma$ vanishing on the boundary (where $\phi_b, l$ are fixed). This is true because we have included   the Gibbons-Hawking boundary term in the action. 
	As a result  we will sum over all Liouville mode fluctuations in the path integral subject to the condition that $\delta \sigma$ vanishes on the boundary.

	Carrying out the integral over the non-zero modes of  $({\hat \nabla}^2-2)$ then gives rise to a factor of $({\text{det}({-\hat \nabla}^2+2)})^{-1}$ in the partition function. 
	The zero mode is fixed by requiring that the boundary has length $l$.
	
	We note that, as for the dilaton above, the  measure $D[\sigma]$ in general has a non-trivial dependence on $\sigma$, since the inner product eq.(\ref{inner sigma}) involves the full metric, eq.(\ref{confg}). However, again,  this dependence  which is the same as for a massless scalar and  can be obtained from the conformal anomaly only gives rise to an overall multiplicative factor in the partition function once we set $\sigma=0$. 
	The resulting determinant $\text{det}(-{\hat \nabla}^2+2)$ then only depends on the metric ${\hat g}_{ab}$.

	Putting all this together gives 
	\be
	\label{format}
	Z_{JT}=\int {D[PV]\over \text{Vol}({\rm sdiffeo})}{1 \over \text{det}({-\hat \nabla}^2 +2)} e^{-S_{JT,\partial}}
	\ee
	Note that the determinant which appears above  depends on the metric ${\hat g}_{ab}$ and therefore on the large diffeomorphisms. 

	\subsection{The asymptotic AdS limit and some remarks}
	\label{asymadslim}
	So far we have been considering the general case of a boundary of finite length. There are two complications in going further with the evaluation of the partition function when the boundary has a  finite length, even for the case when the length is large. First, as was discussed in section \ref{ldiffs} the metric perturbations generated by the small and large diffeomorphisms are not orthogonal and as a result the measure for summing over them is quite complicated to obtain. Second, the dependence of the the determinant $\text{det}(-{\hat \nabla}^2+2)$ on large diffeomorphisms is also not easy to obtain. 
	
	To proceed, we  will  therefore  take the asymptotic AdS limit where we take the length to go to infinity, while also taking the dilaton to diverge at the boundary, as given in eq.\eqref{scaled} and \eqref{scalelength}. 
	Actually this  limit has to be defined more precisely in the path integral where we are  dealing  determinants of various operators. These determinants 	 
	are formally infinite and need to be regulated. We will regulate the determinants by first keeping only the contributions of eigenmodes with finite eigenvalues,  take the asymptotic AdS limit, where $l \rightarrow \infty$,  eq.\eqref{scalelength}, and then finally take the cut-off on the eigenvalues to go to infinity. This order of limits is part of our definition of the asymptotic AdS limit. It will be  responsible for some of the simplification which occurs.
	
	{In the more general case where $l$ is finite we need to regulate the determinants and then take the cut-off introduced for regulating the determinants to zero\footnote{Here we are thinking of the cut-off in the position space. In the momentum space, the cut-off would be taken to infinity.} keeping $l$ fixed. }
	This makes the evaluation of the determinants more complicated. To explain some of the resulting complications  consider evaluating the determinant $\text{det}(-{\hat \nabla}^2+2)$. We cannot use conformal invariance for evaluating this determinant, unlike $\text{det}(-{\hat \nabla}^2)$ which arises for a massless scalar for which at least some information can be obtained,  as we will see in the next subsection.
	A  direct evaluation of $\text{det}(-{\hat \nabla}^2+2)$  is also not easy. For example, consider evaluating this determinant in the  metric eq.(\ref{metads2}). 
	The eigenmodes of $(-{\hat \nabla}^2 +2)$  can also be simultaneously chosen to be eigenmodes of $\partial_\theta$. Denoting  these modes by $\phi_{\lambda,m}$, we have that  $\phi_{\lambda,m} \sim e^{i m \theta}$ and  $({\hat \nabla}^2-2) \phi_{\lambda, m}=- \lambda \phi_{\lambda,m}$.  In the asymptotic AdS limit, as mentioned above,  we first take   the the boundary $r_B\rightarrow \infty$,   keeping $m,\lambda$ fixed,  and then take $m, \lambda \rightarrow \infty$. This means that we are including modes 
	whose wavelength along the $\theta$ direction $\Lambda = r_B/m \gg 1$. Reinstating the radius of AdS, $R_{AdS}$,  in this relation we see that in the asymptotic AdS limit we are only including modes with 
	\be
	\label{inch}
	\Lambda \gg R_{AdS}.
	\ee
	 For such  modes  the asymptotic form for $\phi_{\lambda,m}$ can  be used and this 
	considerably simplifies the analysis. 
	One can then show that the determinant in this limit is  independent of the large diffeomorphisms as discussed in appendix \ref{deterests}. 
	
	In the more general case  when the length is finite, there are modes with $m \ge  r_B$ whose wavelength 
	\be
	\label{inch2}
	\Lambda \le R_{AdS}
	\ee 
	and the  contributions of these modes also need to be included. This is harder to do since we need to include terms with arbitrary number of derivatives  beyond the Schwarzian term. For example these can arise in the action due to the expansion of the extrinsic trace to obtain the analog of  eq.\eqref{nadsqdexk} in the Euclidean AdS disk.
	
	Similarly, there are high wavenumber modes, for  both the  large and small  diffeomorphisms, also with $\Lambda <R_{AdS}$ whose contribution would need to be included at any finite value of $l$. This is again complicated  due to two reasons. First, our estimate that the ratio in eq.(\ref{normalisedjef}), for  the inner product of normalized small and large diffeomorphisms  is suppressed at large $r_B$ is valid only for modes with fixed mode number $m$,   as $r_B\rightarrow \infty$, as discussed further in appendix \ref{offdiagme}. And the mixing between large and small diffeomorphisms  discussed in subsection \ref{ldiffs} above therefore does not vanish for modes with wavelength $\Lambda <R_{AdS}$. Second,  because obtaining the contribution due to such modes,  even after neglecting this mixing,  is not straightforward,  since terms beyond the Schwarzian derivative for the large diffeomorphisms would need to be included, for example for the determinant $\text{det}'(P^\dagger P)$ which arises from the small diffeomorphisms, {see discussion above eq.\eqref{condmoden}. }Let us mention that calculations pertaining to finite value of $l$ have been carried out in the first order formalism \cite{Stanford:2020qhm,Iliesiu:2020zld}\footnote{We thank the referee for pointing this out. }. Although, a perturbative expansion in $\epsilon$ can be done, it is quite challenging for the reasons mentioned above; 
		%including the higher modes and mixing between large and small diffeomporphisms is challenging in the second order formalism
		a careful investigation of the finite cut-off problem is currently underway \cite{Moitra2021}.
	
	As a toy model in appendix \ref{nadsds} we show how these high wavenumber  modes could potentially have a significant effect on the behaviour of the partition function 
	when $l$, the boundary length becomes very big. In the cosmological context which we study below eq.(\ref{inch2}) is replaced by 
	\be
	\label{inch3}
	\Lambda \le H^{-1}
	\ee
	where $H$ is the Hubble constant, and this condition therefore  corresponds to modes which have not yet exited the horizon and ``frozen out". 
	Our analysis shows that such modes  can   significantly affect the wavefunction. 
	
	\subsection{More details on the asymptotic AdS limit} 
	For all  these reasons,  here after  in this subsection we only  consider the asymptotic AdS limit. 	Since the small and large diffeomorphisms become orthogonal in this limit the path integral eq.\eqref{format} gives
	\be
	\label{exzz}
	Z_{JT}=\int D[PV_L] {\sqrt{\text{det}'(P^\dagger P)}\over \text{det}(-{\hat \nabla}^2+2)}e^{-S_{JT,\partial}}
	\ee
The measure for the large diffeomorphisms is given in eq.(\ref{mlarged}). The prime in $\text{det}'(P^\dagger P)$ is to indicate that the zero modes in the space of small diffeomorphisms of the operator $P^\dagger P$ are to be excluded in calculating the determinant. There is in fact one  zero mode for the operator $P^\dagger P$ satisfying the boundary conditions eq.\eqref{veclapdetbc}. More discussion on this zero mode is contained  in the appendix \ref{asyadsdet} above eq.\eqref{zeromodecol}.

In the asymptotic AdS limit as we have defined it above one can show that both $\text{det}'(P^\dagger P)$ and $\text{det}(-{\hat \nabla}^2+2)$ become independent of the large diffeomorphisms as is discussed in appendix \ref{deterests}.
Upto a multiplicative constant which we are not keeping track of we then get  
\be
\label{assays}
Z_{JT}= \int D[PV_L] e^{-S_{JT,\partial}}
\ee
This final expression agrees completely with what has been obtained from the first order formalism, \cite{Saad:2019lba}.

For completeness let us carry out the remaining integral over $V_L$ here. As is well known fact the boundary action  $S_{JT,\partial}$ gives rise to the Schwarzian term involving the time reparametrization generated by the large diffeomorphism as follows. As discussed in appendix \ref{etlds2} the extrinsic curvature for a general boundary curve specified as $(r(u),\theta(u))$ is given by 
	\be
	\label{extcur}
	K=\frac{ r(u)-\epsilon ^2 r''(u)}{({r(u)^2-1}) {\epsilon  } \theta '(u)}
	\ee
	where $r(u)$ is the radial coordinate along the boundary as a function of proper time $u$ and prime denotes a derivative with respect to $u$. 
	From eq.\eqref{element} we get that upto corrections sub-leading in $r_B$
	\be
	\label{condo}
	{d\theta \over d u}= {1\over \epsilon} {1\over r(u)}
	\ee
	where $r(u)$ is given in eq.(\ref{rudisk}) with 
	\be
	\label{valve}
	V_L^r(u)=i r  \sum_m {\hat c}_m m e^{i m \theta(u)}
	\ee
	Substituting gives, 
	\be
	\label{subK}
	K=1+\frac{ \left(-3 \theta ''^2+\theta '^4+2 \theta ^{(3)} \theta '\right)}{2 \theta '^2}\epsilon^2+O\left(\epsilon^{3}\right)
	\ee
	The net result is the path integral eq.\eqref{assays} with action 
	\begin{align}
	\label{acts}
	S_{JT,\partial}  &= -\frac{\phi_B\epsilon}{8\pi G}\int_{0}^{2\pi} du \,\,\text{Sch}\pqty{\tan(\theta(u)\over 2),u}\nonumber\\
	& =-\frac{\pi}{4  GJ\beta}\pqty{1 +2\sum_{m\geq 2 }(m^2-m^4)\hat{c}_{m}\hat{c}_{-m}}
	\end{align}
	and measure eq.\eqref{mlarged}.
	This agrees with the result obtained earlier.
	In particular note that the measure we have obtained from the second order formalism above agrees with the measure obtained in \cite{Stanford:2017thb} see also \cite{Saad:2019lba}. 
	
	The integral over the modes $\hat{c}_m$ is in fact one-loop exact \cite{Stanford:2017thb}. Using the measure eq.\eqref{mlarged}, the action eq.\eqref{acts} and noting eq.\eqref{ldfcoefreal}, we have the path integral as 
	\begin{align}
	Z_{JT}&=e^{\pi(4GJ\beta)^{-1}}\int \prod_{m\geq 2}d\hat{c}_m d\hat{c}_m^*8\pi \abs{ m}(m^2-1)\exp\left[\frac{\pi}{2  GJ\beta}\sum_{m\geq 2 }(m^2-m^4)\hat{c}_{m}\hat{c}_{m}^*\right]
	\nonumber\\
	&=e^{\pi{(4GJ\beta)}^{-1}}\prod_{m\geq 2}\frac{32\pi GJ\beta}{m}\nonumber\\
	&=e^{\pi{(4GJ\beta)}^{-1}}\frac{(32\pi GJ\beta)^{-3/2}}{\sqrt{2\pi}}\label{adspires	}
	\end{align}
	
	Adding the topological term eq.\eqref{Stop} for completeness gives the partition function in the asymptotic AdS limit to be 

	\begin{align}
	Z_{JT}&=\exp\left[{{\Phi_0\over 4G}+{\pi\over 4GJ\beta}}\right]\frac{(32\pi GJ\beta)^{-3/2}}{\sqrt{2\pi}}\label{zschtop}
	\end{align}

	Let us conclude this section with a  remark. 
	As mentioned above for the boundary located at a finite value for  the length $l$ with  dilaton  taking value $\phi_B$, 
	the path integral we have defined is still quite explicit, eq.\eqref{part}, but much harder to fully  evaluate. This is true even when the boundary length $l \gg 1$, which one might expect  is simpler than that of the general case. 
	 We hope to return to this issue and also  to  the analogous one   in   dS JT gravity,  where it is related to computing the wavefunction at late but finite time  after including modes which have not yet exited the horizon, in the future.   In dS JT, the finite time wavefunction was studied in \cite{Iliesiu:2020zld}.

	\section{JT path integral with matter in AdS}
	\label{jtwithmatterads2}
	We shall  next extend the analysis of the previous section to include  additional scalar massless matter fields. This problem is also studied in \cite{Yang:2018gdb}. The path integral is given by 
	\begin{equation}
	Z_{JT+M}=	\int {D}[\phi] \frac{{D}[g_{\mu\nu}]}{\text{Vol}(\text{sdiffeo})}\pqty{\prod_{i=1}^{N}D[{\varphi}_i]} \exp{- S_{JT}-S_M}\label{madspiexp}
	\end{equation}
	where $S_{JT}$ is the same action for the JT theory as before,  the measure ${D} \phi, {D} g_{\mu\nu}$ and the volume of small diffeomorphisms, 
	$\text{Vol}(\text{sdiffeo})$ are the same as above   and $S_M$ is the action for the minimally coupled massless scalar fields, $\varphi_i$, given by 
	\begin{align}
	S_M=\sum_{i=1}^{N}\half \int d^2x\sqrt{g}(\del\varphi_i)^2\label{matact}
	\end{align}
	where $N$ is the number of matter fields.
	As can be seen from the above action, the matter fields do not directly couple to the dilaton. We will carry out the path integral for  fixed boundary values for the scalar fields,
	\begin{align}
	\varphi_i\big\vert_\del=\hat{\varphi}_i (u)\label{matterbc}
	\end{align}
	with $u$ being, upto a multiplicative constant,  the proper length along the boundary, eq.\eqref{elements}.
	The resulting partition function is a functional of ${\hat \varphi}_i(u)$, besides being a function of the length $l$ and the boundary value of the dilaton $\phi_B$, as before. We discuss the general case of finite $l,\phi_B$ first and then turn to the asymptotic AdS limit below. 
	
	Working  in conformal gauge  we can  carry out the integral over the dilaton and the Liouville modes.  
	Since the matter fields do not couple to the dilaton  directly   the dilaton integral will   localize the path integral to constant negative curvature metrics as before and allow us to   set the  the Liouville mode $\sigma$ in eq.(\ref{confg})  to vanish.  After the Liouville mode integral is done we are  then left with the integral over  diffeomorphisms and the matter fields, giving
	\be
	\label{valZ}
	Z_{JT+M}=\int {D[PV] \pqty{\prod_{i=1}^{N}D[{\varphi}_i]}\over \text{Vol}(\text{sdiffeo})}  {1 \over \text{det}(-\hat{\nabla}^2+2)} e^{-S_{JT,\del}- S_M}
	\ee 
	The measure for the scalar fields in this path integral is to be evaluated  using a metric ${\hat g}_{ab}$, with curvature ${\hat {R}}=-2$.
	This measure follows from the standard ultra local inner product for two scalar perturbations   given by 
	\be
	\label{inroads}
	(\delta_1 {\varphi_i}, \delta_2{\varphi_i})=\int d^2x\sqrt{\hat g}  \ \delta_1{\varphi_i}\delta_2{\varphi_i}
	\ee
	Thus   the  
	background geometry for the scalar path integral is hyperbolic space with a boundary determined by the large diffeomorphisms. 
	
	To perform the path integral over the fields $\varphi_i$, we first expand it around the classical solution   obtained by solving the scalar laplacian equation with the boundary condition specified by eq.\eqref{matterbc} and also demanding that the solution is regular everywhere in the interior. 
	
	Let us denote the  resulting solution  to be ${ \varphi}^{(0)}_i(r,\theta)$. 
	Expanding the fields $\varphi_i$ around this solution as 
	\begin{align}
	\varphi_i= \varphi_i^{(0)}+\delta{\varphi}_i\label{matspexp}
	\end{align}
	The boundary condition eq.\eqref{matterbc} translates to the  Dirichlet condition,
	\begin{align}
	\delta{\varphi}_i\vert_\del=0\label{matterspbc}
	\end{align}
	We can write the path integral for the matter fields then as
	\be
	\label{pics}
	Z_M= \int \pqty{\prod_{i=1}^{N}D[{\varphi}_i]}e^{-S_M}= e^{-S_{M,cl}} \int \pqty{\prod_{i=1}^{N}D[{\delta\varphi}_i] e^{\half\int d^2 x \sqrt{\hat g} {\hat g}^{ab} \partial _a {\delta \varphi_i} \partial_b {\delta \varphi_i}}}
	\ee
	where $S_{M,cl}$, the classical contribution resulting from ${ \varphi}^{(0)}_i$,   is given after using the equations of motion, by a boundary term, 
	\be
	\label{class}
	S_{M,cl}= \sum_i\half \int_{\partial} ds \sqrt{\gamma} { \varphi}^{(0)}_i \partial_n { \varphi}^{(0)}_i
	\ee
	with $\partial_n$ being the normal derivative at the boundary. 
	Note that the Laplacian $\hat{\nabla}^2$ has no zero modes for the Dirichlet boundary conditions satisfied by ${\delta \varphi}_i$. Thus the path integral over ${\delta  \varphi}_i$ is straightforward and gives, 
	\be
	\label{pisb}
	Z_M= \int \pqty{\prod_{i=1}^{N}D[{\delta\varphi}_i]}e^{-S_M}= {e^{-S_{M,cl}}\over \left(\text{det}\left(-{\hat{\nabla}^2\over 2}\right)\right)^{N \over 2}}
	\ee
	In much of the discussion later in this paper, we will drop the factor of $\half$ that appears in the determinant in the above expression as it will only change the overall numerical coefficient of the path integral which we are not keeping track of. 
	Both $S_{M,cl}$ and the determinant on the RHS depend on the large diffeomorphisms.	This dependence  is not easy to  obtain, for the general case of a finite boundary of length $l$,   as is discussed in appendix {\ref{deterests}}. The reason, related to the discussion  towards the end of the subsection \ref{asymadslim}, is the presence of high wave number modes with wavelength less than the boundary length,   $\Lambda <R_{AdS}$. The Schwarzian action is no longer sufficient to describe the dependence on the large diffeomorphisms for such modes. In addition, as is also discussed in subsection \ref{ldiffs}, the subsequent step involving the integral over the diffeomorphisms is also  not easy to carry out in  this case. 
	
	Keeping these points  in mind we again  restrict ourselves to
	 the asymptotic AdS limit  for the subsequent evaluation of the path integral.  As discussed in appendix \ref{deterests} the dependence of the large diffeomorphisms in $\text{det}(-{\hat \nabla}^2)$ vanishes in the  asymptotic AdS limit when $r_B \rightarrow \infty$, after a suitable length dependent counter term is added. 
	 If  the boundary values of the scalars ${\varphi}_i^{(0)}$ vanish the path integral is therefore unchanged (upto an overall temperature independent prefactor) by the presence of the matter in the asymptotic AdS limit. And  the thermodynamics essentially does not change,  other than a possible change in the ground state entropy.

	 When the boundary values ${\varphi}_i^{(0)}$ are non-zero, the matter sector does  couple to the large diffeomorphisms. 
	   In appendix \ref{mcttrp}, eq.\eqref{ass4},  we show that  the dependence of $S_{M,cl}$ is given, for  $r_B \rightarrow \infty$,  by  
		\be
	\label{fulcra}
	S_{M,cl}= {1\over 2} \int du_1 du_2 \theta'(u_1) \theta'(u_2) \sum_i{\hat \varphi}_i(u_1) {\hat \varphi}_i(u_2) F(\theta(u_1), \theta(u_2) )
	\ee
	Here $\theta(u)$ specifies the time reparametrization as a function of the boundary proper length $u$, and $\theta'={d\theta\over du}$. 
	${\hat \varphi}_i(u)$  is the value of the scalars along  the boundary  and the function $F$ is defined in eq.\eqref{Fdef}. 
	At linear order in $\delta \theta(u)$ this becomes, 
	\be
	\label{links}
	S_{M,cl}= \int_\del du_1 du_2 \hat{\varphi}(u_1)\hat{\varphi}(u_2)(\delta\theta'(u_1)F(u_1,u_2)+\delta\theta(u_1)\del_{u_1}F(u_1,u_2))
	\ee
	Upto an overall constant  the path integral then  takes the form,
	\be
	\label{new part}
	Z_{JT+M}=\int D[PV_L] e^{-S_{JT,\del}} e^{-S_{M,cl}}
	\ee
	The measure $D[PV_L]$ is defined in eq.\eqref{mlarged}, and the action $S_{JT,\del}$ involves the Schwarzian derivative of $\theta(u)$, 
	  eq.\eqref{acts}. 
	The path integral can then be done by integrating out the large diffeomorphisms perturbatively, including the self interactions from the Schwarzian  term and the interactions with the matter fields, to obtain $Z_{JT+M}$ as a function of the boundary values of the scalar fields ${\hat \varphi}_i(u)$, $\beta$ and $J$.  We will not go into the details here. 
	These calculations are also discussed in \cite{Maldacena:2016upp}. 
	
	\subsection{Further remarks}
	\label{eadsmfr}
	Let us end this section with some comments.
	We have seen that the path integral at finite  values of the boundary length $l$ is  difficult to calculate even when $l\gg 1$. 
	Some of the reasons for this were mentioned above. 
	On the other hand the quantum effects of matter vanish when $l \rightarrow \infty$ since the matter determinant does not couple to large diffeomorphisms in this limit anymore as was also mentioned above.

	One way to  obtain a tractable situation where quantum effects due to matter can be incorporated is to consider a semi-classical limit
	by taking  $G$, the gravitational constant which appears in front of the JT action eq.\eqref{eadsjtact}, to vanish, $G\rightarrow 0$, with the number of scalar fields, $N\rightarrow \infty$  keeping $G N$ fixed, {\cite{Moitra:2019xoj}}. In this limit the measure for  the diffeomorphisms is not important since gravity is classical and  quantum fluctuations over these diffeomorphisms can be ignored, similarly the dependence of $\text{det}(-{\hat \nabla}^2+2)$ on the large diffeomorphisms can be neglected.  	However  the quantum effects of matter  remain. This limit has received considerable attention recently, {\cite{Almheiri:2014cka,Moitra:2019xoj,Almheiri:2019psf,Engelsoy:2016xyb}}. 
	The saddle point equations in this limit for the system we are considering were obtained in \cite{Moitra:2019xoj}. It was found that they can typically be solved only in slowly varying  situations where the excited modes  have wavelengths $\Lambda \gg R_{AdS}$ as discussed in \cite{Moitra:2019xoj}. 
	
	More generally one could consider a system away from the semi-classical limit, with a finite  number of matter  fields, where we are interested in the response to slowly varying sources provided for example by the boundary values $\hat{\varphi}_i^{(0)}$.  The sources are varying at wavelength 
	\be
	\label{condwave}
	\Lambda \gg R_{AdS}.
	\ee
	 In this case one can consider constructing a Wilsonian effective action which will contain the sources coupled to the large diffeomorphisms by integrating out the other degrees of freedom. The determinants which arise must be valued in $\text{Diff}(S^1)/SL(2,R)$ and can be expanded in a derivative expansion. The leading term in this expansion which depends on the large diffeomorphisms is the Schwarzian term, other terms  involve more derivatives and would be suppressed when eq.(\ref{condwave}) is met. 
	   
	The  resulting effective  action, after adding a suitable counter term to cancel a boundary length dependent  term,  is then  given by 
	\be
	\label{pisources}
Z_{JT+M}[\beta,J,{\hat \varphi}_i(u)]= 	\int D[PV_L]  e^{-S}
\ee
where the action, see appendix \ref{deterests},  is 
	\be
	\label{fullacta}
	S= { \epsilon \over  8 \pi G} C  \int du \,\text{Sch}(\tan(\theta(u)/2), u) + S_{M,cl}.
	\ee
	 $S_{M,cl}$  above arises from the quadratic action of the scalar fields and depends on the boundary values ${\hat \varphi}_i(u)$ and time reparametrization $\theta(u)$; to leading order it is  is given  in eq.(\ref{fulcra}), a correction at $O(\epsilon)$ can also be similarly obtained.  The measure in eq.(\ref{pisources}) is the $\text{Diff}(S^1)/SL(2,R)$ invariant measure given in eq.\eqref{mlarged} above.  The coefficient $C$  in front of the 
	 Schwarzian action to begin with, before the  short wavelength modes  have been integrated out,   is given by 
	 \be
	 \label{coeffc}
	 C=-\phi_B+ G (N - 26  +  q_2).
	 \ee
	 The first term on the RHS, $\phi_B$,  is  from the classical JT action. The matter fields  contribute the  second term, $G N$, eq.\eqref{detinlesch};  the third term,  $-26 G$,  comes from $\text{det}'(P^\dagger P)$, eq.\eqref{pdpfinres} and the term with coefficient $q_2$ which we have not been able to determine and should be of order unity,  arises from $\text{det}(-\hat{\nabla}^2+2)$, see eq.\eqref{delsqm2anz}. The factor of $\epsilon$ multiplying the Schwarzian shows that the effect of the matter determinant etc vanishes when $\epsilon \rightarrow 0$.  Note that  the  coefficient $C$  will be renormalized from this starting value  though,  once we integrate  out the short wavelength modes. 
	
	This Wilsonian effective action can then be used for calculating the long-wavelength properties of the system, including computing loop effects from modes meeting eq.(\ref{condwave}).  
	If necessary a  renormalization procedure  can be  carried out  to make such calculations optimal. We leave a further analysis along these lines for the future.

\section{Double trumpet in AdS}
\label{dbads}
In this section we will extend our discussion to consider  the path integral over connected geometries with two boundaries in Euclidean AdS.
 These spaces  have Euler character   $\chi=0$. Such a spacetime  is often referred to as the double trumpet geometry. The path integral, denoted as $Z_{DT}$ reads
\begin{align}
Z_{DT}=\int  \frac{{D}[\phi]{D}[g_{\mu\nu}]}{\text{Vol}(\Omega)} \exp{- S_{JT}}\label{dbpiexp}
\end{align}
The action for the path integral is given as in eq.\eqref{eadsjtact} above, with boundary terms at both boundaries.
 Note that in this case the boundary contribution to the path integral  will arise from both boundaries. 
The non-trivial part of the calculation, like for the disk, is to correctly identify the metric configurations   which need to be included in the path integral and obtain a measure for  summing over them. 
We discuss  this issue    first here and thereafter in the next subsection will carry out the path integral in a systematic manner, analogous to section \ref{purejtads2}, by first summing over the dilaton,  the conformal mode and then the diffeomorphisms. 

The boundary conditions we impose are that $\phi$ takes values $ \phi_{B,1}, \phi_{B,2},$ at  the two boundaries which are taken to have lengths $l_1, l_2$ respectively. 
It will be convenient to keep in mind the following background metric for the double trumpet, which has curvature $R=-2$, with two   boundaries,  at $r \rightarrow  \infty$, $r\rightarrow -\infty$ (henceforth also referred to right and left boundaries respectively):
\begin{align}
ds^2={dr^2\over r^2+1}+(r^2+1)d\theta^2,\quad \theta\sim \theta+b\label{dbmet}
\end{align}
Note that this metric can be obtained from  eq.\eqref{metads2}  by the analytic continuation,  
\begin{align}
r\rightarrow -ir, \theta\rightarrow i\theta\label{rthac},
\end{align}
however now the periodicity of $\theta$ is a free parameter $b$. This parameter  actually corresponds to a modulus and we will integrate over it in  the path integral, as we will see below. 

While we  first set up the path integral with general boundary conditions, as in the case of the disk, it will turn out to be difficult to carry out the calculations all the way through in this general case. As a result,  at some point in the discussion below we will    specialize to the asymptotic AdS limit.  This limit is  defined for two boundaries by taking  the length of both  boundaries to go to infinity and  also taking $\phi_B \rightarrow \infty$, while keeping the ratio $\phi_B/l $  to be fixed at each boundary. The ratio $\phi_B/l$ takes  an independent value at both boundaries, i.e., we take $\epsilon \rightarrow 0$ with
\begin{align}
\label{limeades2}
\phi_{B,1}=  & {2\pi\over J \beta_1 \epsilon_1},   l_1 = {2 \pi \over \epsilon_1}\nonumber \\
\phi_{B,2}= & {2\pi\over J \beta_2 \epsilon_2},    l_2 = { 2 \pi \over \epsilon_2}
\end{align} 
The resulting answer will then depend on both the parameters ${1\over J \beta_1}$ and ${1\over J \beta_2}$. The asymptotic limit corresponds to taking $\epsilon_1,\epsilon_2\rightarrow 0$. 

The path integral for the disk topology can be interpreted as the partition function of the boundary theory, as was mentioned above. 
After a suitable rescaling of $\epsilon_1, \epsilon_2$ we can take the renormalized  lengths of the two boundaries in the double trumpet to be $\beta_1,\beta_2$ so that the
path integral for the double trumpet geometry can be interpreted as giving  a contribution to the connected   two point   function of the partition functions $\langle Z(\beta_1)Z(\beta_2)\rangle $ for two boundary theories. We will have more to say about this interpretation in section \ref{dbtwm} below. 

In general metric perturbations about a metric $g_{ab}$ can be decomposed, similar to eq.\eqref{metpersplit} in subsection \ref{pjtadsstrategy},  as
\be
\label{decomdt}
\delta g_{ab}= \delta \sigma g_{ab} \oplus \delta {\tilde g}_{ab}
\ee
Here  $\delta {\tilde g}_{ab}$ are traceless metric perturbations which will include, for the double trumpet,  perturbations produced by small diffeomorphisms, large diffeomorphisms and moduli. 
We  describe all three types of perturbations below. 

Perturbations produced by small diffeomorphisms are generated by vector fields $V_s$ and of the form 
\be
\label{small}
\delta {\tilde g}_{ab}= P V_s\ee
where the operator $P$ is given in eq.\eqref{popdef} . The vector fields $V_s$ satisfy the boundary conditions, eq.\eqref{veclapdetbc} at both  the boundaries. These perturbations  describe the same spacetime after a coordinate transformation and therefore give rise to gauge transformations. Their volume $\text{Vol}(\text{sdiffeo})$ is the factor $\text{Vol}(\Omega)$  in  the denominator of eq.\eqref{dbpiexp}. 

The perturbations produced by large diffeomorphisms  describe fluctuations of both boundaries in double trumpet case. They correspond, as in the disk, to zero modes of the operator $P^\dagger P$, with the operator $P^\dagger$ as defined in eq.\eqref{P1daggerdef} acting on traceless metric perturbations.
Denoting a vector field which generates such a transformation by $V_L$ the condition $V_L$ satisfies is 
\be
\label{condee}
P^\dagger P V_L =0
\ee
and the  metric perturbation it produces is  $\delta {\tilde g}_{ab}= (PV_L)_{ab}$. 
Taking a cue from the disk case,  in  identifying   these  we look for diffeomorphisms  which  reduce to reparametrizations of the two boundary circles, in the limit where both boundaries  have large length. Such diffeomorphisms can be obtained by setting 
\be
\label{steel}
V_L  = * d \psi
\ee
with $\psi$  being a scalar field satisfying the eq.\eqref{laps} . 

Solutions to this equation for a background metric eq.\eqref{dbmet} can be obtained from eq.\eqref{psiform} after  noting the analytic continuation eq.(\ref{rthac}) and are given by 
\begin{align}
\psi(r,\theta)=\sum_{|m|>0}e^{i\tilde{m}\theta}\pqty{A_m(r+\tilde{m})\pqty{r+i\over r-i}^{i\tilde{m}\over 2}+B_m(r-\tilde{m})\pqty{r+i\over r-i}^{-i\tilde{m}\over 2}}\label{dblapsol2}
\end{align}
where 
\begin{align}
\tilde{m}=\frac{2\pi m}{b}\label{tmmrel}
\end{align}
Note that we have twice the number of solutions compared to the disk case, since there are two modes for every value of $m$. And  
unlike the disk, there is no condition of regularity in the interior which cuts down the number of solutions since the coordinate system in which the metric  in eq.(\ref{dbmet}) is written is non-singular everywhere. 
We should also mention  that the sum in eq.\eqref{dblapsol2} does not include an  $m=0$ mode. We will have more to say about this sector shortly. 

Before proceeding let us note that the solution in eq.(\ref{dblapsol2}) has functions  involving $(r+i)$ and $(r-i)$ raised to various powers. These are defined, for a general exponent $a$, as follows
\be
\label{defat}
(r+i)^{i a} = \exp[i a \log (r+i)], \ (r-i)^{i a}= \exp[i a \log(r-i)]
\ee
with  the log function in both cases  being  defined to have a branch cut along the negative real axis, i.e.,
\begin{align}
\ln z=\ln \abs{z}+i \text{Arg}(z),\quad \text{Arg}(z)\in [-\pi,\pi]\label{dnlnz}
\end{align}
Also  in our definition, 
 \be
 \label{defat}
\left( {r+i\over r-i}\right)^{i {\tilde m}\over 2}= {(r+i)^{i {\tilde m}\over 2}\over (r-i)^{i {\tilde m}\over 2}} = \exp[i \left({{\tilde m}\over 2} \log(r+i) - {{\tilde m}\over 2}\log(r-i)\right) ]=\exp[- {{\tilde m}\over 2}\left( \text{Arg}(r+i) - \text{Arg}(r-i)\right) ]
 \ee
With these definitions we see that  the reality of  
 $\psi$ imposes the condition 
 \be
 \label{contra}
 A^*_{-m}=B_m
 \ee
 on the coefficients in eq.(\ref{dblapsol2}).

We will sometimes find it convenient to work with linear combinations of the basis elements used in the expansion in eq.(\ref{dblapsol2}), and  rewrite $\psi$ as follows,  
\begin{align}
\psi(r,\theta)=&\sum_{\abs{m}>0}e^{i\tilde{m}\theta}{\gamma_m\pqty{(r+\tilde{m})\pqty{r+i\over r-i}^{i\tilde{m}\over 2}-(r-\tilde{m})\pqty{r+i\over r-i}^{-i\tilde{m}\over 2}}}\nonumber\\
&+\sum_{\abs{m}>0}e^{i\tilde{m}\theta}{\delta_m \pqty{(r+\tilde{m})\pqty{r+i\over r-i}^{i\tilde{m}\over 2}e^{\tilde{m}\pi}-(r-\tilde{m})\pqty{r+i\over r-i}^{-i\tilde{m}\over 2}e^{-\tilde{m}\pi}}}\label{dblapsolrw}
\end{align}
The coefficients $\gamma_m, \delta_m$ are related to $A_m, B_m$ by 
\begin{align}
\gamma_m+e^{\tilde{m}\pi}\delta_m=A_m,\quad \gamma_m+e^{-\tilde{m}\pi}\delta_m=-B_m\label{dbabgdre}.
\end{align}
It is easy to see that the radial functions multiplying   $\gamma_m, \delta_m$  vanish respectively  as $r\rightarrow  \infty$ and $r\rightarrow -\infty$ respectively, to leading order . In this basis it is therefore manifestly clear that the expansion eq.(\ref{dblapsol2})  includes independent perturbations at the two ends.  Using {eq.\eqref{steel}} it is  also easy to see that the resulting diffeomorphisms become independent reparametrizations of the $\theta$ direction at $r\pm \infty$. 

We now turn to the ${ m}=0$ sector. There are two  solutions to eq.\eqref{laps} in this sector, these  are independent of $\theta$ and are given by 
\begin{align}
\psi_1 & = s \, r\label{dbldzm1}   \\
\psi_2 & =   \, {t\over 2\pi}\,   \pqty{2 +i r \log({r+i\over r-i}) }\label{dbldzm2}
\end{align}
where $t, s$ are arbitrary coefficients. 
It is easy to see that $\psi_1$ corresponds to the $U(1)$ isometry under which $\theta\rightarrow \theta-s$. Since it keeps the metric and boundary unchanged it does not correspond a  distinct spacetime, and we must not sum over it in the path integral. 
On the other hand $\psi_2$ give rises to the diffeomorphism
$V_{ tw}=* d \psi_2$. The subscript $m$ is to indicate that it is a modulus and $tw$ is to denote that this vector field introduces a relative twist between the two boundaries. It is easy to see that  $V_{ tw}^r=0$ and 
\be
\label{bee}
V_{tw}^\theta={t\over 2\pi}  \left[ -{2 r \over r^2+1}-i \log({r+i\over r-i})\right]
\ee
As $r\rightarrow \infty$, $V^\theta_{tw}\rightarrow 0$, while as $r\rightarrow -\infty$, $V^\theta_{tw} \rightarrow  t $. Thus this diffeomorphism produces a relative twist between the  $\theta$ variables parametrizing the circles at the two boundaries in the limit when the boundaries have length $l \rightarrow \infty$. 
It is in fact one of the two moduli associated with this geometry. The corresponding metric perturbation is given by 
\begin{align}
(PV_{tw})_{\mu\nu}=-{2t\over\pi  (r^2+1)}\pqty{\begin{matrix}
	0 & 1\\
	1& 0 
	\end{matrix}}\label{dbtwper}
\end{align}

The other modulus for this space is related to the  parameter $b$ which is the size of the $\theta$ circle, eq.\eqref{dbmet}. Consider a vector field 
\be
\label{defame}
V_{b}^\mu= {\delta b \over b}(-r,\theta)
\ee 
under which $\theta \rightarrow \theta+\frac{\delta b}{b}\theta$, so that the periodicity of $\theta$ changes. The subscript $b$ is to denote that this is a vector field corresponding to the modulus parameter $b$.
This vector field is not single valued on the circle, however the metric perturbation it gives rise to,  is  well-defined and single valued, 
\begin{align}
(PV_{b})_{\mu\nu }=\frac{2\delta b}{b}\pqty{ \begin{matrix}
	-\frac{1}{(1+r^2)^2} & 0 \\
	0& 1
	\end{matrix}}.\label{dbmodper}
\end{align}
We  will  also include this metric perturbation in the sum over all configurations in the path integral. Note that both $V_{b}, V_{tw}$ satisfy
\begin{align}
P^\dagger P V_{b}=0=P^\dagger P V_{tw}\label{pdpvbtw}
\end{align}

To summarize the discussion so far  then we have argued  that  general metric deformations which we sum over include changes in the conformal factor and deformations associated with small diffeomorphisms which preserve the boundaries, large diffeomorphisms which changes the boundaries, and two moduli. 

Let us now turn to describe the measure in the space of all metric deformations. This measure arises from the inner product in the space of metric perturbations, eq.\eqref{metnrm} as described in subsection \ref{purejtads2}. 
The decomposition  in eq.(\ref{decomdt}) is an orthogonal one with respect to the inner product eq.\eqref{metnrm}. 
Thus the measure in eq.(\ref{piexp}) can be written as $D[g_{ab}]  = D[\sigma]D[{\tilde g}_{ab}]$ where $D[\sigma]$ is the measure for the sum over conformal factors following from the inner product, eq.(\ref{inner sigma}).

As discussed in appendix \ref{offdiagme} in general the large and small diffeomorphisms   included in $D[{\tilde g}_{ab}]$ are not orthogonal to each other and the resulting measure is hard to obtain, even when the lengths $l_1,l_2$ of the two boundaries are large but finite. This is because modes of wave number ${\tilde m}$ coming from small and large diffeomorphisms, meeting the condition
\be
\label{conddta}
{\tilde m}/l  \ge O(1)
\ee can mix with each other. The resulting complications for the disk topology are discussed in subsection \ref{asymadslim} and there are similar issues in the double trumpet as well, see discussion after eq.\eqref{vsvlmrb}.

In order to avoid  these complications we will therefore finally have to resort to the asymptotic AdS limit as described above.
In this limit there are no modes which meet the condition eq.(\ref{conddta}) (since the $ l \rightarrow \infty$  limit is taken while keeping the mode number ${\tilde m}$ fixed). 
The small and large diffeomorphisms, and moduli are all orthogonal to each other in this limit, see discussion after eq.\eqref{dbldtact},  and the  measure then splits up into a product 
\be
\label{measures}
D[{\tilde g}_{ab}]= D[PV_s] D[PV_L] D[PV_{mod}]
\ee
where the three terms  on the RHS denote the measures for summing over  the small and  large diffeomorphisms and the two moduli with $V_{mod}$ denoting $V_{tw}, V_{b}$. 

We will describe these three measures in more detail next. 
The definition of  $D[PV_s]$ is the same as in the disk case and we get in an analogous way,  after carrying out the integral over the small diffeomorphisms that 
\be
\label{diska}
\int {D[PV_s]\over \text{Vol}(\text{sdiffeos})} = \sqrt{\text{det}'(P^\dagger P) }
\ee
Note that there is a zero mode of $ P^\dagger P$ which corresponds to the $U(1)$ isometry of eq.(\ref{dbmet}).
 
The measure in the space of large diffeomorphisms is described in appendix \ref{dbldfiscact}. Expressing the complex modes $\gamma_m,\delta_m$ appearing in eq.\eqref{dblapsolrw} in terms of the real variables $p_m, q_m, r_m, s_m$, as
\begin{align}
\gamma_m=p_m+iq_m,\quad \delta_m=r_m+is_m\label{dbgdpqrs}
\end{align}
we find that the measure in terms of the coefficients above is given by eq.\eqref{dbldfmes2}
\begin{align}
\int  D[PV_L]=\int \prod_{m\geq 1}dp_m\,dq_m\,dr_m\,ds_m (16 b \left(\tilde{m}^3+\tilde{m}\right) \sinh ^2(\pi  \tilde{m}))^2 \label{dbldifmes}
\end{align}
Finally for the moduli,  from eq.\eqref{dbtwper} and eq.\eqref{dbmodper}, it is straightforward to evaluate the inner products using eq.\eqref{metnrm} and we find 
\begin{align}
\langle PV_{tw}, PV_{tw}  \rangle =\frac{4t^2 b}{\pi};\quad \langle  PV_{b}, PV_{b} \rangle ={4\pi (\delta b)^2\over b};\quad \langle  PV_{tw}, PV_{b} \rangle =0\label{moduliinp}
\end{align}
  the measure for moduli after integrating over the twist modulus  turns out to be 
\be
\label{mm}
\int D[PV_{tw}]D[PV_{b}]= 4  \int dt \, db = 4 \int b db
\ee
Here we have used the fact that the range of $t $ is $[0,b]$, since a  twist between the two ends which is bigger than $b$ in magnitude can always be brought to lie in this range using the periodicity  $\theta\simeq \theta+b$. 
%(** Sunil: why does $\zeta$ take the values $[0,b]$? **)

\subsection{Carrying out the path integral}
\label{dsdiskpi}
We now turn to a systematic evaluation of the path integral eq.\ref{dbpiexp}.
 As for the disk, we will  first carry out the dilaton path integral, then the integral over the Liouville mode and finally, after taking the asymptotic AdS limit, the integral  over the diffeomorphisms and moduli. 
 
It is convenient, but not essential, to expand the dilaton about a background ${\phi_0}$,
\be
\label{backing}
\phi={\phi_0} + \delta \phi
\ee
where 
the background 
\be
\label{bald}
{ \phi_0}= \alpha r + i \beta\left[ r\log({r+i\over r-i}) - 2i \right]
\ee
The coefficients $\alpha, \beta$ can be adjusted so that ${ \phi_0}$ takes the values $\phi_{B1}, \phi_{B2}$ at the two boundaries in the metric, eq.\eqref{dbmet} which   are located at $r_{B1}, r_{B2}$. For the boundary lengths $l_1,l_2\gg 1$ we have that 
\be
\label{calebdt}
r_{B1}\simeq -{l_1\over b }, \ \ r_{B2}\simeq {l_2\over  b }
\ee
The fluctuation $\delta \phi$ about the background then satisfies Dirichlet boundary conditions at the two ends. 
Let us  also note that there is no  solution to the equations of motion where the metric has the required form eq.(\ref{dbmet}) so as to meet the condition $R=-2$, 
and  the dilaton takes the values $\phi_{B1}, \phi_{B2}$ at both ends with   $\phi_{B1,2}>0$.

We  carry out the  dilaton integral by taking a contour along which  $\delta \phi$ is  purely imaginary, as in the disk topology. This yields, 
\be
\label{parted}
Z_{DT}= \int {D[\sigma] D[{ {\tilde g}_{ab}}] \over \text{Vol}(\text{sdiffeo})} \delta (R+2) e^{-S_{JT,\partial_1} - S_{JT,\partial_2}}
\ee
(as in the disk we will not keep track of any overall constants in $Z$ from here on carefully.)
Note that the delta function imposes the constraint $R=-2$ and there are contributions from both boundaries in the exponent on the RHS now.

For the integral over $\sigma$ we write the metric as 
\be
\label{metexp}
g_{ab}=e^{2\sigma} {\hat g}_{ab},
\ee
where ${\hat g}_{ab}$ is a metric obtained from eq.\eqref{dbmet} after  carrying out changes due to small and large diffeomorphisms as well as the moduli.
We will  impose Dirichlet boundary conditions on $\sigma$ at the two ends. This condition is needed to obtain a well defined variational principle in the presence of the Gibbons-Hawking boundary terms at boundaries. 

This gives
\be
\label{partfdt}
Z_{DT}=\int {D[{\tilde g}_{ab}] \over \text{Vol}(\text{sdiffeo}) \text{det}({-\hat \nabla}^2+2) } e^{-S_{JT,\partial_1} - S_{JT,\partial_2}}
\ee
Note that with the Dirichlet boundary conditions we are imposing on the Liouville mode $\text{det}({-\hat \nabla}^2+2)$ has no zero modes. 

At this point to simplify the measure and also deal with $\text{det}({-\hat \nabla}^2+2)$, which in general depends on the large diffeomorphisms and also moduli,  in  a tractable manner,
we take the asymptotic AdS limit described earlier, eq.\eqref{limeades2}. 
The measure $D[{\tilde g}_{ab}]$ then becomes a product, eq.\eqref{measures} and proceeding as discussed at the end of the previous subsection we get
\be
\label{partgdt}
Z_{DT}=\int  b db\,\,D[PV_L]  {\sqrt{\text{det}'(P^\dagger P)}\over \text{det} (-{\hat \nabla}^2+2)}     e^{-S_{JT,\partial_1} - S_{JT,\partial_2}}
\ee

As argued in appendix {\ref{deterests}} both $\sqrt{\text{det}'(P^\dagger P)}$ and $\text{det} (-\hat{\nabla}^2+2)$ become independent of the large diffeomorphisms in this limit. Furthermore their ratio has important cancellations. In particular, the exponential divergences that we shall discuss shortly,  cancel in this ratio, see appendix {\ref{adsdtcoldets}}. 
The action, obtained in eq.\eqref{dbldtact}, is given by 
\be
\label{bcd}
S_{JT,\partial_1}+S_{JT,\partial_2} =\frac{b^2}{16\pi G J}\pqty{\frac{1}{\beta_1}+\frac{1}{\beta_2}}+\sum_{m\geq 1}\frac{1}{2\pi G}\frac{b^2}{J} \tilde{m}^2 (\tilde{m}^2+1)\sinh^2(\tilde{m}\pi) \pqty{\frac{p_m^2+q_m^2}{\beta_1}+\frac{r_m^2+s_m^2}{\beta_2}}
\ee
where $p_m,q_m,r_m,s_m$ are related to the modes $\gamma_m,\delta_m$ appearing in eq.\eqref{dblapsolrw} by the relation eq.\eqref{dbgdpqrs}.
The measure for summing over the large diffeomorphisms is given in eq.\eqref{dbldifmes}.  All this then leads to 
\begin{align}
\label{parthdt}
Z_{DT}=
\int  b db\,\,&\prod_{m\geq 1}dp_m \, dq_m {16b \tilde{m}(\tilde{m}^2+1)\over\text{csch}^2(\pi \tilde{m})} \exp[-\frac{b^2}{16\pi G J\beta_1}\pqty{1+\sum_{m\geq 1}{8 \tilde{m}^2 (\tilde{m}^2+1)\over\text{csch}^2(\pi \tilde{m})}\pqty{p_m^2+q_m^2}}]\nonumber\\
\times &\prod_{m\geq 1}dr_m \, ds_m {16b \tilde{m}(\tilde{m}^2+1)\over\text{csch}^2(\pi \tilde{m})} \exp[-\frac{b^2}{16\pi G J\beta_2}\pqty{1+\sum_{m\geq 1}{8 \tilde{m}^2 (\tilde{m}^2+1)\over\text{csch}^2(\pi \tilde{m})}\pqty{r_m^2+s_m^2}}]
\end{align}
which agrees with eq.(127) of \cite{Saad:2019lba}. Note that we have cancelled terms in the action  which are proportional to length that  arise from the various determinants by adding counterterm with suitably chosen coefficient, see discussion around eq.\eqref{btJT}. 

Doing the integrals over $p_m, q_m,r_m,s_m$ in eq.(\ref{parthdt}) we get 
\be
\label{patriot}
Z_{DT}=\int b db\,\,e^{-\frac{b^2}{16\pi G J}\pqty{{1\over \beta_1} +{1\over \beta_2}}}\frac{1}{16\pi ^2 G J\sqrt{\beta_1\beta_2}}
\ee
which further yields 
\begin{align}
Z_{DT}=\frac{1}{\pi}{\sqrt{\beta_1 \beta_2}\over (\beta_1+\beta_2)}\label{dbmodinre}
\end{align}
 as the final result in agreement with eq.(135) of \cite{Saad:2019lba} (we have not been careful about the overall numerical factor as discussed above). 
 
We note that away from the asymptotic AdS limit the additional modes, meeting condition eq.(\ref{conddta}) would enter in the calculation and
 one would have determine to their dependence in both $\text{det}'(P^\dagger P)$ and $\text{det} (-\hat{\nabla}^2+2)$.
This dependence is not easy to obtain and would involve an infinite number of higher derivative terms beyond the Schwarzian. 
Similarly the measure for summing over such modes is not easy to calculate. For all these reasons we will not attempt a calculation of $Z_{DT}$ in this more general case here.

\subsection{Adding matter}
\label{dbtwm}

Next we turn to adding matter to the theory and consider its effect in the path integral while summing over connected geometries with two boundaries. 
To begin we take a free bosonic massless scalar field. It's action is given by
\be
\label{mattdt}
S_M=\half \int d^2 x \sqrt{g}  (\partial\varphi)^2
\ee
and is the same as 
eq.\eqref{matact} with $N=1$. We will also consider fermionic matter subsequently.

The matter field does not couple to the dilaton and we can carry out the integral over $\phi$ and thereafter over the conformal factor $\sigma$ as before, leading to the partition function, $Z_{DT+M}$
being given by 
\be
\label{jtmdt}
Z_{DT+M}=\int {D[{\tilde g}_{ab}] D[\varphi]\over \text{Vol}(\text{sdiffeo}) \text{det}(-{\hat \nabla}^2+2)} e^{-(S_{JT,\partial_1}+S_{JT,\partial_2}+S_M)}
\ee
where $S_{JT,\del_{1,2}}$ is given by eq.\eqref{dbbjtact} at each of the boundaries.

First let us consider the case where the matter vanishes at the boundaries. Carrying out the path integral over $\varphi$ then gives, 
\be
\label{matt2}
Z_{DT+M}= \int {D[{\tilde g}_{ab}]\over \text{Vol}(\text{sdiffeo}) \text{det}(-{\hat \nabla}^2+2) \,(\text{det}(-{\hat \nabla}^2))^{1/2}}  e^{-S_{JT,\partial_1}-S_{JT,\partial_2}}
\ee
The factor $ (\text{det}(-{\hat \nabla}^2))^{1/2}$ in the denominator arose from the integral over the matter field and with  the matter field vanishing at both boundaries 
is obtained from the product of eigenvalues of the laplacian ${\hat \nabla}^2$ with Dirichlet boundary conditions. 
It is easy to see that with these boundary conditions the operator has   no zero modes.  
In general  $ (\text{det}(-{\hat \nabla}^2))^{1/2}$  will depend both on the moduli, see appendix \ref{dbmatapx}, and the large diffeomorphisms as discussed in appendix {\ref{deterests}}. 

To proceed we now take the asymptotic AdS limit, eq.\eqref{limeades2}. In this limit the dependence on the large diffeomorphisms of $ (\text{det}(-{\hat \nabla}^2))^{1/2}$ vanishes,
as discussed in appendix \textcolor{red}{\ref{deterests}}. However there is still an important dependence on the modulus $b$ as  we discuss shortly below and  in appendix \ref{dbmatapx}. 
Also the measure breaks up into a measure over the small and  large diffeomorphisms and moduli as mentioned in eq.\eqref{measures}. 
Carrying out the integral over the small diffeomorphisms then gives, 
\be
\label{mattdt3}
Z_{DT+M}= \int b db\, D[PV_L]  {\sqrt{\text{det}'(P^\dagger P)}\over \text{det}(-{\hat \nabla}^2+2) (\text{det}(-{\hat \nabla}^2))^{1/2}} e^{-S_{JT,\partial_1} -S_{JT,\partial_2}}
\ee
Here the measure for summing over the large diffeomorphisms $D[PV_L]$ is given in eq.\eqref{dbldifmes} and the two boundary actions, $S_{JT,\partial_{1}},S_{JT,\partial_{2}}$ are given in eq.\eqref{dbbjtact12}. 
Note that in the asymptotic AdS limit both $\sqrt{\text{det}'(P^\dagger P)}$ and $  \text{det}(-{\hat \nabla}^2+2)$ are independent of the large diffeomorphisms. 	Again, as mentioned earlier, their ratio has crucial cancellations as discussed in appendix \ref{adsdtcoldets}.

The  matter determinant depends on the modulus $b$ and this dependence is given by 
\be
\label{matted}
Z_M[b]={1\over (\text{det}(-{\hat \nabla}^2))^{1/2}}= {e^{-\frac{b}{6}} \over \eta({ib\over 2\pi})}
\ee
$\eta(\tau)$ on the RHS is the Dedekind eta function. 

Keeping all these facts in mind and carrying out the integral over the large diffeomorphisms then gives, 
\be
\label{finaldtz}
Z_{DT+M}=\int b db e^{-{b^2\over 16 \pi G}\left({1\over \beta_1}+{1\over \beta_2}\right)} \frac{1}{16\pi ^2 G J\sqrt{\beta_1\beta_2}} Z_M[b]
\ee
Now we come to a rather interesting consequence of having added the matter. Using the well known properties of $\eta(\tau)$ under modular transformations
 it is easy to see, as discussed in appendix \ref{dbmatapx}, eq.\eqref{dbflpart},  that as $b\rightarrow 0$
\be\label{behzm}
Z_M[b]\rightarrow \sqrt{ b \over 2 \pi} e^{\pi^2\over 6 b}
\ee
As a result   the integral over the modulus $b$ diverges  as $b \rightarrow 0$\footnote{This divergence is more general, see for example eq.\eqref{mdelb0}, and arises in $(\text{det}(-\hat{\nabla^2}+m^2))^{-1}$ for any non-negative $m^2$. } and the partition function $Z_{DT+M}$ is  in fact  not well defined. {To examine the behaviour of the wavefunction as $b\rightarrow\infty$, we note from the results for determinants evaluated in appendix \ref{dbmatapx},\ref{adsdtcoldets}, see eq.\eqref{dbfuldetrs}, \eqref{lndetdel}, that in this limit, the contribution from various determinants can atmost go as $e^{x b}, x>0$. However the boundary terms of the JT theory, after integrating over the large diffeomorphisms has the behaviour $e^{-y b^2}, y>0$, see eq.\eqref{finaldtz}. Thus the wavefunction is convergent at the other end of the $b$-integral as $b\rightarrow\infty$. }

Why does the divergence as $b\rightarrow0$ arise ? We see from eq.\eqref{dbmet} that as $b \rightarrow 0$ the ``neck"  of the double trumpet gets thinner and thinner. More precisely the geometry, eq.\eqref{dbmet} has a geodesic winding around the $\theta$ direction with minimum length $ b$ and  the length of this geodesic goes to zero when $b$ vanishes.  The divergence is related to  the quantum stress tensor of matter giving   a negative contribution due to the Casimir effect which blows up as the size of the neck vanishes. 
 
The result for the double trumpet partition function in the absence of matter, eq.\eqref{dbmodinre} can be  interpreted as a two point correlation between the partition functions of both boundary theories $\langle Z(\beta_1) Z(\beta_2)\rangle$, \cite{Saad:2019lba}, which could arise  for example in a boundary theory  with random couplings. The divergence, once bosonic matter is added, suggests that the dominant contribution in the sum over geometries will arise when the neck goes to zero size resulting in the two ends not being connected at all and the connected two point function for the partition function vanishing. This suggests  that in the presence of bosonic matter one is describing a more conventional system without random couplings. To put it another way, the theory with matter is ill-defined due to the divergence above.  To make it well-defined,  one possibility  could be to   take  the result from the double trumpet which peaks at $b=0$  as a clue and 
  simply disallow all topologies except the disk.   

However, it could well be that this is not the only possibility, it is certainly not a very elegant one. Instead, perhaps  further study  will show that the path integral can be made well defined in various ways and the  resulting dynamics  would then determine  whether wormholes are  allowed  or not depending on how the divergence is  tamed \footnote{We are grateful to the members of the TIFR String Theory group for a lively discussion on this question resulting in the more nuanced comments presented above!}.  We leave  a more detailed investigation along these lines  for the future.

We can also consider what happens if fermionic matter is added instead of the bosonic matter we considered above. Let us take as an example 
 one complex free fermion field $\psi$ with central charge $c=1$ and action
\be
\label{complex}
S_{M,f}=\int d^2x\sqrt{g} {\bar \psi} \gamma^\mu\partial_\mu \psi
\ee
where the subscript $f$ in $S_{M,f}$ in to indicate the fermionic nature of matter.
Since we are thinking of $\theta$ direction as the Euclidean time direction, or  as the temperature directions, 
  we  impose anti-boundary conditions along it. In addition let us also impose anti-periodic conditions, i.e. NS boundary conditions,  in the radial  direction.
  
  The partition function, $Z_{M,f}$ as a function of the moduli $b$ can then be easily written down and is given by 
  \be
  \label{zmaf}
  Z_{M,f}= \text{Tr}_{(NS)} e^{-b H}
  \ee
  where $H$ the Hamiltonian is given by 
  \be
  \label{Ham}
  H=\sum_{r=n+1/2} r c_{-r}^\dagger c_{r}-{1\over 24}
  \ee
  In the notation used in \cite{Polchinski:1998rr} eq.(10.7.8a), for  the boundary conditions above,  
  \be
  \label{valpart}
  Z_{M,f}=Z^0_0(\tau)
  \ee
  where $\tau=  {ib \over 2\pi}$.
  $Z_{M,f}$ will replaces the factor $Z_{M}[b]$ in eq.(\ref{finaldtz}). 
  To understand the  $b\rightarrow 0$ limit we can do a modular transformation. Since $Z^0_0(\tau)=Z^0_0(-1/\tau)$, (eq.(10.7.14)  of \cite{Polchinski:1998rr}),  we learn that 
  \be
  \label{smallbz}
  Z_{M,f}\rightarrow e^{\pi^2\over 6 b}.
  \ee
 As a result,  once again the integral over $b$ diverges. 
  
  We could also consider imposing periodic (Ramond) instead of anti-periodic boundary conditions in the radial  direction at the two ends of the double trumpet, while still  keeping the boundary conditions along the temperature direction to be anti-periodic. This gives $Z_{M,f}= \text{Tr}_{(R)} e^{-bH}$, which in the $b\rightarrow 0$ limit diverges
  in the same way as eq.(\ref{smallbz}). However suppose we sum over both the NS and Ramond sectors in the path integral with an opposite relative sign, then 
  \be
  \label{matters}
  Z_{M,f}=\text{Tr}_{(NS)}e^{-b H}- \text{Tr}_{(R)}e^{-b H}
  \ee
  and now the leading divergence at small $b$ would cancel. From eq.(10.7.14) of \cite{Polchinski:1998rr} we see that $\text{Tr}_{(R)}e^{-bH} \rightarrow Z^0_1(-1/\tau)$ under the modular transformation, $\tau\rightarrow -{1\over \tau}$. Thus,   we get that after this modular transformation  
  \be
  \label{news}
  Z_{M,f}= Z^0_0(-1/\tau) - Z^0_1(-1/\tau) = Tr_{(NS)}[(1-(-1)^F) {\tilde q}^{H}]
  \ee
  where ${\tilde q}= e^{-4 \pi^2\over b}$, and $F$ denotes the  fermion number operator, under which the NS vacuum has charge $0$ and the $\psi,\psi^\dagger$ operators have charge $\pm 1$ respectively. It is now easy to see that the vacuum state after the modular transformation is projected out and the leading contribution on the RHS in eq.(\ref{news}) arises from the first excited states with $F=1$. These have $H=\half-{1\over 24}= {11\over 24}$, so that 
  as $b\rightarrow 0$ now 
  \be
  \label{news2}
  Z_{M,f}\simeq e^{-11\pi^2\over 6 b}\rightarrow 0
  \ee
  We see therefore that $Z_{M,f}$ now decays very rapidly as $b\rightarrow 0$ and this renders the integral over $b$ convergent in the region where $b \rightarrow 0$. 
  
    We have not explored in detail whether the relative minus sign between the NS and R sector in eq.(\ref{matters}) gives a consistent theory.  In string theory (where there is additional matter and world sheet supersymmetry)  the contributions of the   NS and R sectors 
    do come with relative opposite sign, as in eq.\eqref{matters} above, and this is easy to understand   on the basis of spacetime spin statistics. However,  our problem  is different and in particular  spacetime itself is  two dimensional here.
    
    Finally, we can consider imposing periodic boundary conditions   for the fermion along the $\theta$ direction \footnote{We are grateful to Shiraz Minwalla for bringing this example to our attention  and for emphasizing its importance.}. This would correspond to calculating not the partition function but an index $\text{Tr}[(-1)^F e^{-\beta H}]$. For the disk topology it  would not be possible to impose this boundary condition since the $\theta$ direction shrinks to zero size and going around it is a $2 \pi$ rotation under which the fermion must be anti-periodic. But we can do so for the double trumpet since the $\theta$ circle has a finite size everywhere in the geometry. 
    Imposing NS boundary conditions along the radial direction in the periodic case would  give  $Z_{M,f}= Z^0_1(\tau)$ which after a modular transformation becomes  $Z^1_0(-1/\tau)$. In the limit $b\rightarrow 0$ this means 
    \be
    \label{mewl}
    Z_{M,f} \sim e^{-\pi^2\over 3 b}. 
    \ee
    We see that this now vanishes as $b \rightarrow 0$, and the divergence goes away. The  periodic boundary conditions have reversed the sign of the Casimir energy and there is no obstruction to a wormhole connecting the two ends now. If we impose periodic boundary conditions along the radial direction as well as along the $\theta$ direction, the partition function continues to behave like eq.(\ref{mewl}) as $b\rightarrow 0$ resulting in no divergence. In both these examples we would conclude that the two point function of the index on the two boundaries is non-zero.
    
    The main purpose of the last few examples above was to show that for suitable matter,  added in a consistent manner, one can avoid the divergence seen in the bosonic case\footnote{An investigation of the higher genus surfaces where such a divergence may reappear is currently under progress\cite{Moitra2021}. }. The integral over $b$ should then converge, the double trumpet geometry connecting the two ends would contribute to the path integral and the dual theory would involve  averaging   over coupling constants in some way.

   %%%%%%%%%%%%%%%%%%%%%%%%%%%%%%%%%%%%%%%%%%%%%%%%%%%%%%%%

   \section{ de Sitter JT gravity and the no-boundary wavefunction}
   \label{dsjtpi}
   \subsection{Basic set-up}
   \label{dsbasic}
   In this section we consider JT gravity in dS space. This corresponds to the action  for gravity and the dilaton  
   \be
   \label{actsds}
   S_{JT,dS}= {-i \over 16 \pi G} \pqty{\int d^2x\sqrt{-g} \phi (R-2)- 2 \int_{\del} dx \sqrt{\gamma}\phi K}
   \ee
   In comparison with eq.\eqref{eadsjtact} we see that the cosmological constant is positive here  and we are working in units where the Hubble constant  $H$ is given by 
   \be
   \label{defH}
   H^2=1
   \ee
   Note also that the boundary term above differs from the corresponding one in AdS. In the AdS case there is a term proportional to the length of the boundary, going like   $\int dx \sqrt{\gamma} $, which is absent here.    For the AdS case this term can be thought of as a counter term which is  added, with a suitable coefficient,  to remove a divergence which arises in the path integral when we take the limit where the length of the boundary diverges. However, in the dS case the dependence of the wavefunction  on the length of the boundary   is of physical significance and we should not be adding such a  term\footnote{In addition with  our choice of conventions the relative sign between the bulk and boundary terms is opposite to that in eq.\eqref{eadsjtact}, for a space-like boundary.}.
   
   We also note that the path integral in our conventions is given by 
   \be
   \label{pic}
   Z_{JT,dS}=\int {D[g_{\mu\nu}] D[\phi]\over \text{Vol}(\Omega)} e^{-S_{JT,dS}}
   \ee
   the measure etc. which appears above will be discussed in more detail below. 
   Matter can also be added to the system. Later on we  will consider conformal matter, specifically scalar fields with action eq.\eqref{matact} or fermionic fields with action eq.\eqref{fermfield}.

   We will study the wavefunction of the universe as given by the no boundary proposal. This wavefunction gives the probability amplitude for a  universe  which has length $l$ when  the dilaton takes the value $\phi_B$ and it is given by  the partition function 
   \be
   \label{zz}
   \Psi_{dS}[\phi_B, l] = Z_{JT,dS}=\int {D[g_{\mu\nu}] D[\phi]\over \text{Vol}(\Omega)} e^{-S_{JT,dS}}
   \ee
   For a single connected universe this partition function needs to be calculated over geometries which have one boundary with length $l$ where  $\phi=\phi_B$. 
   One can  think of $\phi$ as providing a clock  for the universe and the wavefunction as giving the amplitude for the universe to have different lengths at time $\phi_B$.

   A key new element in the calculation, in comparison to the AdS case with disk topology, is that the path integral involves metrics of different spacetime signatures. 
   There are two contours which have been suggested  to calculate the no-boundary wavefunction. In the conventional Hartle- Hawking proposal, \cite{PhysRevD.28.2960},  the contour studied involves Euclidean dS which is a sphere, $S^2$, with metric of signature $(2,0)$, which is then connected along the contour, at the equator of the $S^2$,
   to Minkowski dS with signature $(1,1)$.  To implement the no-boundary proposal the contour starts at say the north pole of the $S^2$. We will refer to this as the Hartle-Hawking (HH) contour below. 	In contrast, in the Maldacena contour, \cite{Maldacena:2019cbz},   we start at the north pole but  evolve along $-AdS_2$ which is a Euclidean geometry of signature $(0,2)$, eventually then continuing to Minkowski dS.  We will explore these contours in the second order formalism here. These contours are elaborated more after eq.\eqref{minds}.
   
   As far as our analysis below will reveal, we find no difference between the two contours for the resulting wavefunction. The reason for this, which will become clearer as we proceed is that the fluctuations over which we sum while carrying out the path integral are analytically continued in going from one signature to another,
   and we will not encounter any singularities while carrying out these continuations.

   For signature $(2,0)$ or $(0,2)$ the action, denoted by $S_{JT,edS}$ is given by 
   \begin{align}
   S_{JT,edS}=-\frac{1}{16\pi G}\pqty{\int d^2x\sqrt{g}\phi (R-2)+2\int dx \sqrt{\gamma}\phi K}\label{neadsjtact}, 
   \end{align}
   For a contour which passes through regions of different signature we will calculate piece-wise the contribution to $S_{JT}$ and add them to get the full result, keeping track of boundaries which arise when the different signature pieces are glued together. 
   
   Note that there is actually an additional topological term in the action, eq.(\ref{Stop}). This term is also present in the de Sitter case we are considering here.
   For $(2,0)$ and $(0,2)$  signature cases this term is given by 
   \be
   \label{stope}
   S_{top,1}= -{\Phi_0 \over 16 \pi G} \pqty{\int{\sqrt{|g|}} R +2 \int_\del\sqrt {|\gamma|}  K }
   \ee
   and for $(1,1)$ signature by 
   \be
   \label{stopmi}
   S_{top,2}= -{i \Phi_0 \over 16 \pi G}  \pqty{ \int{\sqrt{|g|}} R -2 \int_\del\sqrt {|\gamma|}  K }
   \ee
   where  $\Phi_0$ is a parameter which suppresses topological fluctuations. When dS JT gravity arises from higher dimensions it is related to the volume of the extra dimensions and the topological term gives a contribution proportional to the higher dimensional dS entropy. 
   For a contour which passes through regions of different signature we will again calculate pieces wise the contribution to $S_{top}$ and add them.
   
   In this section we focus on the case with a single boundary. In this case $S_{top}=- {\Phi_0\over 4 G}$ with the contribution coming from the part of the contour which has $(2,0) \text{ or } (0,2)$ signature. This yields 
   \be
   \label{stoph}
   e^{-S_{top}}= e^{{\Phi_0\over 4 G}}
   \ee
   Note that  the higher dimensional de Sitter entropy is given by 
   \be
   \label{hdds}
   S_{dS}= {\Phi_0\over  2 G}
   \ee
   and is twice in magnitude compared to $S_{top}$. 
   
   The three spacetimes mentioned above can be described with the metric
   \be
   \label{meta}
   ds^2={dr^2\over (1-r^2)} + (1-r^2) d\theta^2
   \ee
   the region $r<1$  this describes the $S^2$, and $r>1$,  $-AdS_2$. Taking 
   \be
   \label{anaconda}
   r\rightarrow \pm i  r 
   \ee
   gives dS space with signature $(1,1)$ which can be written as 
   \be
   \label{minds}
   ds^2=-{dr^2\over r^2+1} + (r^2+1) d\theta^2
   \ee. 
   \begin{figure}
   	\centering
   	\begin{tikzpicture}
   	\draw[->,ultra thick] (-5,0)--(5,0) node[right]{};
   	\draw[->,ultra thick] (0,-5)--(0,5) node[above]{};
   	\draw[snake=zigzag, color=red, thick] (1,0) -- (-5,0);
   	\draw [ color=blue, thick, middlearrow=latex] (1,0.1)
   	{[rounded corners]--(0.2,0.1)--(0.2,5)};
   	\draw [ color=green, thick, 3multiarrow=latex] (1,0.1)
   	{[rounded corners]--(5,0.2)arc(0:90:4.6cm)};
   	\draw [ color=blue, thick, middlearrow=latex] (1,-0.1)
   	{[rounded corners]--(0.2,-0.1)--(0.2,-5)};
   	\draw [ color=green, thick, 3multiarrow=latex] (1,-0.1)
   	{[rounded corners]--(5,-0.2) arc(0:-90:4.6cm)};
   	\draw node[ anchor=south west] at	(3,5) {$r$};
   	\draw (3.05,5.3) -- (3.05,5)--(3.4,5);
   	\fill[black] (1,0) circle (0.1cm);
   	\fill[black] (-1,0) circle (0.1cm);
   	\draw node[ anchor=north ] at	(1,0) {$1$};
   	\draw node[ anchor=north ] at	(-1,0) {$-1$};	
   	\draw node[ anchor=south ] at	(1,0) {$P$};
   	\draw node[ anchor=west ] at	(0.2,5) {$Q$};
   	\draw node[ anchor=south east ] at	(0,0) {$O$};
   	\draw node[ anchor=west ] at	(0.2,-5) {$S$};
   	\draw node[ anchor=west ] at	(5,0) {$T$};
   	\end{tikzpicture}
   	\caption{Analytic continuations in the complex r-plane for the no-boundary wavefunction}
   	\label{eefig}
   \end{figure}
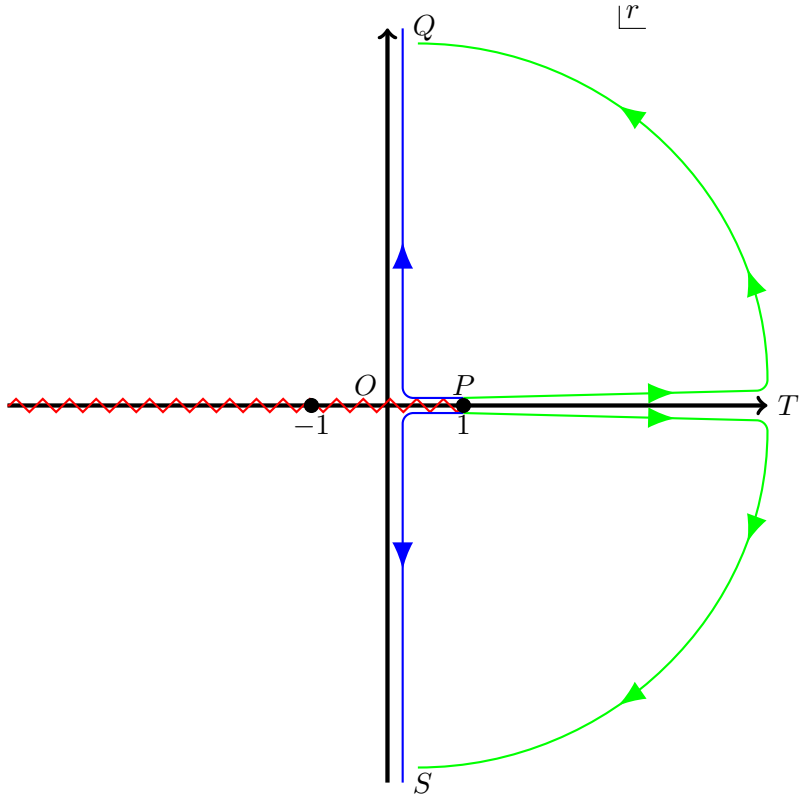

   We  consider  below a value of the length of the boundary $l > 2 \pi $ which is  classically allowed for dS spacetime with  $H=1$. Using the  $(r,\theta)$ coordinate system mentioned above, we can understand the $HH$ and Maldacena contours in Fig.\ref{eefig}. The branch cut in the figure corresponds to the branch points at $r=\pm1,-\infty$ of the associated Legendre function $P^\mu_{\nu}(r)$ that arises as a solution to the eigenvalue equation of  the operator $(-\hat{\nabla}^2+2)$ in Euclidean AdS, see eq.\eqref{adsresol},\eqref{delaltbs}. The $HH$ contour, shown in blue corresponds to starting at $r=1  (P)$, going till $r=0 (O)$ and then continuing till either $Q$ or $S$ depending on whether we started at $r=1$ just above or below the real axis respectively. The Maldacena contour, shown in green, starts at $r=1(P)$ in the $-AdS$ metric and proceeds along real axis to $r\gg 1 (T)$ and then is analytically continued to $Q$ or $S$.

   The steps to calculate the path integral have many  similarities  to the  AdS case considered previously and we will mainly emphasize some of the key new points below. 
   
   We first find a classical solution meeting our boundary conditions, obtain its contribution to the action, then expand around it and compute the contributions due to the quantum fluctuations. 
   
   For the first contour,  the classical equations are solved for the $(2,0)$ signature  case by the  $S^2$ with metric  eq.(\ref{meta})
   and dilaton, 
   \be
   \label{valpa}
   {\phi}_0 = i A r 
   \ee
   On continuing to $(1,1)$ signature using eq.(\ref{anaconda}) we get Minkowski dS space with 
   \be\label{valphic}
   {\phi}_0 =A r. 
   \ee
   Note that if  the boundary value $\phi_B$ is real,  $A$ must be  real  and therefore in the Euclidean signature region the dilaton is imaginary. To meet the boundary conditions we take the boundary to be at  
   $r=r_B$, where  
   \be
   \label{caleb}
   l = 2 \pi \sqrt{r_B^2+1}  \ee
   and fix $A$ to be 
   \be
   \label{fixa}
   A=\phi_B/r_B
   \ee
   
   For the second contour, in the $(0,2)$ signature part the solution is also  given by eq.(\ref{valpa}) (but now for $r>1$) and again continuing to $(1,1)$ signature gives the same result for the dilaton and the same values for $r_B, A$ as above. 
   
   The resulting classical action gives rise to a saddle point contribution of the wavefunction \cite{Maldacena:2019cbz}
   \be
   \label{scla}
   S_{cl}= {i \phi_B \over 4 G}  \sqrt{\left( {l\over 2\pi}\right)^2 -1 }
   \ee
   For $l\gg 1$, the leading term denoted by $\hat{S}_{cl}$, is given by 
   \be
   \label{scl}
   \hat{S}_{cl}={i \phi_B l  \over 8 \pi   G}
   \ee
   This saddle point contribution corresponds to a universe which is expanding with increasing time\footnote{In the rest of the section also we will mostly analyze this branch of the wavefunction.}.
   
   This  contribution to the wavefunction is the same in the two contours {\cite{Maldacena:2019cbz, PhysRevD.28.2960}}. For example, in figure \ref{eefig}, starting from the point P and reaching Q either along the blue contour (HH contour) or along the green contour (Maldacena contour) gives the same contribution to the wavefunction.  This is because the value of the on-shell classical action only depends on the  data close to the boundary ($\phi_B$ and the extrinsic curvature of the surface of length $l$) and these are determined by the dS part of the contour\footnote{More detailed analysis of the various saddle point contributions as well as the classically disallowed region will be discussed in \cite{umstsknext}.}. However, this contribution will in general differ from the contribution to the wavefunction obtained by starting from P and ending at S.

   %When $l< 2 \pi$ we are in the classically disallowed region. The classical solution of the dilaton is now taken to be $\phi=A r$, where $A$ is given in eq.\eqref{fixa}. The boundary value of the dilaton,$\phi_B$ is taken to be real and so $A$ is real. The relation between the length of the boundary and the radial location of the boundary, $r_B$, is now given by 
   %\begin{align}
   %l=2\pi \sqrt{1-r_B^2}\label{lclasdis}
   %\end{align}
   % The action is given by 
   %\be
   %\label{classdis}
   %S_{cl}=  {\phi_B \sqrt{1- \left({l \over 2 \pi}\right)^2} \over 4 G}
   %\ee
   %Note that the wavefunction which  in the WKB approximation is given by 
   %\be
   %\label{wf}
   %\Psi_{dS} \sim e^{-S_{cl}}
   %\ee
   %is exponentially suppressed in the classically disallowed region, with the exponential suppression growing with time, $\phi_B$, and is   classically oscillating in the allowed region.
  
   \subsection{Quantum fluctuations}
   \label{dsqfluc}
   Next, expanding about the classical part  we  compute the contributions due to the quantum fluctuations in the path integral. The calculation is closely related to that in section \ref{dilint}. We write  the dilaton as
   \be
   \label{word}
   \phi= {\phi}_0 + \delta \phi
   \ee
   where $\delta \phi$ is the fluctuation which we will integrate over.
   And we work in conformal gauge by writing the metric as 
   \be
   \label{metal}
   g_{ab}= {\hat g}_{ab}e^{2 \sigma} 
   \ee
   where ${\hat g}_{ab}$ is a conformally flat metric of appropriate signature with curvature ${\hat R}=2$.
   
   The Liouville mode fluctuations can be expanded in eigenmodes of the operator $\hat{\nabla}^2+2$ .
   \begin{align}
   \delta \sigma  & =  \sum c_\lambda \phi_\lambda \nonumber\\ 
   (\hat{\nabla}^2+2 )\phi_\lambda & =  -\lambda \phi_\lambda\label{dsdsmodex}
   \end{align}
   where the modes  $\phi_\lambda$ are  regular at the north pole and the coefficients $c_\lambda, \phi_\lambda$ are chosen so that $\delta \sigma$ is real for the $(2,0)$ or $(0,2)$ signature parts of the contour. For the Minkowski $(1,1)$ part we take the modes to be given by analytic continuation from the value it takes in the $(2,0)$ or $(0,2)$ segments of the geometry, depending on whether we are considering the HH or Maldacena contours and impose  Dirichlet boundary conditions,
   \be
   \label{abc}
   \phi_\lambda \big\vert_\partial=0
   \ee
   at the boundary $r=r_B$. This ensures 
   $\delta \sigma=0$  at the boundary. 
   For the dilaton fluctuation we do a similar expansion
   \be
   \label{dileep}
   \delta \phi =  \sum  d_\lambda \phi_\lambda
   \ee
   in terms of the same mode functions $\phi_\lambda$, 
   except that the expansion coefficients  $d_\lambda$ are chosen to be ``phase mismatched" compared to $c_\lambda$ so that in the $(2,0)$ or $(0,2)$ parts the dilaton fluctuation is purely imaginary. Again by analytic continuation we obtain $\delta \phi$ in the $(1,1)$ region and impose Dirichlet boundary conditions on it as well.
   
   The dilaton and Liouville mode path integrals can then be done in a way very similar to the AdS case, leading to
   \be 
   \Psi_{dS}[\phi_B,l]=  e^{-\hat{S}_{cl}}\int {D[PV]\over \det(\hat{\nabla}^2+2) \text{Vol}(\text{sdiffeo})} e^{-S_{JT,\partial}}\label{dsjtparpi}
   \ee
   where $S_{JT,\del}$, which arises entirely from the boundary term in eq.\eqref{actsds} and $\hat{S}_{cl}$ is given in eq.(\ref{scl}). 
   Note that when the boundary length $l\gg 1$ so that  $r_B$, eq.(\ref{caleb}), meets the condition,  $r_B\gg 1$, we get, as  discussed in appendix \ref{etds2gc},  that the boundary action is given by 
   \begin{align}
   S_{JT,\del}=-\frac{i\phi_B \epsilon}{8\pi G}\int du\,\text{Sch}\pqty{\tan({\theta(u)\over 2}),u}\label{dsschac}
   \end{align}
   where the line element along the boundary is
   \be
   \label{le}
   ds^2={du^2 \over \epsilon^2}\simeq -{dr^2 \over r^2} + r^2 d\theta^2
   \ee
   with $\epsilon\sim 1/r_B \ll 1$. 
   
   Let us also mention that since  the contour goes over  metrics with  different signatures    care needs to be taken in how we define the operator 
   $(\hat{\nabla}^2 +2)$  (similarly comments also apply to to $(P^\dagger P)$ and ${\hat {\nabla}}^2$ which will appear below). We  do this by analytic continuation as follows.
   Suppose we are working in the metric eq.(\ref{meta}) and its analytic continuation eq.(\ref{minds}). We promote the radial variable which appears in the metric and in the operator $(\hat{\nabla}^2 +2)$ to be a complex coordinate and analytically continue the operator as we go along the contour. The eigenmodes $\phi_\lambda$ are constructed to be regular at the north pole or the origin of the disk for the $(2,0)$ or $(0,2)$ signature regions respectively  and then analytically continued to dS spacetime.
   And the eigenvalues are then determined by imposing Dirichlet boundary conditions at the boundary.
   Let us  also note that  the fluctuations $\phi_\lambda$ will not be real everywhere along the two contours. We will take them to be real in the $(2,0)$ or $(0,2)$ regions,
   this will result in them being complex, in general, in the dS region of the path integral.

   The vector fields $V$ which appear in eq.(\ref{dsjtparpi}) include large and small diffeomorphisms, $V_L$ and $V_s$,  respectively as in the AdS case.  The small diffeomorphisms $V_s$,  which satisfy the boundary conditions eq.\eqref{veclapdetbc}, generate  the gauge transformations whose volume is in the denominator in eq.\eqref{dsjtparpi}. 
   The large diffeomorphisms $V_L$ are zero modes of $P^\dagger P$ and correspond to different ways in which the boundary wiggles with $\phi=\phi_B$ on the boundary. The operators   $P, P^\dagger $  are given in eq.\eqref{popdef} and \eqref{P1daggerdef}. These vector fields  can be defined in the $(2,0)$, $ (0,2)$ signature regions as in the AdS case and  are  also then analytically continued   to the $(1,1)$ region. 
   
   More specifically,  $V_L$ can be written  as $V_L=*d \psi$ where $\psi$ is a zero mode of   $(\hat{\nabla}^2 + 2 )$,
   satisfying 
   \be
   \label{zm}
   (\hat{\nabla}^2 + 2 )\psi=0.
   \ee
   In the $(2,0)$ or $(0,2)$ segments of the contour  $\psi$ is given from eq.(\ref{psiform}) to be 
   \be
   \label{psiforme}
   \psi=\sum_{\abs{m}>1} {\hat c}_m e^{i m \theta} (r+\abs{m}) \left({r-1 \over r+1}\right)^{\abs{m}\over 2}
   \ee
   the $m=0,1-1$ cases give rise to the Killing vectors.
   The solution for the dS segment of signature $(1,1)$ is then obtained by analytic continuation after taking $(r\rightarrow -i r, \theta \rightarrow \theta)$, and becomes, 
   \be
   \label{psici}
   \psi= \sum_{\abs{m}>1} {\hat c}_m e^{im\theta}  (-ir+\abs{m})\left({r-i\over r+i}\right)^{\abs{m} \over 2}
   \ee
   with the coefficients ${\hat c}_m$  being  chosen so that $\psi$ is real in the region $r\gg 1$ of the dS part of the contour, which means the relation
   \begin{align}
   {\hat{c}_{-m}=-\hat{c}_m^*}\label{dsldifmrel	}.
   \end{align}
   In general the vector field $V_L$ obtained from  $\psi$ is  then complex in this region. It is given by
   \begin{align}
   V_{L,m}=\hat{c}_m e^{i \theta  m} \left(\frac{r-i}{r+i}\right)^{\frac{\left| m\right| }{2}} \pqty{ m (r+i \left| m\right| ),\frac{i  \left(i r \left| m\right| -m^2+r^2+1\right)}{r^2+1}}\label{dsldif}
   \end{align}
   and the corresponding metric perturbations are given by 
   \begin{align}
   (PV_L)_{ab}=
   2 \hat{c}_me^{i \theta  m}\frac{\left(m^2-1\right)}{(r^2+1)}  \left(\frac{r-i}{r+i}\right)^{\frac{\left| m\right| }{2}}\pqty{\begin{matrix}
   	{ m(r^2+1)^{-1} }&{ \left| m\right|   }\\
   	\left| m\right|  & m (r^2+1)
   	\end{matrix}}\label{dsldifmetper}
   \end{align}
   
   Now we come to a   complication similar to what we found in the AdS case. While the path integral is quite clearly defined as we have seen, evaluating it explicitly,  for a fixed $\phi_B$ and $l$ is difficult even when $ l$ is  large. This is because the metric perturbations resulting from the small and large diffeomorphisms are not orthogonal resulting in a  measure  that is difficult to calculate and  also because evaluating the determinant
   $(\hat{\nabla}^2 +2)$ is  non-trivial.
   
   To simplify things we therefore consider  the asymptotic limit, in which  
   \be
   \label{limes}
   \phi_B\rightarrow {2\pi \over \beta J \epsilon}, l \rightarrow {2\pi \over \epsilon}. 
   \ee with 
   \be
   \label{limestwo}
   \epsilon \rightarrow 0
   \ee 
   In this limit the inner product between metric perturbations generated by small and large diffeomorphisms become orthogonal and the measure simplifies, as in section \ref{asymadslim} above. Doing the integral over the small diffeomorphisms then gives, 
   \be
   \label{ef}
   \Psi_{dS}  = e^{-\hat{S}_{cl}} \int D[P V_L] {  \sqrt {\det'(P^\dagger P) } \over \det(\hat{\nabla}^2+2)} e^{-S_{JT,\del}}
   \ee
   Moreover both $\det'(P^\dagger P)$ and $\det(\hat{\nabla}^2+2)$ do not depend on $V_L$ in this limit and can therefore be taken out of the integral over $V_L$ - the arguments leading to this conclusion are analogous to the AdS case, see also appendix \ref{deterests}. 
   The measure for summing over the diffeomorphisms is obtained from the measure for metric perturbations. This is given in  eq.\eqref{metnrm} with $\delta g_{ab}$ being given by eq.\eqref{crdgvc} for the $(2,0)$ or $(0,2)$ segment of the contour and obtained in the $(1,1)$ segment by continuation. 
   The resulting inner product is then given by 
   \be
   \label{inprodlds}
   \langle PV_{L,m_1} PV_{L,m_2}\rangle = \delta_{m_1, -m_2} \frac{-8 i\pi  \hat{c}_{m_1} \hat{c}_{m_2}\abs{m_1} \left(m_1^2-1\right)}{r_B^2+1} \left(\frac{r_B-i}{r_B+i}\right)^{\abs{m_1} } \left(2 i\abs{m_1} r_B-2 m_1^2+r_B^2+1\right)
   \ee
   which in the $r_B \rightarrow \infty$ limit becomes, 
   \begin{align}
   \langle PV_{L,m} PV_{L,-m}\rangle\simeq-8 i \pi  \hat{c}_m \hat{c}_{-m} \left(m^2-1\right) \abs{m} \label{dsldifinp}
   \end{align}
   Thus the integral over the large diffeomorphisms reduces to the integral over the modes $\hat{c}_m$ with the standard well known measure \cite{Stanford:2017thb,Saad:2019lba}, 
   \be
   \label{mlarged2}
   D[PV_L] = \prod_{|m|\ge 2} d{\hat c}_m d{\hat c}_{m}^* 8 \pi \abs{m} \left(m^2-1\right).
   \ee
   The action in terms of the modes  $\hat{c}_m$ is obtained from eq.\eqref{dsschac} by noting that $\theta(u)$ is specified by eq.\eqref{rudisk} and eq.\eqref{condo}. So we see that $\theta(u)\simeq u+\sum_{{m} \geq 2}m\hat{c}_m e^{imu}$. The action then becomes
   \begin{align}
   S_{JT,\del}=-\frac{i\phi_B \epsilon}{8G}\left[1+2\sum_{m\geq 2}\hat{c}_m \hat{ c}_{-m}(m^4-m^2)\right]\label{dsschact}
   \end{align}
   The integral of the Schwarzian action with the above measure is shown to be one-loop exact \cite{Stanford:2017thb}, leading to the wavefunction in this limit being 
   \be
   \label{eha}
   \Psi_{dS}=\exp [ -{ i\phi_B l\over 8 \pi G}  +   {i \pi \phi_B  \over 4 G l} +  {\phi_0\over 4 G} ]  {(-32i\pi G J\beta)^{-3/2}\over \sqrt{2\pi}} { \sqrt {\text{det}'(P^\dagger P) } \over \text{det}({\hat \nabla}^2 +2)}
   \ee
   where we have also added the contribution from $S_{top}$, and $\hat{S}_{cl}$, eq.\eqref{stoph} and \eqref{scla}.
   Putting in eq.(\ref{limes})  gives
   \be
   \label{eh}
   \Psi_{dS} = \exp[ -{ i  \pi \over  2 G \beta J  \epsilon^2}  + {i\pi \over 4 G  \beta J}   + {\phi_0 \over 4 G}]     \ {(- 32i\pi G J\beta)^{-3/2}\over \sqrt{2\pi}}{ \sqrt {\text{det}'(P^\dagger P) } \over \text{det}({\hat \nabla}^2 +2)}
   \ee
   More correctly, this is the value of $\Psi$ upto an overall coefficient which we have not fixed. 
   Note that in the limit we are considering the first term in the exponent which arises from eq.(\ref{scl})
   \be
   \label{limdcl}
   S_{cl}=  { i \pi  \over  2 G \beta J  \epsilon^2} \rightarrow \infty
   \ee
   and thus the  wavefunction has very rapid fluctuations in its phase.
   
   It is worth mentioning that   the two determinants on the RHS  above  in general  depend on $l$ and can also give rise to a term diverging linearly like $l$ in the exponent of $\Psi$, as is discussed especially for the AdS case in appendix \ref{asyadsdet} in considerable detail. There are some subtle issues which arise in this context having to do with how the determinants are regulated in the UV, and related to  the order of limits involved while  taking the asymptotic dS limit. This is also connected  to the discussion below. 
   
   It is worth emphasizing that while we have considered the asymptotic  limit eq.(\ref{limes}), eq.(\ref{limestwo}) since it is analogous to the asymptotic AdS limit which was also tractable, 
   in the context of cosmology one really wants to obtain  $\Psi$ for fixed values of  $l,\phi_B$. 
   The $l \rightarrow \infty$ then is  the the limit of $\Psi$ obtained first for such fixed values. 
   
   The  case with $l$ fixed  is  considerably more complicated, as was  emphasized in the AdS spacetime, and we unfortunately have to postpone such an investigation for the future. 
   It is however worth noting that the different order of limits  required when we work at $l$ fixed and take the cut-off  on the eigenvalues, which regulates the determinants, to infinity first can yield a significantly different result. In this  limit modes of the form $ e^{i m \theta}$ with  mode number $ m \ge {1\over l}$  have a physical wavelength which lies within the universe and such modes can play an important role in determining the behaviour of the determinants and the resulting behaviour of the wavefunction. In contrast in the asymptotic limit, since $l \rightarrow \infty$  first, all modes which are kept in the determinants have a diverging physical wave length.	To illustrate how the behaviour at  fixed and large $l$ might be different  we evaluate the integral over the large diffeomorphisms  with a measure obtained from the inner product eq.(\ref{inprodlds}). We find that the behaviour of $\Psi$ changes quite dramatically at large $l$ and begins to decay exponentially going like 
   \be
   \label{dec}
   \Psi_{dS} \sim e^{- {l \over 4}}
   \ee
   due to the presence of the extra modes with 
   \begin{align}
   m\ge {1\over l}\label{mlhigh}
   \end{align}
   in the path integral, see eq.\eqref{nadslartau} in appendix \ref{nadsds}. 
   We hasten to reiterate here that this calculation is not really self consistent because these modes mix with small diffeomorphisms, since we are at finite $l$ (or non-zero $\epsilon$) , eq.(\ref{normalisedjef}), and this mixing needs to be included in obtaining the correct measure while integrating over them.  Our purpose in presenting the discussion of appendix \ref{nadsds} is mainly to emphasize that a different result can be potentially obtained with the different order of limits, due to such modes  once they are correctly included in the path integral. 
   %\textcolor{red}{A calculation of the wavefunction at finite $l$ is done in \cite{Iliesiu:2020zld} in the first order formalism}.  

   \subsection{Adding matter}
   \label{dsdwmat}
   We end this section by making some comments about the case with matter. We again consider $N$ free bosonic scalar fields,  as in the AdS case with the action 
   \begin{align}
   S_M=i\sum_{i=1}^{N}\half \int d^2x\sqrt{-g}(\del\varphi_i)^2\label{matact}
   \end{align}
   Although we will only discuss about bosonic fields for now, we mention the action for a free fermionic field for completeness, which is given by
   \begin{align}
   \label{fermfield}
   S_{M,f}=i\int d^2x\sqrt{-g} {\bar \psi} \gamma^\mu\partial_\mu \psi
   \end{align}
   
   At late times for $r_B \gg 1$ the classical action for the bosonic fields as a functional of their boundary values is given by 
   \be
   \label{classism}
   S_{M,cl}={-1\over 2} \int du_1 du_2 \,\theta'(u_1) \theta'(u_2) {\hat \varphi}(u_1) {\hat \varphi}(u_2) F(\theta(u_1), \theta(u_2) )
   \ee
   where $u$ is the rescaled proper length along the boundary, eq.(\ref{le}), and ${\hat \varphi}_i(u)$  is the late time value of $\varphi_i$. The details of the above result can be found in appendix \ref{mcttrp}, see eq.\eqref{dsmcoup}
   The quantity $F$ in eq.(\ref{classism}) is given by eq.\eqref{Fdef} of appendix \ref{mcttrp}. To obtain the behaviour of the wavefunction in the asymptotic dS limit we would couple the Schwarzian action to the matter action above and integrate the large diffeomorphisms to obtain the wavefunction as a functional of the 
   boundary values ${\hat \varphi_i}(u)$. This was studied in considerable detail in \cite{Maldacena:2019cbz}. 
   We will not pursue this line of investigation further here. 
   
   One can also include quantum corrections due to the matter fields. The quantum corrections come from    ${ (\text{det}({-\hat \nabla}^2))^{-N/2}}$  which arises when one integrates out the matter fields. The dependence on the large diffeomorphisms of this determinant is suppressed at large $l$ going  like  $ O(1/r_B)\sim O(\epsilon)$ {analogous to the case of AdS disk, see discussion after eq.\eqref{detinlesch}}. The resulting term to quadratic order in the diffeomorphisms vanishes in the asymptotic dS limit. One can include  the quantum effects of matter and neglect those due to the other degrees of freedom, which are difficult to obtain at finite $\epsilon$, as mentioned above by  working in the semi-classical limit where we take $N\rightarrow \infty$ and $G\rightarrow 0$ keeping $GN$ fixed. 
   Solving the resulting saddle point equations now with the additional quantum effective action to $O(\epsilon)$ yields the wavefunction as a function of $\phi_B$ for large values of  $l$ in the theory. We leave further investigation of this interesting limit  for the future. 
   
   One can also try to go beyond the semi-classical limit and include the quantum effects of matter as well as the gravitational degrees by working at fixed and large $l$. However, now modes within the horizon meeting the condition eq.(\ref{mlhigh}) will need to be included and this is more challenging as discussed above. 
   We also leave an investigation of this interesting case  for the future. 
   
   Before ending let us give a  few more details on the matter determinant calculation. To calculate the matter  path integral
   \be
   \label{mattpath}
   \int D\varphi e^{-S_M}
   \ee
   with  the  scalar $\varphi$ being subject to Dirichlet boundary conditions at the dS boundary we proceed as follows. We consider the operator $\hat{\nabla}^2$ 
   which is obtained in different regions of spacetime along the contour by analytic continuation as per our discussion of the  operator $(\hat{\nabla}^2+2)$ above. 
   And expand $\phi$ in terms of the complete set of eigenmodes of this operator which satisfy the equation $\hat{\nabla}^2 \varphi_\lambda=-\lambda \varphi_\lambda$.   
   The  eigenmodes are analytically continued from the $(2,0)$ or $(0,2)$ regions to the $(1,1)$ region. 
   Specifically in the $(2,0)$ or $(0,2)$ regions  these modes, which satisfy the   regularity condition,  at $r=1$ ($r$ being the radial coordinate in eq.\eqref{meta}) are given by 		 $P_{v-1/2}^{-|m|} (r) e^{i m \theta}$ where $P_{v-1/2}^{-|m|}(r)$ are associated Legendre functions, with eigenvalue $\lambda= v^2-\frac{1}{4}$. After analytic continuation to dS they become $P_{v-1/2}^{-|m|}(\pm i r) e^{i m \theta}$. The eigenvalues are obtained by imposing Dirichlet boundary conditions on these modes. 
   
   In the  complex $r$ plane shown in Fig.\ref{eefig}, $P_{v-1/2}^{-|m|}(r)$ has singularities at $\pm 1$. The contours shown in Fig.\ref{eefig3} illustrates how the analytic continuations we have in mind are to be carried out. It also shows why the HH and Maldacena contours will agree since both avoid any singularities and the solutions along these contours can be analytically continued to each other. For completeness we should also mention that the inner product eq.\eqref{metnrm} which goes into defining the measure of the path integral, \eqref{mlarged2}, should be analytically continued  as well along the contour. 
   
   %%%%%%%%%%%%%%%%%%%%%%%%%%%%%%%%%%%%%%%%%%%%%%%%%%%%

\section{de Sitter double trumpet}
\label{desitdobt}
Let us now turn to discussing  the analogue of the double trumpet spacetime in the context of de Sitter space. 
More specifically, as in section \ref{dsjtpi} we will consider the no-boundary proposal for calculating the  wavefunction but now we  ask about the amplitude for two disconnected universes of length $l_1,l_2$ to arises  when the dilaton takes values $\phi_{B1}, \phi_{B2}$ respectively. 
The result for this amplitude  is suppressed by a factor of $e^{S_{dS}/2}$ where $S_{dS}$ is given by eq.(\ref{hdds}), 
compared to the amplitude for producing one universe. We will find that the amplitude to produce two disconnected universes is non-zero in pure JT gravity.
Once matter is included the result for the double trumpet space can be finite, or have a divergence of the kind we found in the AdS case which arises when the neck of the wormhole shrinks to zero size due to quantum effects of the matter stress-tensor. 

We will start by considering  pure dS JT gravity and then add matter. 

\begin{figure}
	\centering
	\begin{tikzpicture}
	\draw (2,2.5)  .. controls (0.5,0) and (-0.5,0) .. (-2,2.5);
	\draw (3,2.5)  .. controls  (0.5,-1) and (-0.5,-1) .. (-3,2.5);
	\draw (2.5,2.5) ellipse (15pt and 2pt);
	\draw (-2.5,2.5) ellipse (15pt and 2pt);
	\draw [dashed] (2,2.5) .. controls(2.1,2.6) .. (2.2,3.2);
	\draw [dashed] (3,2.5) .. controls(3,2.6) .. (3.5,3.2);
	\draw [dashed] (2.8,3.2) ellipse (18pt and 2pt);
	\draw [dashed] (-2,2.5) .. controls(-2.1,2.6) .. (-2.2,3.2);
	\draw [dashed] (-3,2.5) .. controls(-3,2.6) .. (-3.5,3.2);
	\draw [dashed] (-2.8,3.2) ellipse (18pt and 2pt);	
	\draw  node[ anchor=north east] at (-0.3,3) {$r\rightarrow \pm i r$};	
	\draw  node[ anchor=north east] at (5.2,3) {$r\rightarrow \pm i r$};	
	\end{tikzpicture}
	\caption{dS double trumpet from AdS double trumpet}
	\label{eefig2}
\end{figure}
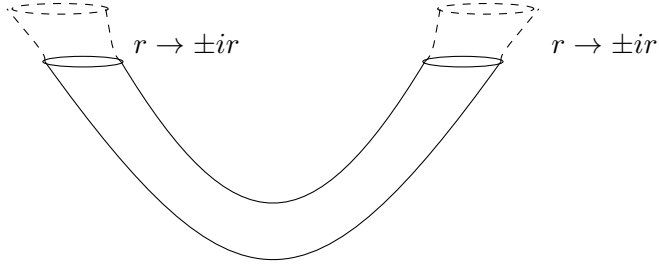

Let us note that the pure JT theory does not have a classical solution with the double trumpet topology and the dilaton meeting its boundary conditions. 
For carrying out the path integral in this case we have to    use the Maldacena contour. Along this contour  the geometry  has a segment with  $-AdS_2$ metric of signature $(0,2)$ which then connects to dS space ending in two boundaries as shown in  Fig.\ref{eefig2}. The $-AdS_2$ segment is now   described by the double trumpet geometry  with signature $(0,2)$. 
One can think of doing  the path integral by starting with a fiducial metric of the form eq.\eqref{metads2}, with $ds^2\rightarrow -ds^2$ for $-AdS_2$ space, and incorporating fluctuations about this fiducial metric. To join this spacetime to dS space with two disconnected boundaries we  continue the two ends of the  double trumpet taking $r\rightarrow \pm i r$. This gives at each end the metric
\be
\label{metered}
ds^2\equiv\hat{g}_{ab}dx^a dx^b=-{dr^2\over r^2-1} +(r^2-1) d\theta^2, \quad\theta \sim \theta+ b
\ee
which can be easily seen to be  a metric with curvature $\hat{R}=2$ describing the ``Milne"   region of dS space. By taking the  boundary to lie at  $r=r_{B1,2}$ and choosing $r_{B1,2}$ suitably at the two ends we can impose the condition that the two universes have lengths $l_1,l_2$. To carry out the path integral for the dilaton we expand it about a background value, $\phi_0$, which is the analogue of eq.(\ref{bald}), and  which takes values, $\phi_{B1}, \phi_{B2}$ at the two ends of de Sitter space. The fluctuation about this background is then given by $\delta \phi$ and both  $\phi_0$ and $\delta \phi$, are analytically continued  across the different regions  of the spacetime with signatures $(0,2)$ and $(1,1)$. Similarly the fluctuations in the conformal factor $\delta \sigma$ are also defined across the two regions with different signature by analytic continuation. Both $\delta \phi, \delta \sigma$ are expanded in terms of the eigenmodes of the operator $(\hat{\nabla}^2+2)$, with a relative factor of $i$ between their expansion coefficients, as for the dS integral with disk topology.

The result of the path integral over the dilaton and the conformal factor then gives for the wavefunction, $\Psi_{DDT}$, when  $l_1,l_2\gg 1$, 
\be
\label{pis}
\Psi_{DDT}= e^{-i\left(  {\phi_{B1} l_1\over  8 \pi G} + {\phi_{B2} l_2 \over 8 \pi G}\right) } \int{  D[{\tilde g}_{ab}]\over \text{Vol}(\text{sdiffeo}) \text{det}({\hat {\nabla}}^2 +2) } 
e^{-S_{JT,\partial_1} - S_{JT, \partial_2}}
\ee
where 
\be
\label{defsjt}
S_{JT,\partial_{1,2}}=-\frac{i\phi_B \epsilon}{8\pi G}\int_{\del_{1,2}} du\,\text{Sch}\pqty{\tan({\theta(u)\over 2}),u}
\ee
and $D[{\tilde g}_{ab}]$ is the measure for summing over traceless metric deformations which includes small, large diffeomorphisms and the moduli. 

As discussed in section \ref{dsjtpi}  earlier in the dS case, the case of physical interest is one where the lengths $l_1,l_2$ are finite. However  this is a difficult situation in which to make progress since there are modes meeting the condition, eq.\eqref{mlhigh}, which have not yet exited the horizon. The  dynamics  of such short distance modes is difficult to evaluate. To make progress we therefore  consider the asymptotic limit where eq.(\ref{limeades2}) is met.  In this asymptotic limit all modes have exited the horizon and are frozen out by the exponential expansion of the universe.  It could be, as suggested in \cite{Maldacena:2019cbz} eq.(2.12), that in fact this asymptotic limit is the more fundamental quantity in dS space and the finite length case should  be thought of  as arising from it by integrating back in some of the modes which are yet to exit the horizon. 

In the asymptotic limit, 
reasoning analogous to the AdS case in subsection \ref{asymadslim}, and for the disk topology in dS, subsection \ref{dsqfluc},   then leads to the conclusion that 
\be
\label{valpsi2}
\Psi_{DDT}= \exp[- {i\phi_{B1} l_1+i\phi_{B2} l_2 \over 8 \pi G} ] Z_{\partial,DDT}
\ee
where 
\begin{align}
\label{defzzdt}
Z_{\partial, DDT}&= \int b db\, e^{- {ib^2 \over 16 \pi G J}\left({1\over \beta_1}+{1\over \beta_2}\right)}\nonumber\\ &\times\prod_{m\geq 1}dp_m \, dq_m {16b \tilde{m}(\tilde{m}^2+1)\over\text{csch}^2(\pi \tilde{m})} \exp[-\frac{i b^2}{2\pi G J\beta_1}{\sum_{m\geq 1}{ \tilde{m}^2 (\tilde{m}^2+1)\over\text{csch}^2(\pi \tilde{m})}\pqty{p_m^2+q_m^2}}]\nonumber\\
\times &\prod_{m\geq 1}dr_m \, ds_m {16b \tilde{m}(\tilde{m}^2+1)\over\text{csch}^2(\pi \tilde{m})} \exp[-\frac{i b^2}{2\pi G J\beta_2}{\sum_{m\geq 1}{ \tilde{m}^2 (\tilde{m}^2+1)\over\text{csch}^2(\pi \tilde{m})}\pqty{r_m^2+s_m^2}}]
\end{align}
Doing the integral over $p_m,q_m$ etc in eq.(\ref{defzzdt}) gives, 
\be
\label{defzzdt}
Z_{\partial,DDT}= \int b db e^{- {ib^2 \over 16 \pi G J}\left({1\over \beta_1}+{1\over \beta_2}\right)} \frac{1}{16i\pi^2 GJ\sqrt{\beta_1\beta_2}}
\ee
which in turn gives, 
\be
\label{defzzdt3}
\Psi_{DDT}= \exp[- {i\phi_{B1} l_1+i\phi_{B2} l_2 \over 8\pi  G } ]\frac{2}{\pi}{\sqrt{\beta_1 \beta_2}\over (\beta_1+\beta_2)}
\ee

We can also consider adding matter. Consider   a massless scalar field with action eq.\eqref{matact}. 
We impose vanishing boundary conditions for the scalars at the two ends. Other boundary conditions can also be similarly  dealt with by a straightforward extension of the methods discussed in this paper, but we will not do so here. These  cases  would give rise to the wavefunction for two disconnected universes with the scalar field taking some specified values in these universes.  

The path integral can be carried out along the lines described above for this case too. It follows from the previous section that in the asymptotic limit discussed above 
\be
\label{zzmdt}
Z_{DDT+M}=\int b db\, e^{- {ib^2 \over 16 \pi G J}\left({1\over \beta_1}+{1\over \beta_2}\right)} \frac{1}{16i\pi^2 GJ\sqrt{\beta_1\beta_2}} Z_{M,s}[b]
\ee
where 
\be
\label{valmatds}
Z_{M,s}[b]={1\over (\text{det}(-{\hat \nabla}^2))^{1/2}}
\ee

As discussed in appendix \ref{dsdbdets} when $b\rightarrow 0$,  $Z_{M,s}$ diverges like eq.\eqref{behzm}. 
As in the AdS case this divergence arises due to the Casimir effect for  matter which results in a diverging stress tensor  when the size of the neck  in the wormhole goes to zero.

Now let us turn to the fermionic case with action \eqref{fermfield}.
 For AdS the anti-periodic boundary condition along the $\theta$ direction is needed for computing the thermal partition function, however here it is upto us to specify whether the fermion satisfies  periodic or anti-periodic boundary conditions along the $\theta$ direction. 
When the boundary condition is anti-periodic  the disk topology path integral corresponding to producing one universe gives a well-defined result but the double trumpet has a divergence of the form eq.\eqref{smallbz}. 
When the boundary condition is periodic (for both universes)  the disk topology does not contribute, since the $\theta$ circle shrinks to zero size and therefore  the fermion must 
necessarily having anti-periodic boundary conditions along it. The leading contribution then arises from the double trumpet topology for  two connected universes.
In this case one cannot produce one universe from nothing through quantum tunnelling they must come in at least a pair!

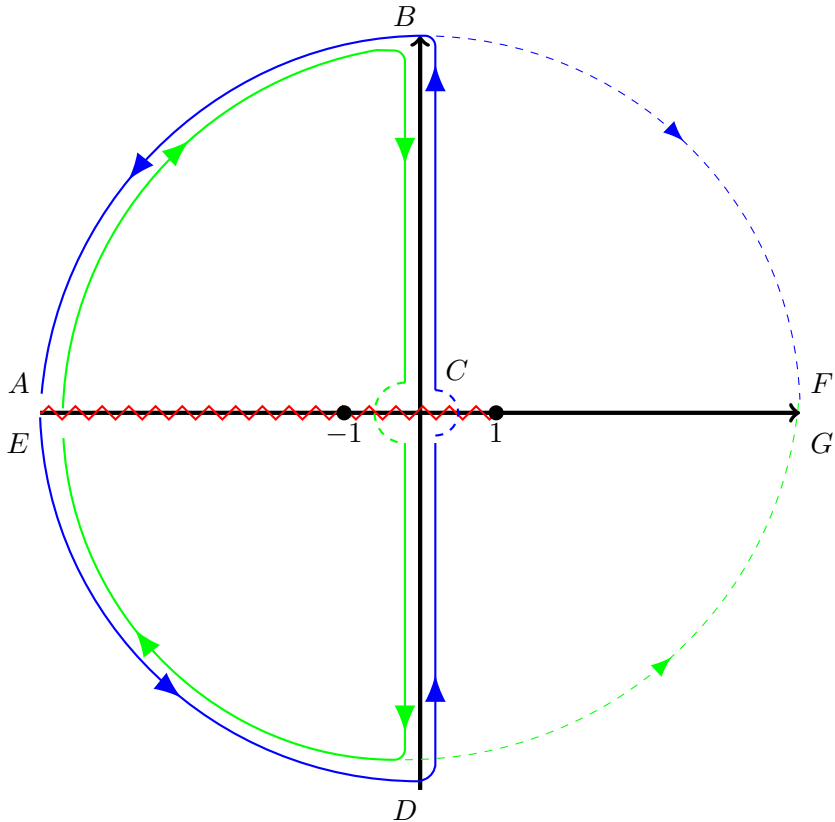
\begin{figure}
	\centering
	\begin{tikzpicture}
	\draw[->,ultra thick] (-5,0)--(5,0) node[right]{};
	\draw[->,ultra thick] (0,-5)--(0,5) node[above]{};
	\fill[black] (1,0) circle (0.04cm);
	\draw[snake=zigzag, color=red, thick] (1,0) -- (-5,0);
	\draw [color=blue,thick, 2multiarrow=latex] (-5,-0.06) arc (-178:-90:5cm){[rounded corners] --(0.2,-4.8)--(0.2,-0.4)};
	\draw[dashed,color=blue, thick]  (0.2,-0.3) arc (-90:90:0.3cm);
	\draw[color=blue,thick,2multiarrow=latex] {[rounded corners] (0.2,0.3)--(0.2,5) arc (90:175:5.2cm)} ;
	\draw [color=blue,dashed, middlearrow=latex] (0,5) arc (90:2:5cm); 
	\draw [color= green,thick,2multiarrow=latex] (-4.7,0.06) arc (178:100:5cm){[rounded corners] --(-0.2,4.8)--(-0.2,0.4)};
	\draw [color=green, thick, dashed] (-0.2,0.4) arc (90:270:0.4cm);
	\draw [color=green,thick,2multiarrow=latex]{[rounded corners] (-0.2,-0.4) -- (-0.2,-4.6) arc (-90:-177:4.5cm)} ;
	\draw [color=green,dashed, middlearrow=latex] (-0.2,-4.6) arc (-90:-5:5.2cm); 
	\fill[black] (1,0) circle (0.1cm);
	\fill[black] (-1,0) circle (0.1cm);
	\draw node[ anchor=north ] at	(1,0) {$1$};
	\draw node[ anchor=north ] at	(-1,0) {$-1$};	
	\draw node[ anchor=east ] at	(-5,0.4) {$A$};
	\draw node[ anchor=south ] at	(-0.2,5) {$B$};	
	\draw node[ anchor=south west ] at	(0.2,0.3) {$C$};	
	\draw node[ anchor=north ] at	(-0.2,-5) {$D$};	
	\draw node[ anchor=east ] at	(-5,-0.4) {$E$};	
	\draw node[ anchor=west ] at	(5,0.4) {$F$};	
	\draw node[ anchor=west] at	(5,-0.4) {$G$};	
	\end{tikzpicture}
	\caption{dS double trumpet continuations from AdS double trumpet}
	\label{eefig3}
\end{figure}

Let us end with one comment. The continuation from the  $-AdS_2$ to $dS_2$ can be done by taking $r\rightarrow \pm ir$. We have not specified, either in the case of the disk or the double trumpet, which of the two analytic continuations needs to be carried out. In the disk topology, as far as we can tell, it does not make a difference and both are allowed. 
For the double trumpet there are a total of four possibilities, see Fig.\ref{eefig3}, since we have two choices at either end. More precisely, as can be seen from the figure, we can start at $r=-\infty$ from either above the real axis at $A$ (green contour) or below the real axis at $E$ (blue contour). In each case there are again two possibilities after going along the imaginary axis. For example, for the blue contour, at $B$, we can continue to either $A$ or $F$. All of these seem to be allowed, as far as we can tell.  When dealing with eigenmodes for various determinants, e.g. a scalar laplacian, the modes need to be analytically continued, we find it is possible to do so for all four continuations. A proper understanding of this issue is also left for the future.

	\section{Conclusion}
\label{conclusion}
In this paper we have formulated the path integral for JT gravity in the second order formalism working directly with the metric and the dilaton. 
This allows one to incorporate matter easily as well. It   also allows one to  investigate whether spacetimes with different topologies can contribute to the path integral. 
We considered both AdS and dS spacetimes in our analysis. For matter, we have mostly  considered free bosons or fermions, but many of our conclusions extend more generally  to conformal matter and some  even more generally to non-conformal matter. 

Many questions remain to be followed up. 

We found agreement with the first order formalism for the  pure JT theory in the asymptotic AdS or dS limits. Away from this limit the path integral is more complicated due to the presence of  modes whose  wavelength along the boundary is short  and in particular   smaller than the radius of AdS or dS. Understanding the dynamics of these modes and carrying out the path integral more generally, is an important open question. One might hope that  some of the determinants which  arise can be made more tractable by a general analysis,  of the kind  that leads to a simplification  in the ratio of determinants $\sqrt{\text{det}'(P^\dagger P)}\over \text{det}(-\hat{\nabla}^2+2)$  for  compact manifolds (\cite{Witten:1991we,sarnak1987determinants,d1986determinants}),   and such simplifications might help with the problem.

In dS space the path integral is more non-trivial to define since the no-boundary proposal, which is what we explored here, requires one to include regions of spacetime with different signatures along the contour of  the integral. In our formulation we  continue modes analytically between  these regions while  carrying out the path integral. While this means the general metric, dilaton  and matter configurations we sum over are complex, the resulting path integral is  well defined, as best as we can tell. However, this needs to be investigated  more carefully further. Another issue for the dS case is how to deal with the  divergences which arise when we take the asymptotic limit. In  AdS space  these are dealt with by the standard procedure of holographic re-normalization  after adding suitable counter-terms which are local on the boundary. But it is less clear if such a procedure is the correct one to adopt in dS space. 
It would be worth understanding this issue  better as well. Calculating the path integral away from the asymptotic limit is especially important in the dS case, since we are interested in the wavefunction for a universe of finite size and at  finite values of the dilaton.  More  generally, it will be  worth  establishing firmly whether  a precise and sensible formulation of JT gravity can be given in dS space, as a start  even for spacetimes with the topology of the disk. 

Adding matter introduces some interesting new facets. Most important, as we have seen above, is the divergence which arises in the integral over moduli space while computing the path integral for the double trumpet topology  in the presence of matter.   This divergence is due to the Casimir effect leading to  a negative stress tensor  which diverges when the neck of the wormhole goes to zero size and is the analogue of the tachyon divergence in Bosonic string theory. While the precise result we get is simply that the path integral   is ill-defined, the divergence suggests that perhaps the presence of matter would cause the spacetime to disconnect into two pieces each of disk topology with one boundary. 	In fact, taking a cue from the divergence one could simply posit  that to get a well-defined theory one should  disallow higher topologies and only keep spacetimes with the  topology of  the disk. 
However this is  clearly too premature. Rather, further analysis is needed to see what are the possible ways to make  the path integral  well-defined and 
whether the divergence can be cured while still allowing the two boundaries to stay connected. It could well be that the fate of the wormhole depends on the details of how the   divergence is tamed. This is an important issue which we hope to investigate more fully in later work

The double trumpet geometry in AdS can be thought of as a contribution to the connected two point function of the partition functions of the two boundary theories, $\langle Z(\beta_1) Z(\beta_2)\rangle $. If the double trumpet and more generally wormholes survive in the path integral it would suggest that the boundary theories dual to JT gravity  involve some unconventional features, for example random couplings over which one needs to sum, resulting in this connected correlation. 
JT gravity can be obtained by dimensional reduction from higher dimensional  near-extremal black holes, \cite{Nayak_2018}\cite{Moitra:2019bub} and one expects that the higher dimensional systems which arise for example in string theory are more conventional with a Hamiltonian with fixed coupling constants. The dimensional reduction gives rise typically to a lot of extra matter.  It would be satisfying if the presence of  this matter itself  causes wormholes to pinch off and  the two boundaries to disconnect.  However, as was mentioned above,  this needs to be investigated further. It could also be that the dimensional reduction to two dimensions removes essential degrees of freedom in the system and thereafter  wormholes are allowed in the two dimensional theory \footnote{Another possibility is that the higher dimensional theory itself has wormholes due to averaging over the various ground states of the extremal system. We are grateful to Ashoke Sen for emphasizing this possibility to us.}. 

It is also worth drawing attention to the fact that the divergence mentioned above does not always arise. For example, in the double trumpet if one consider fermions instead of bosons with periodic boundary conditions along the two boundaries  then the Casimir effect  reverses its sign and the divergence goes away. 
The periodic  boundary conditions in the AdS context would be appropriate for computing an index $\text{Tr}((-1)^F e^{-\beta H})$ instead of the partition function $\text{Tr}(e^{-\beta H})$. Investigating the behavior of the divergence as  we vary the kind of matter and the boundary conditions we impose on it is another important direction to pursue.  

A similar  divergence for the double trumpet also arises in the dS case.  And as in the AdS case with fermions, by imposing periodic boundary conditions along the spatial boundary the  divergence  goes away. 
This suggests that for appropriate matter and  boundary conditions the wavefunction of the universe in the no-boundary proposal   can have an amplitude  to create multiple universes. It is clearly important to understand this more deeply. In this context one would also like to study the ``pants diagram" which corresponds to one universe tunnelling  into two, and more generally study the role of higher topologies. If wormhole do arise, the proper setting for quantum cosmology would be the third quantized one, where one is dealing with a multiverse. 

We have focussed on the no-boundary proposal in dS space here. There are other possibilities that are also  worth investigating {\cite{Vilenkin:1986cy,Vilenkin:1994rn,Vilenkin:2018dch}} .
One would also like  to add an inflaton to the theory and study the resulting dynamics of the system, including how it behaves in the presence of a potential for the inflaton, with metastable minima. It would also be interesting to explore the role of ``bra-ket" wormholes and their contribution to density matrices\cite{Chen:2020tes,PhysRevD.34.2267,Anous:2020lka}.

Finally, we have not explored the Lorentzian AdS theory here.  The fact that the quantum effects of matter only arise when we are away from the asymptotic AdS limit, with a boundary of finite length, is true for the Lorentzian case  as well, as was discussed in \cite{Moitra:2019xoj}. Thus, for discussing the effects of Hawking radiation by coupling the JT system to external radiation, as has been done quite extensively in recent literature,\cite{Moitra:2019xoj, Engelsoy:2016xyb,Almheiri:2019hni, Almheiri:2019psf, Penington:2019npb},  one needs to work away from the asymptotic limit. To  incorporate corrections beyond the leading semi-classical analysis (obtained with  $N$ matter species by taking  $N \rightarrow \infty$) one would  then also need to include the effects of the short wave length modes mentioned in the first few paragraphs above. The dynamics of these modes might in fact play a key role in the recovery of information during the evaporation process. A discussion of the Lorentzian theory will also be of interest from the point of view of potentially taming moduli space divergences that were discussed above, see \cite{Stanford:2020wkf}.

Clearly two dimensional gravity is a rich and fascinating playground. While results obtained in  lower dimensional settings might not always apply  to higher dimensions  one can hope to  gain some important insights from them.  We   look forward eagerly to exploring some of the questions mentioned above in the future.

	\acknowledgments
	\label{acknw}
	We thank Onkar Parrikar, Ashoke Sen and the TIFR String theory group including Abhijit Gadde, Gautam Mandal and  Shiraz Minwalla, for useful discussions.
	We acknowledge the support of the Government of India, Department of Atomic Energy, under Project No. 12-R{\&}D-TFR-5.02-0200 and support from the Infosys Foundation in form of the Endowment for the Study of the Quantum Structure of Spacetime. S. P. T. acknowledges support from a J. C. Bose Fellowship, Department of Science and Technology, Government of India. Most of all, we are grateful to the people of India for generously supporting research in String Theory.

	\newpage

	\appendix
	\section{Coordinate transformations}
	\label{coordtrsfs}

	\subsection{Euclidean AdS  disk}
	\label{eads2d}
	The metric for the Euclidean $AdS_2$ disk geometry is given by 
	\begin{equation}
	ds^2=(r^2-1)dt^2+\frac{dr^2}{(r^2-1)}\label{edkr}
	\end{equation}
	Redefining the coordinate $t,r$ as 
	\begin{equation}
	r=\cosh(\rho),\quad t=\theta\label{edkrrho}
	\end{equation}
	we get
	\begin{equation}
	ds^2=d\rho^2+\sinh^2(\rho)d\theta^2\label{edklerho}
	\end{equation}
	Defining the coordinate $r_{*}$ as 
	\begin{equation}
	r_{*}=\log(\tanh(\rho\over 2))\label{edkrsrho}
	\end{equation}
	The metric then becomes
	\begin{equation}
	ds^2=\frac{d\theta^2+dr_{*}^2}{\sinh^2(r_{*})}\label{edslers}
	\end{equation}
	Writing in term of the complex coordinates defined as
	\begin{equation}
	\zeta=\frac{\theta-i r_{*}}{2}\label{edszrs}
	\end{equation}
	we find that the metric is given by 
	\begin{equation}
	ds^2=-\frac{4d\zeta \,d\bar{\zeta}}{\sin^2(\zeta-\bar{\zeta})}\label{edslezeta}
	\end{equation}
	Futher doing a  coordinate transformation
	\begin{equation}
	\zeta=\arctan(x), \bar{\zeta}=\arctan(\bar{x})\Rightarrow x=\tan(\zeta),\bar{x}=\tan{\bar{\zeta}}\label{edspcze}
	\end{equation}
	it is easy to see that the metric becomes
	\begin{equation}
	ds^2=-\frac{4 dx d\bar{x}}{(x-\bar{x})^2}=\frac{dt^2+dz^2}{z^2} \quad \text{where} \,\, x=\frac{t+i z}{2}\label{edslepc}
	\end{equation}
	which can further be written in terms of $\hat{r}$ defined by
	\begin{equation}
	\hat{r}=\exp(r_{*})=\sqrt{r-1\over r+1}\label{edsplrrs}
	\end{equation}
	as
	\begin{equation}
	ds^2=4\frac{d\hat{r}^2+\hat{r}^2 d\theta^2}{(1-\hat{r}^2)^2}\label{edspolar}
	\end{equation}

	\subsection{Euclidean AdS double trumpet}
	\label{eads2dtg}
	The line element for this geometry is given by 
	\begin{equation}
	ds^2=(r^2+1)dt^2+\frac{dr^2}{(r^2+1)}\label{edtrt}
	\end{equation}
	The two boundaries correspond to the limits $r\rightarrow \infty$ and $r\rightarrow-\infty$. Performing the coordinate transformations
	\begin{equation}
	r=\sinh(\rho),\quad t=\theta,\label{edtrrho}
	\end{equation}
	we find that the metric is given by
	\begin{equation}
	ds^2=d\rho^2+\cosh^2(\rho)d\theta^2\label{edtlerho}
	\end{equation}
	It has to be noted that the $\theta$ direction is periodic with period $2\pi$.
	Defining $r_{*}$ coordinate as 
	\begin{equation}
	dr_{*}=\frac{d\rho}{\cosh(\rho)}\rightarrow r_{*}=2\arctan(\tanh(\frac{\rho}{2}))\Rightarrow \tan(r_*)=\sinh(\rho)\label{edtrsrho}
	\end{equation}
	In term of the $r_*$ coordinates the metric becomes
	\begin{equation}
	ds^2=\frac{dr_*^2+d\theta^2}{\cos^2(r_*)}\label{edletrs}
	\end{equation}
	This can be written in complex coordinates as 
	\begin{equation}
	\zeta=\frac{\theta-i r_*}{2},\bar{\zeta}=\frac{\theta+i r_*}{2}\label{edtrsze}
	\end{equation}
	the metric becomes
	\begin{equation}
	ds^2=\frac{4d\zeta\, d\bar{\zeta}}{\cosh^2(\zeta-\bar{\zeta})}\label{edtleze}
	\end{equation}
	To get it to the Poincare form, consider the further coordinate transformation
	\begin{equation}
	x=\coth(\zeta),\bar{x}=\tanh\bar{\zeta}\label{edtpcze}
	\end{equation}
	and hence the metric becomes
	\begin{equation}
	ds^2=-\frac{4 dx d\bar{x}}{(x-\bar{x})^2}\label{edtlepc}
	\end{equation}

	\subsection{de Sitter}
	\label{ds2cr}
	The metric for the de Sitter spacetime in 2 dimensions, for Lorentzian signature, is given in the global coordinates as
	\begin{equation}
	ds^2=-d\tau^2+\cosh^2(\tau)d\theta^2\label{dslerho}
	\end{equation}
	As before, we define the coordinate $r_{*}$ as 
	\begin{equation}
	dr_{*}=\frac{d\tau}{\cosh(\tau)}=2\arctan(\tanh(\tau\over 2))\Rightarrow \cosh\tau \cos r_*=1\label{dsrsrho}
	\end{equation}
	and the metric in the $r_*$ becomes
	\begin{equation}
	ds^2=\frac{d\theta^2-dr_{*}^2}{\cos^2(r_*)}\label{dslers}
	\end{equation}
	Doing the transformation
	\begin{align}
	r_*=\arctan(r)\label{dsrsr}
	\end{align}
	gives us
	\begin{align}
	ds^2=(1+r^2)d\theta^2-\frac{dr^2}{r^2+1}\label{dsrthcs}
	\end{align}
	From eq.\eqref{dsrsrho} and \eqref{dsrsr}, we find
	\begin{align}
	r=\sinh\tau\label{dsrthtaucs}
	\end{align} 
	Defining the null coordinates
	\begin{equation}
	\zeta^{\pm}=\frac{\theta\pm r_{*}}{2}\label{dsrsze}
	\end{equation}
	the line element becomes in these coordinates,
	\begin{equation}
	ds^2=\frac{4d\zeta^{+}d\zeta^{-}}{\cos^2(\zeta^{+}-\zeta^{-})}\label{dsleze}
	\end{equation}
	The coordinate transformation 
	\begin{equation}
	x^+_p= \cot\zeta^+ , x^-_p=-\tan \zeta^-\label{dspcgl}
	\end{equation}
	The line element  eq.\eqref{dsleze} becomes
	\begin{equation}
	ds^2=4 \frac{dx^+_p dx^-_p}{(x^+_p- x^-_p)^2}\label{dspclc}
	\end{equation}
	
	\section{Extrinsic Trace and Schwarzian action}
	\label{etlds2}
	\subsection{Euclidean AdS disk}
	\label{eadsextr}
	Consider the metric 
	\begin{equation}
	ds^2=\frac{dr^2}{r^2-1}+(r^2-1) d\theta^2\label{adsexkgb}
	\end{equation}
	The boundary is located in the region where $r\rightarrow \infty$. The general boundary curve is specified by $(r(u),\theta(u))$ where $u$ is the proper boundary time. The line element on the boundary is 
	\begin{equation}
	ds^2\bigg\vert_\del={du^2\over \epsilon^2}\label{adsbm1}
	\end{equation}
	where $\epsilon\ll 1$. The tangent vector to the boundary curve is given by 
	\begin{equation}
	{\bf V}\equiv \del_u=r'(u)\del_r+\theta'(u)\del_\theta\label{adsexkgbv}
	\end{equation}
	The unit normalized normal vector is then given by 
	\begin{equation}
	n^r=\frac{(r^2-1)^\frac{3}{2}\theta'}{\sqrt{\theta'^2(r^2-1)^2+r'^2}}\,\quad n^\theta=-\frac{r'}{\sqrt{(r^2-1)}\sqrt{\theta'^2(r^2-1)^2+r'^2}}\label{adsexkgbn}
	\end{equation}
	The extrinsic trace is given by 
	\begin{align}
	K&={\nabla_\mu n^\mu}=\del_r n^r=\frac{\del_u n^r}{r'}\nonumber\\
	&=\frac{\sqrt{r^2-1} \left(\left(r^2-1\right) r' \theta ''(u)+\theta ' \left(r''+r \left(-r r''+3 r'^2+\left(r^2-1\right)^2 \theta '^2\right)\right)\right)}{\left(r'^2+\left(r^2-1\right)^2 \theta '^2\right)^{3/2}}\label{adsexkgbgc}
	\end{align}
	where it is understood that $r,\theta$ are functions of $u$.
	Differentiating the line element relation gives
	\begin{align}
	2 \left(r r' \theta '^2-\frac{r r'^3}{\left(r^2-1\right)^2}+\frac{r' r''}{r^2-1}+\left(r^2-1\right) \theta ' \theta ''(u)\right)=0\label{lineeldi}
	\end{align}
	Using this to simplify the eq.\eqref{adsexkgbgc}, we get
	\begin{align}
	K=\frac{ r-\epsilon ^2 r''}{({r^2-1}) {\epsilon  } \theta '}\label{fulsadspexk}
	\end{align}
	Noting from eq.\eqref{adsexkgb} and eq.\eqref{adsbm1} that to leading order
	\begin{equation}
	r^2{\theta'^2}\simeq{\epsilon^{-2}}\Rightarrow r\simeq\frac{1}{\epsilon\theta'} \label{adsexkrtu}
	\end{equation}
	The above relation need to be extended to one higher to obtain the leading schwarzian term in the extrinsic trace. Doing so gives 
	\begin{align}
	r\simeq \frac{1}{\epsilon  \theta '(u)}+\epsilon  \left(\frac{\theta '(u)}{2}-\frac{\theta ''(u)^2}{2 \theta '(u)^3}\right)\label{rthadsqd}
	\end{align}
	With this relation between $\theta , r$, expanding the extrinsic trace to quadratic order in $\epsilon$, we get
	\begin{align}
	K&=1+\epsilon^2\frac{ \left(-3 \theta ''(u)^2+\theta '^4+2 \theta ^{(3)}(u) \theta '\right)}{2\theta '^2}+O\left(\epsilon^{3}\right)\nonumber\\
	&=1+\epsilon^2\pqty{\text{Sch}\left(\tan({\theta(u)\over 2}),u\right)}+\order(\epsilon^{3})\label{adsexkschu}
	\end{align}
	We will now provide some formulae that will be useful later on in appendix \ref{nadsds}.
	
	\subsection{Euclidean AdS double trumpet}
	\label{dbexk}
	The line element is given by eq.\eqref{dbmet}. The geometry now has two boundaries in the asymptotic region, i.e near $r\rightarrow\pm\infty$. We need to be a bit careful when evaluating  the boundary terms in the $JT$ action as the relative signs play a crucial role in the final result of the path integral. Consider a curve given by $(r(u),\theta(u))$ where $u$ is proportional to the boundary proper time. We have
	
	\begin{equation}
	\del_u=r'(u)\del_r+\theta'(u)\del_\theta\label{dbexkgbv}
	\end{equation}
	The unit normalized normal vector is then given by 
	\begin{equation}
	n^r=\pm\frac{(r^2+1)^\frac{3}{2}\theta'}{\sqrt{\theta'^2(r^2+1)^2+r'^2}}\,\quad n^\theta=\mp\frac{r'}{\sqrt{(r^2+1)}\sqrt{\theta'^2(r^2+1)^2+r'^2}}\label{dbexkgbn}
	\end{equation}
	where the upper sign corresponds to the right boundary and the lower sign to the left boundary.
	The extrinsic trace is given by 
	\begin{align}
	K&={\nabla_\mu n^\mu}=\del_r n^r=\frac{\del_u n^r}{r'}\nonumber\\
	&=\frac{\sqrt{r^2+1} \left(\left(r^2+1\right) r' \theta ''+\theta ' \left(r \left(3 r'^2+\left(r^2+1\right)^2 \theta '^2\right)-\left(r^2+1\right) r''\right)\right)}{\left(r'^2+\left(r^2+1\right)^2 \theta '^2\right)^{3/2}}\label{dbexkgbgc}
	\end{align}
	where it is understood that $r,\theta$ are functions of $u$.  Consider the situation when both the boundaries of the double trumpet geometry have the line element as given in eq.\eqref{adsbm1} with the same parameter $\epsilon$. The coordinate $u$ is chosen such that its range is same as the range of the $\theta$ coordinate. From eq.\eqref{dbmet}, we have to leading order
	\begin{equation}
	r^2{\theta'^2}\simeq{1\over \epsilon^{2}}\Rightarrow r\simeq\pm\frac{1}{\epsilon\theta'} \label{dbexkrtu}
	\end{equation}
	where we need to use $+$ sign at the right boundary and $-$ at the left boundary. More generally, the parameter $\epsilon$ need not be same at both boundaries. 
	The above relation need to be extended to one higher to obtain the leading Schwarzian term in the extrinsic trace. Doing so gives 
	\begin{align}
	r\simeq \pm \pqty{\frac{1}{\epsilon \theta '(u)}-\epsilon \left(\frac{\theta '(u)}{2}+\frac{\theta ''(u)^2}{2 \theta '(u)^3}\right)}	\label{dbrthqd}
	\end{align}
	With this relation between $\theta , r$, expanding the extrinsic trace to quadratic order in $\epsilon$, we get
	\begin{align}
	K&=1-\epsilon^2\frac{ \left(3 \theta ''^2+\theta '^4-2 \theta ^{(3)} \theta '\right)}{2\theta '^2}+O\left(\epsilon^3\right)\nonumber\\
	&=1+\epsilon^2{\text{Sch}\left(\tanh({\theta(u)\over 2}),u\right)}+\order(\epsilon^3)\label{dbexkschu}
	\end{align}
	We get the above action for both the signs in eq.\eqref{dbrthqd}, or in other words, at both the boundaries. This relative plus sign in between the boundary terms at both the boundaries is important because of the dependence on the moduli of the integral over the large diffeomorphisms as we shall see later in appendix \ref{dbapdx}. The boundary term of the $JT$ action eq.\eqref{bjtact}, at either boundary,  then becomes
	\begin{align}
	S_{JT,\del}=-\frac{\phi_B\epsilon}{8\pi G}\int du\,\text{Sch}\left(\tanh({\theta(u)\over 2}),u\right)\label{dbbjtact}
	\end{align}
	Denoting the boundaries $\del_1,\del_2$ and	writing the two boundary terms explicitly we have
	\begin{align}
	S_{JT,\del}=S_{JT,\del_1}+S_{JT,\del_2}=-\frac{\phi_B\epsilon}{8\pi G}\pqty{\int_{\del_1} du\,\text{Sch}\left(\tanh({\theta(u)\over 2}),u\right)+\int_{\del_2} du\,\text{Sch}\left(\tanh({\theta(u)\over 2}),u\right)}\label{dbbjtact12}
	\end{align}

	\subsection{de Sitter}
	\label{etds2gc}
	Consider the metric in global coordinates given by 
	\begin{equation}
	ds^2=-d\tau^2+\cosh^2\tau d\theta^2\label{dsexkgb}
	\end{equation}
	 The general boundary curve is specified by $(\tau(u),\theta(u))$ where $u$ is the proper boundary time. The line element on the boundary is given by
	\begin{equation}
	ds^2\bigg\vert_\del={du^2\over \epsilon^2}\label{dsbm1}
	\end{equation}
	 The tangent vector to the boundary curve is given by 
	\begin{equation}
	{\bf V}\equiv \del_u=\tau'(u)\del_\tau+\theta'(u)\del_\theta\label{dsexkgbv}
	\end{equation}
	The unit normalized normal vector is then given by 
	\begin{equation}
	n^\tau=-\frac{\theta'(u)\cosh\tau}{\sqrt{\theta'(u)^2\cosh^2\tau-\tau'(u)^2}}\,\quad n^\theta=-\frac{\tau'(u)}{\cosh\tau\sqrt{\theta'(u)^2\cosh^2\tau-\tau'(u)^2}}\label{dsexkgbn}
	\end{equation}
	The extrinsic trace is given by 
	\begin{align}
	K&=\frac{-\theta '' \tau ' \cosh \tau +\theta ' \left(\tau '' \cosh \tau -2 \tau '^2 \sinh \tau \right)+\theta '^3 \sinh \tau  \cosh ^2\tau }{\left(\theta '^2 \cosh ^2\tau -\tau '^2\right)^{3/2}}\label{dsexkgbgc}
	\end{align}
	where it is understood that $\tau,\theta$ are functions of $u$.
	Noting from eq.\eqref{dsexkgb} and eq.\eqref{dsbm1} that to leading order
	\begin{equation}
	{e^{2\tau}\over 4}{\theta'(u)^2}\simeq{\epsilon^{-2}}\Rightarrow \tau (u)\simeq-\log \left(\frac{  \epsilon\theta '}{2} \right) \label{dsexkrtu}
	\end{equation}
	Correcting this relation to one higher order in $\epsilon$, we have
	\begin{align}
	\tau(u)=-\ln({\epsilon\theta'\over 2})+\frac{\epsilon^2}{2}\pqty{{\theta''^2\over \theta'^2}-{\theta'^2\over 2}}\label{dsgbtthrel}
	\end{align}
	and expanding to quadratic order in $\epsilon$, we get
	\begin{align}
	K&=1-\epsilon^2\frac{ \left(-3 \theta ''^2+\theta '^4+2 \theta'''\theta '\right)}{2 \theta '^2}+O\left(\epsilon^{3}\right)
	=1-\epsilon^2{\text{Sch}\left(\tan({\theta(u)\over 2}),u\right)}+\order(\epsilon^{3})\label{dsexkschu}
	\end{align}
	Also, we can simplify the extrinsic trace formula eq.\eqref{dsexkgbgc} as follows using eq.\eqref{dsexkgb} and eq.\eqref{dsbm1}. 
	Doing so, we get
	\begin{align}
	K=\frac{\tau''+(\epsilon^{-2}+\tau'^2)\tanh\tau}{\epsilon^{-1}\sqrt{\epsilon^{-2}+\tau'^2}}\label{exactexk}
	\end{align}

	We now provide some useful formulae of the same calculations in the metric eq.\eqref{minds}.  The unit normal vector components are given by 
	\begin{align}
	n^r=\frac{(\sqrt{r^2+1})^3 \theta '}{\sqrt{\left(r^2+1\right)^2 \theta '^2-r'^2}},\quad n^\theta=\frac{ r '}{\sqrt{r^2+1}\sqrt{\left(r^2+1\right)^2 \theta '^2-r'^2}}\label{dsnrnt}
	\end{align}
	The extrinsic trace is then given by 
	\begin{align}
	K=\frac{\sqrt{r^2+1} \left(\theta ' \left(\left(r^2+1\right) r''+r \left(\left(r^2+1\right)^2 \theta '^2-3 r'^2\right)\right)-\left(r^2+1\right) r' \theta ''\right)}{\left(\left(r^2+1\right)^2 \theta '^2-r'^2\right)^{3/2}}\label{dsekrth}
	\end{align}
	Expanding $r(u)$ in terms of $\theta(u)$ as 
	\begin{align}
	r(u)=\frac{1}{\epsilon\theta'}+\frac{\epsilon  }{2 \theta '(u)}\left(\frac{\theta ''(u)}{\theta '(u)}\right)^2-\frac{1}{2} \epsilon  \theta '(u)+\order{(\epsilon^2)}\label{dsrthbdre}
	\end{align}
	we get the extrinsic trace as in eq.\eqref{dsexkschu}.
	We now derive some formulae that will be used in appendix \ref{nadsds}. Taking 
	\begin{align}
	\tau\rightarrow\tau-i\frac{\pi}{2}\label{taunads}
	\end{align}
	we see that the metric in eq.\eqref{dsexkgb} becomes
	\begin{align}
	ds^2=-(d\tau^2+\sinh^2\tau d\theta^2)\label{nadsgbmet}
	\end{align}
	which is negative of the metric for global $AdS_2$. Using eq.\eqref{taunads} then becomes
	\begin{align}
	K=-\frac{\tau' \sinh (\tau) \theta ''+\theta ' \left(\cosh (\tau) \left(2 \tau'^2+\sinh ^2(\tau) \theta '^2\right)-\tau'' \sinh (\tau)\right)}{\left(\tau'^2+\sinh ^2(\tau) \theta '^2\right)^{3/2}	}\label{nadsexk}
	\end{align}
	Now, taking the line element on the boundary to be
	\begin{align}
	ds^2=-\sinh^2\tau_0\, du^2\label{nadsbdele}
	\end{align}
	where $\tau_0$ is an arbitrary fixed value of $\tau$. The boundary relation then reads
	\begin{align}
	-\tau'^2-\sinh^2\tau\,\,\theta^2=-\sinh^2\tau_0\label{nadsbdrel}
	\end{align}
	where $\tau_0$ is an arbitrary value.	Expanding $\tau$ and $\theta$ as 
	\begin{align}
	\tau=\tau_0+\delta\tau,\quad \theta=u+\delta\theta\label{nadsexp}
	\end{align}
	and solving for $\delta\tau$ iteratively to quadratic order in $\delta\theta$, we get
	\begin{align}
	\delta \tau(u)=-\tanh \tau_0 \delta \theta '(u)+\frac{1}{4}  \tanh \tau_0 \text{sech}^2\tau_0 \left((\cosh (2 \tau_{0})+3) \delta \theta '(u)^2-2 \delta \theta ''(u)^2\right)+O\left(\kappa ^3\right)\label{nadetaudtheqo}
	\end{align}
	Using eq.\eqref{nadetaudtheqo} to expand the extrinsic trace eq.\eqref{nadsexk} to quadratic order in $\delta\theta$, we find
	\begin{align}
	K=&-\coth \tau_0-  \text{csch}\tau_0 \text{sech}\tau_0 \left(\delta \theta ^{(3)}(u)+\delta \theta '(u)\right)\nonumber\\
	&-\frac{1}{4}  \frac{\text{csch}\tau_0} {\text{cosh}^3\tau_0} \left(4 \delta \theta'''^{2}+4 \delta \theta ^{(4)} \delta \theta ''-(3 \cosh (2 \tau_{0})+5) \delta \theta ''^2+2 \sinh ^2\tau_0 \delta \theta'^2-2 (\cosh (2 \tau_{0})+3) \delta \theta ^{(3)} \delta \theta '\right)\label{nadsqdexk}
	\end{align}
	\section{Zeta-function regularization}
	\label{zetaregapp}
	In this appendix, we mention some useful formulae pertaining to Zeta-function regularization that are used in this work. The Riemann-zeta function, denoted by $\zeta(s)$ is, given by 
	\begin{equation}
	\zeta(s)=\sum_{m=1}\frac{1}{m^s}\label{rzetad},\quad \quad 
	\end{equation}
	and it has the specific values
	\begin{equation}
	\zeta(0)=-\frac{1}{2},\zeta'(0)=-\ln\sqrt{2\pi}.\label{rzeta0va}
	\end{equation}
	The generalized Zeta function, $\zeta(s,m_0)$ is given by 
	\begin{equation}
	\zeta(s,m_0)=\sum_{m=0}\frac{1}{(m+m_0)^s},\quad  \quad \zeta'(0,m_0)=\ln\frac{\Gamma(m_0)}{\sqrt{2\pi}}\label{gzetadef}
	\end{equation}
	Consider the sum $\sum_{m>1}\ln(\alpha\over m)$. Defining $\zeta_A(s)$ as 
	\begin{equation}
	\zeta_A(s)=\sum_{m=1}\frac{1}{\lambda_m^s}\Rightarrow\sum_{m=2}\log\lambda_m=-\zeta_A'(0)-\ln\lambda_1\label{zetaa}
	\end{equation}
	and using $\lambda_m=\frac{\alpha}{m}$ in the above, we get that
	\begin{align}
	\zeta_A(s)=\sum_{m=1}\frac{m^s}{\alpha^s}=\frac{\zeta(-s)}{\alpha^s}
	\Rightarrow \zeta_A'(0)=-\zeta'(0)-\zeta(0)\ln\alpha=\ln(\sqrt{2\pi\alpha})\label{zetaazeta}
	\end{align}
	Using the result eq.\eqref{zetaazeta} in eq.\eqref{zetaa}, we get
	\begin{equation}
	\sum_{m=2}\ln({\alpha\over m})=-\ln\sqrt{2\pi\alpha}-\ln\alpha=-\ln\sqrt{2\pi\alpha^3}\label{aovermzeta}
	\end{equation}
	Now consider the sum $\sum_{m=2}\ln(m-m_0)$. Defining $\zeta_B(s,m_0)$ as
	\begin{equation}
	\zeta_B(s,m_0)=\sum_{m=0}\frac{1}{\lambda_m^s}\Rightarrow\sum_{m=2}\log\lambda_m=-\zeta_B'(0,m_0)-\ln\lambda_1-\ln\lambda_0\label{zetab}
	\end{equation}
	Using the value of $\lambda_m=m-m_0$, we see that 
	\begin{align}
	\zeta_B(s,m_0)=\sum_{m=0}\frac{1}{(m-m_0)^s}=\zeta(s,-m_0)\Rightarrow \zeta_B'(0,m_0)=\zeta'(0,-m_0)=\ln{\Gamma(-m_0)\over \sqrt{2\pi}}\label{zetabgzeta}
	\end{align}	
	Using the result of eq.\eqref{zetabgzeta} in eq.\eqref{zetab}, we get
	\begin{equation}
	\sum_{m=2}\ln(m-m_0)=-\ln{\Gamma(-m_0)\over \sqrt{2\pi}}-\ln(1-m_0)-\ln(-m_0)=-\ln({\Gamma(2-m_0)\over \sqrt{2\pi}})\label{lnm0}
	\end{equation}
	Generalizing the above results, we note here a general formula for the zeta function-regularized product,
	\begin{align}
	\prod_{m\geq 2}\frac{\alpha}{m}\frac{(m-m_1)\dots (m-m_p)}{(m-\tilde{m}_1)\dots (m-\tilde{m}_q)}\rightarrow \frac{(2\pi)^{\frac{p-q}{2}}}{\sqrt{2\pi \alpha^3}}\frac{\Gamma(2-\tilde{m}_1)\dots \Gamma(2-\tilde{m}_q)}{\Gamma(2-{m}_1)\dots \Gamma(2-{m}_p)}\label{zetareggen}
	\end{align}

	\section{Conformal Killing Vectors in the Euclidean AdS Disk }
	\label{ckvdisk}
	In this appendix, we shall explicitly evaluate the conformal Killing vectors for the Euclidean AdS disk topology with the metric given in polar coordinates as in eq.\eqref{metads2}. First, note that conformal Killing vectors satisfy the condition $PV=0$. This immediately implies that $P^\dagger PV=0$ for a CKV. Thus, we only look for CKVs in the sector of zero modes of the operator $P^\dagger P$. For a general zero mode of this operator, we can write the vector field as
	\begin{align}
	V^a_{m}=k_1 \nabla^a\psi_m +k_2 \epsilon^{ab}\nabla_b\psi_m\label{genzero}
	\end{align}
	where $\psi_{m}$ is given in eq.\eqref{psiform} and the condition that the vector field be real means that $k_1, k_2$ are real. The metric components are then given by 
	\begin{align}
	(PV_m)_{rr}&=\frac{2 \hat{c}_{m} \left(m^2-1\right) e^{i \theta  m} \left(\frac{r-1}{r+1}\right)^{\frac{\abs{m} }{2}} (k_1 \abs{m} +i k_2 m)}{\left(r^2-1\right)^2}\nonumber\\
	(PV_m)_{r\theta}&=\frac{2 i \hat{c}_{m} \left(m^2-1\right) e^{i \theta  m} \left(\frac{r-1}{r+1}\right)^{\frac{\abs{m} }{2}} (k_1 m+i k_2 \abs{m} )}{r^2-1}\nonumber\\
	(PV_m)_{
		\theta\theta}&=-2 i \hat{c}_{m} \left(m^2-1\right) e^{i \theta  m} \left(\frac{r-1}{r+1}\right)^{\frac{\abs{m} }{2}} (k_2 m-i k_1 \abs{m} )\label{ckvmetper}
	\end{align}
	It can be seen from the above that if all the metric components $\delta g_{ab}=(PV_{m})_{ab}$ were to vanish, the possibilities are 
	\begin{align}
	m=0,1,-1 ,\quad k_1\neq 0, k_2=0\label{globalckv}
	\end{align}
	or 
	\begin{align}
	k_2=i k_1\text{sign}(m)\label{localckv}
	\end{align}
	 {The diffeomorphisms corresponding to $k_2\neq 0, k_1=0$ in eq.\eqref{globalckv} are exact isometries of  $AdS_2$. This is straightforward to see since if $m=0\pm1$ it follows from eq.(\ref{ckvmetper}) that $PV=0$, and with $k_1=0$ is follows that $\nabla\cdot V=0$, leading to the conclusion that $\nabla_a V_b+\nabla_b V_a=0$.}	
	
	Among the set of CKVs those   given by  eq.\eqref{globalckv}  correspond to vector fields of the form
	\begin{equation}
	V_\mu=q_0 \nabla_\mu\hat{\psi}_0+q_1\nabla_\mu \hat{\psi}_1+q_2\nabla_\mu \hat{\psi}_2
	\end{equation}
	where 
	\begin{equation}
	\hat{\psi}_0=r,\hat{\psi}_1=\sqrt{r^2-1} \cos \theta ,\hat{\psi}_2= \sqrt{r^2-1} \sin \theta 
	\end{equation}
	and $q_i$ are arbitrary real constants. These give rise to an $SL(2,R)$ algebra. The above functions $\hat{\psi}_i$ are in fact  linear combinations of the solutions appearing in eq.\eqref{psiform} for $m=1,-1,0$ modes.
	
{	An important observation to note here is the following. The CKVs both in eq.\eqref{globalckv} and eq.\eqref{localckv} do not satisfy the boundary conditions eq.\eqref{veclapdetbc} corresponding to that of  allowed small diffeomorphisms. Thus on the disk  the operator $P^\dagger P$ has no zero modes. }

	\section{Estimate of the inner product of  metric perturbations arising from large and small diffeomorphism}
	\label{offdiagme}
	We are interested in calculating the inner product
	\begin{align}
	\langle PV_s,PV_L\rangle\over \sqrt{\langle PV_s,PV_s\rangle\langle PV_L, PV_L\rangle}\label{wantinp}
	\end{align}
	Let us calculate each of the terms in the above expression. In terms of scalar field, $\psi_0$,  $\xi_\lambda, \psi_\lambda$, the vector fields corresponding to large and small diffeomorphisms are given by 
	\be
	\label{larsminsc}
	V_L^a=\epsilon^{ab}\nabla_b\psi_0
	\ee
	and 
	\begin{align}
	V_{s,\lambda,m}^a=\nabla^a\xi_{\lambda,m}+\epsilon^{ab}\nabla_b\psi_{\lambda,m}\label{smalldifexp}
	\end{align}
	respectively. 
	
	The large diffeomorphism  $V_L^a$ being a zero mode of the $P^\dagger P$ translates into the condition that 
	\begin{align}
	\nabla^2\psi_0=2\psi_0\label{lareq}
	\end{align}
	and the small diffeomorphism being an eigenmode of $P^\dagger P$ with eigenvalue $\lambda$, i.e $P^\dagger P V_s=\lambda V_s$ translates into
	\begin{align}
	\nabla^2\psi_{\lambda, m}=(2+\lambda)\psi_{\lambda,m}\nonumber\\
	\nabla^2\xi_{\lambda,m} =(2+\lambda)\xi_{\lambda,m}\label{smdifsceq}
	\end{align}
	The first of the boundary conditions for  the small diffeomorphisms in eq.\eqref{veclapdetbc} just becomes
	\begin{align}
	V\cdot n=0&\Rightarrow V^r_{s,\lambda,m}=0\nonumber\\&\Rightarrow  g^{rr}\del_r\xi_{\lambda,m}+\del_\theta\psi_{\lambda,m}=0
	\label{rsqmval}
	\end{align}
	which gives
	\be
	\label{qmval}
	g^{rr}\del_r\xi_{\lambda,m}+i m \psi_{\lambda,m}=0
	\ee at the boundary $r=r_B$.
	The second condition becomes
	\begin{align}
	t^a n^b (PV_{s,\lambda,m})_{ab}=0\Rightarrow (PV_{s,\lambda,m})_{\theta r}=0\nonumber\\
	\Rightarrow \nabla_\theta (V_{s,\lambda,m})_r+\nabla_r (V_{s,\lambda,m})_\theta=0\nonumber\\
	\Rightarrow\del_r V^\theta_{s,\lambda,m}=0\label{vsmseco}
	\end{align},
	at $r=r_B$, 
	where the last line is obtained by using the first condition in  eq\eqref{qmval}. Expressed in terms of the $\xi_{\lambda,m},\psi_{\lambda,m}$, this condition becomes
	\begin{align}
	\del_r V^\theta_{s,\lambda,m}=0\Rightarrow im\del_r(g^{\theta\theta}\xi_{\lambda,m})-{\del_r^2\psi_{\lambda,m}}=0\label{qminal}
	\end{align}
	To understand  how these conditions can be met in the asymptotic AdS limit when $r_B \rightarrow \infty$
	more clearly, let us first examine the scalar field equation for $\psi_{\lambda,m}$ carefully. The general solution is given by 
	\begin{align}
	\psi_{\lambda,m}=e^{im\theta}\pqty{c_1 P_{ v -\half}^{m}(r)+c_2 Q_{ v -\half}^{m}(r)}\label{psimlam}
	\end{align}
	where
	\begin{align}
	 v =\half\sqrt{9+4\lambda}\label{nuval}
	\end{align}
	and $P_\alpha^m, Q_\alpha^m$ are the associated Legendre functions of the first and second kind respectively.
	Regularity at the origin of the above solution forces us to choose $c_2=0$. Then behaviour of this solution for $r\gg 1$ can be immediately obtained from the asymptotic forms of the associated Legendre function and is given by 
	\begin{align}
	\psi_{\lambda,m}(r)=c_1(r^{-\half- v }f_{1}(\lambda,m)+r^{-\half+ v }g_{1}(\lambda,m))\label{psilmexp}
	\end{align}
	where $f_1(\lambda,m),g_1(\lambda,m)$ are some specific expressions which can be read off from the asymptotic behaviour of the associated Legendre functions and the $\theta$ dependence is not explicitly shown. From the above, it is clear that if $ v $ is imaginary, the expression has the functional form
	\begin{align}
	\psi_{\lambda,m}(r)=\frac{c_1}{\sqrt{r}}F\cos(w\log r-\beta)\label{psilmase}
	\end{align}
	where $w=-i v $ and $F,\beta$ are given by 
	\begin{align}
	F=2\sqrt{f_1 g_1},\tan\beta=\frac{i(g_1-f_1)}{g_1+f_1}\label{Fbetaex}
	\end{align}
	It is clear from the expression eq.\eqref{psilmase} that the magnitude of the scalar field solution $\psi_{\lambda,m}\sim \frac{c_1}{\sqrt{r}}$.
	The same analysis holds for $\xi_{\lambda,m}$, albeit with a different constant instead of $c_1$, say $d_1$. 
	It is now clear that one way to satisfy the conditions eq.\eqref{qmval} and eq.\eqref{qminal} is to choose the constants $c_1,d_1$  such that $d_1\sim \frac{c_1}{r_B}$ so that in  eq.\eqref{qmval}, the two terms are comparable and cancel each other whereas in eq.\eqref{qminal}, the second term dominates,
	giving rise to the condition
	\be
	\label{evalset1}
	\partial_r^2 \psi_{\lambda,m}\bigg\vert_{r=r_B}\simeq 0 
	\ee
	which determines eigenvalue $\lambda$.
	The alternate way is to choose the constants $c_1,d_1$ such that $c_1\sim \frac{d_1}{r_B}$ so that in  eq.\eqref{qminal}, the terms are comparable and cancel each other whereas in eq.\eqref{qmval}, the first term dominates giving rise to the condition 
	\be
	\label{evalset2}
	\del_r\xi_{\lambda,m}\bigg\vert_{r=r_B}\simeq 0
	\ee
	thus determining the eigenvalue $\lambda$.
	These two ways of meeting the conditions eq.(\ref{qmval}) and eq.(\ref{qminal})  give rise to two sets of eigenvalues and in fact exhaust all the possibilities.

	We now proceed to evaluate the various inner products. We shall first evaluate the expressions in general and then take the asymptotic AdS limit to get the estimates.
	Consider the inner product of two metric perturbations, one with a large diffeomorphism and one with a small diffeomorphism.
	\begin{align}
	\langle PV_{s,\lambda,m}, PV_{L,-m} \rangle&=\langle V_{s,\lambda,m}, P^\dagger P V_{L,-m}\rangle +\int_\del ds n^a V_{s,\lambda,m}^b (PV_{L,-m})_{ab}\nonumber\\
	&=\int_{r=r_B} d\theta \sqrt{\gamma}n^r V_{s,\lambda,m}^\theta (PV_{L,-m})_{r\theta}\nonumber\\
	&=2\pi r_B^2 V_{s,\lambda,m}^\theta (PV_{L,-m})_{r\theta}\bigg\vert_{r=r_B}\label{smlarpvinp}
	\end{align}
	In the large $r_B$  limit, from eq.\eqref{crdgvc}, we have
	\begin{align}
	(PV_{L,-m})_{r,\theta}=-{2\hat{c}_{-m}\abs{m}(m^2-1)\over r_B^2}\label{pvlrtbv}
	\end{align}
	From the eq.\eqref{larsminsc}, we get
	\begin{align}
	V^\theta_{s,\lambda,m}&= g^{\theta\theta }im\xi_{\lambda,m} - \del_r\psi_{\lambda,m}\nonumber\\
	&\simeq {im\over r^2}\xi_{\lambda,m} - \del_r\psi_{\lambda,m}\label{vstase}
	\end{align}
	which for either set of eigenvalues determined by eq.\eqref{evalset1}  or  eq.\eqref{evalset2} becomes, using the equations of motion, 
	\begin{align}
	V^\theta_{s,\lambda,m}\bigg\vert_{\del}&\simeq - \frac{(\lambda +2)\psi_{\lambda,m}(r_B)}{2r_B}\label{vsthaads2}
	\end{align}
	Then, using eq.\eqref{pvlrtbv} and eq.\eqref{vsthaads2} in eq.\eqref{smlarpvinp}, we get
	\begin{align}
	\langle PV_{s,\lambda,m}, PV_{L,-m} \rangle&={2\pi  (\lambda +2)\psi_{\lambda,m}(r_B)\hat{c}_{-m}\abs{m}(m^2-1)\over r_B}\label{smldinpd}
	\end{align}
	The inner product of two metric deformations both corresponding to large diffeomorphisms has already been obtained in eq.\eqref{innerlarged} which in the asymptotic AdS limit becomes
	\begin{align}
	\langle PV_{L,m},PV_{L,-m}\rangle \simeq 4\pi \hat{c}_m\hat{c}_{-m}\abs{m}(m^2-1)\label{pvlpvlasy}
	\end{align}
	We now are left to calculate inner product of two metric perturbations corresponding to small diffeomorphisms. This is given by 
	\begin{align}
	\langle PV_{s,\lambda_1,m_1},PV_{s,\lambda_2,m_2}\rangle=\lambda_1\delta_{\lambda_1,\lambda_2}\delta_{m_1,-m_2}\langle V_{s,\lambda_1,m_1},V_{s,\lambda_2,m_2} \rangle\label{pvspvs}
	\end{align}
	The inner product of two small diffeomorphisms can be manipulated as follows.
	\begin{align}
	\langle V_{s,\lambda,m},V_{s,\lambda,-m} \rangle&=\int d^2x\sqrt{g}( \nabla^a\xi_{\lambda,m}(V_{s,\lambda,-m})_a+\epsilon^{ab}\nabla_b\psi_{\lambda,m} (V_{s,\lambda,-m})_a)\nonumber\\
	&=-\int d^2x\sqrt{g}( \xi_{\lambda,m}g^{ab}+\epsilon^{ab}\psi_{\lambda,m} )\nabla_b(V_{s,\lambda,-m})_a+\int_{\del}ds \,\,n_b(g^{ab} \xi_{\lambda,m} +\epsilon^{ab}\psi_{\lambda,m})(V_{s,\lambda,-m})_a\nonumber\\
	&=-\int d^2x\sqrt{g}( \xi_{\lambda,m}\nabla^2\xi_{\lambda,-m}+\psi_{\lambda,m} \nabla^2\psi_{\lambda,-m})+\int_{\del}d\theta \,r_B\,\,n_b(\epsilon^{ab}\psi_{\lambda,m}(V_{s,\lambda,-m})_a)\nonumber\\
	&=-(\lambda+2)\pqty{\int d^2x\sqrt{g}( \xi_{\lambda,m}\xi_{\lambda,-m}+ \psi_{\lambda,m}\psi_{\lambda,-m})}+2\pi \,r_B^2\,\,\psi_{\lambda,m}(V_{s,\lambda,-m})^{\theta}\bigg\vert_\del\nonumber\\
	&=-(\lambda+2)\pqty{\pi \,r_B\,\,\psi_{\lambda,m}\psi_{\lambda,-m}\bigg\vert_\del+\int d^2x\,\,\sqrt{g}(\xi_{\lambda,m}\xi_{\lambda,-m}+ \psi_{\lambda,m}\psi_{\lambda,-m})}\label{vsvsinp}
	\end{align}
	For either set of eigenvalues, only one of the term in the bulk integral dominates. Since, the bulk integral is positive definite
	we see that
	\begin{align}
	\abs{\langle PV_{s,\lambda,m},PV_{s,\lambda,-m}\rangle}\geq\abs{(\lambda+2)r_B\psi_{\lambda,m}(r_B)\psi_{\lambda,-m}(r_B)}\label{pvpvsads}
	\end{align}
	Putting together  eq.\eqref{smldinpd},\eqref{pvlpvlasy} and eq.\eqref{pvpvsads}, we find in the asymptotic AdS limit, that
	\begin{align}
	{\langle PV_s,PV_L\rangle\over \sqrt{\langle PV_s,PV_s\rangle\langle PV_L, PV_L\rangle}}\leq \order(r_B^{-\frac{3}{2}})\label{aadssmin}
	\end{align}

	Let us now consider the case when there are some modes such that 
	\be
	\label{condma}
	m\sim r_B\gg 1.
	\ee
	This modes will need to be included  when we are  considering
	the general case with a boundary of finite length. Our estimates need to be revised for such modes.  
	We will examine below the case where the eigenvalue $\lambda \ll  m^2$ while $m$ becomes big meeting eq.(\ref{condma}). 
	
	We start with the analysis of the scalar field solution.  For $r\sim\order(1)$, and $\lambda$ such that ${m^2}\gg{\lambda}$ , the equation eq.\eqref{smdifsceq} for the scalar field $\psi_{\lambda,m}$, with $\theta$ dependence being $e^{im\theta}$, is
	\begin{align}
	\del_r((r^2-1)\del_r\psi_{\lambda,m})-\frac{m^2}{r^2-1}\psi_{\lambda,m}=0\label{nhmrbsc}
	\end{align}
	the solution for which we take to be
	\begin{align}
	\psi_{\lambda,m}=\pqty{\frac{r-1}{r+1}}^{\abs{m}\over 2}\label{nhscmrs}
	\end{align}
	This solution can be extended till the region where $r$ is such that 
	\begin{align}
	1\ll r^2\ll  {m^2\over \lambda}\label{matcreg}
	\end{align}
	So, in this region, the solution eq.\eqref{nhscmrs} becomes
	\begin{align}
	\psi_{\lambda,m}=\exp(-\frac{\abs{m}}{r})\label{dpmor}
	\end{align}
	The scalar field equation, eq.\eqref{smdifsceq}, for $r\gg 1$ is  
	\begin{align}
	\del_r(r^2\del_r\psi_{\lambda,m})-\frac{m^2}{r^2}\psi_{\lambda,m}=(2+\lambda)\psi_{\lambda,m}\label{mrbscfar}
	\end{align}
	the solution for which is given by
	\begin{align}
	\psi_{\lambda,m}=\sqrt{\frac{\abs{m}}{{r}}}(c_1\Gamma(1+ v )I_{v}\pqty{\abs{m}\over r}+c_2\Gamma(1- v )I_{- v }\pqty{\abs{m}\over r})\label{frmrbssol}
	\end{align}
	where $I_a$ is the modified Bessel function of the first kind and $ v $ is as defined in eq.\eqref{nuval}.
	Matching this with the solution eq.\eqref{dpmor} in the region $1\ll r\ll m$, we find
	\begin{align}
	c_2={\Gamma( v )\over \sqrt{2\pi}},\,c_1=-c_2\frac{\Gamma(1- v )}{\Gamma(1+ v )}\label{mrbfscs}
	\end{align}
	and so the scalar solution becomes
	\begin{align}
	\psi_{\lambda,m}(r)=\sqrt{\frac{\abs{m} \pi}{2r}}\text{cosec}(\pi v )\left(I_{- v }\pqty{\frac{\abs{m}}{r}}-I_{ v }\pqty{\frac{\abs{m}}{r}}\right)=\sqrt{\frac{2\abs{m}}{\pi r}}K_{ v }\pqty{\abs{m}\over r}\label{scbdmrs}
	\end{align}
	where $K_{ v }$ is the modified Bessel function of the second kind.
	The scalar $\xi_{\lambda.m}$ will also behave in the same way. We also see from   eq.\eqref{scbdmrs}  that the scalar field solution $\psi_{\lambda,m}$  is a function of the combination $\pqty{m\over r}\sim\order{(1)}$ near the boundary. We also   see from eq.(\ref{evalset1}), eq.(\ref{evalset2}) that the boundary conditions can be met when the  relative coefficient between $\psi_{\lambda,m}, \xi_{\lambda,m}$ is  order unity.

	We can now estimate the magnitudes of the required quantities. First, from eq.\eqref{pvlrtbv} and \eqref{pvlpvlasy}, we find that 
	\begin{align}
	\frac{(PV_{L,-m})_{r\theta}}{\sqrt{\langle PV_{L,m},PV_{L,-m}\rangle}}\sim \order\left(\frac{1}{\sqrt{m}}\right)\label{vlmrb}
	\end{align}
	To estimate the magnitude of $V_{s,\lambda,m}$, we note as mentioned above that  scalar fields  $\psi_{\lambda,m},\xi_{\lambda,m}$  in eq.\eqref{scbdmrs} are a function of the combination $\pqty{m\over r}\sim\order{(1)}$ near the boundary. This gives,
	\begin{align}
	V_{s,\lambda,m}^\theta=\pqty{{im\over r}{\xi_{\lambda,m}\over r}-\frac{\psi_{\lambda,m}'}{m}}\sim \order\pqty{1\over m}\label{vstlmrb}
	\end{align}
	where the prime is a derivative with respect to the quantity $\left(r\over m\right)$.
	To estimate the value of $\langle PV_{s,\lambda,m},PV_{s,\lambda,-m}\rangle$, we show using eq.\eqref{vsvsinp} that this quantity is of $\order{\left({\sqrt{m}}\right)}$. To see this, consider the bulk integral in eq.\eqref{vsvsinp}
	\begin{align}
	\int d^2x \sqrt{g}\psi_{\lambda,m}\psi_{\lambda,-m}&=2\pi\pqty{\int_1^{r_c}\pqty{r-1\over r+1}^\frac{\abs{m}}{2}+\int_{r_c}^{r_B}\psi_{\lambda,m}\psi_{\lambda,-m}\left({m\over r}\right)dr}\nonumber\\
	&\simeq 2\pi m\int_{r_c\over m}^{r_B\over m}dx \psi_{\lambda,m}(x)\psi_{\lambda,-m}(x)\nonumber\\
	&\simeq \order{(m)}\label{vsvsbulmrb}
	\end{align}
	where $r_c$ is such that $1\ll r_c\ll m$. It is also easy to see, by noting eq.\eqref{vstlmrb} that the boundary term is also of the same order as above and so
	\begin{align}
	\sqrt{\langle PV_{s,\lambda,m},PV_{s,\lambda,-m}\rangle}\simeq \order(\sqrt{m})\label{inpsmsmmrb}
	\end{align}
	Thus putting together eq.\eqref{inpsmsmmrb}, eq.\eqref{vstlmrb} and eq.\eqref{vlmrb} and noting eq.\eqref{smlarpvinp}, we find that 
	\begin{align}
	{\langle PV_s,PV_L\rangle\over \sqrt{\langle PV_s,PV_s\rangle\langle PV_L, PV_L\rangle}}\simeq \order{(1)}\label{vsvlmrb}
	\end{align}

	We can easily extend the above analysis for the case of Double trumpet topology in Euclidean AdS spacetime discussed in \ref{dbads}. To evaluate the quantity in eq.\eqref{wantinp}, we would proceed as before. The discussion till eq.\eqref{evalset2} continues to hold true except that the eigenvalues are now determined by imposing either eq.\eqref{evalset1} or eq.\eqref{evalset2} at both the boundaries. Taking the left and right boundaries to be located at $r=-r_{B1}$ and $r=r_{B2}$ respectively. Since the vector fields for small and large diffeomorphisms can be chosen so that the modes corresponding to left and right boundaries can be decoupled, the corresponding boundary terms in eq.\eqref{smlarpvinp} will be independent of each other. The inner product of two large diffeomorphisms in the basis in eq.\eqref{dblapsolrw} is calculated in detail in appendix \ref{dbldfiscact}, the final result appearing in eq.\eqref{dbtotm}. The calculation  of  inner product of small diffeomorphisms is again analogous to the disk case with the  expression in eq.\eqref{vsvsinp} interpreted as having two boundary terms which again are independent of each other. Thus it immediately follows that we will have the result analogous to eq.\eqref{aadssmin} in the present case of double trumpet topology also.

	\section{Estimation of various determinants in Euclidean AdS disk}
	\label{deterests}
	
	In this section we shall discuss  the computation of  various determinants in detail in AdS spacetime for the Euler characteristic $\chi=1$ corresponding to the disk topology. To begin with, we will compute the determinant of the scalar Laplacian. This requires the specification of appropriate boundary conditions which we take to be  Dirichlet boundary conditions.
	We will mostly consider the case when the boundary has large length $l\sim {1\over \epsilon} \gg 1$ and obtain the dependence of the determinant on the large diffeomorphisms discussed above.  In the asymptotic AdS limit, where $\epsilon \rightarrow 0$,  we will find that   a length dependent counter-term needs to be    added to get a finite result,  and that the   dependence  on large diffeomorphisms vanishes. We will make essential use of the conformal anomaly in the analysis. 
	
	Similar results   will also be obtained for $\text{det}'(P^\dagger P)$.
	For $\text{det}(- \hat{\nabla}^2+2)$, on general grounds,  upto $O(\epsilon)$, the dependence of the large diffeomorphisms will be shown to be of the form of the Schwarzian action with a coefficient  which  is linear in   $\epsilon$, but we will not be able to obtain the precise  value of this coefficient.

	Coming back to the scalar case, we are interested in the dependence of $\text{det}(-\hat{\nabla}^2)$ on the large diffeomorphisms. On general grounds this dependence should be a functional 
	of $\text{Diff} (S^1)/SL(2,R)$, since it is easy to see that diffeomorphisms lying in the $SL(2,R)$ isometry group of $AdS_2$ must leave the determinant unchanged \footnote{
	To be very explicit, suppose the initial metric is taken as eq.\eqref{metads2} with the boundary specified as $r=r_B$. Let the new coordinates after the $SL(2,R)$ transformation be $\tilde{r},\tilde{\theta}$. The boundary is now specified to be $\tilde{r}=r_B$. Since neither the metric nor the specification of the boundary in terms of the coordinate has changed, it naturally follows that the value of the determinant of the laplacian operator will not change.}.  This imposes a strong restriction  on the kind of terms that appear in the final result for the determinant. The simplest such term that one can write is proportional to length of the boundary. The next term, which involves two derivative with respect to $u$, eq.(\ref{elements}) - the rescaled proper time along the boundary- is uniquely given by the Schwarzian action. On dimensional grounds its coefficient must go like $\epsilon$, and by using the conformal anomaly we can  obtain the coefficient in front of this action as we show  below.  Beyond this, in general additional terms will also be present - these will involve additional derivatives of $u$ and correspondingly additional powers of $\epsilon$. If we consider modes whose mode number $m$, eq.\eqref{vecask} is small enough meeting the condition 
	\be
	\label{condmoden}
	m \epsilon \ll 1,
	\ee
	so that their wavelength  meets the condition,  $\Lambda\gg R_{AdS}$, eq.(\ref{inch}), then these additional terms will be suppressed. For modes of higher mode number  where  eq.(\ref{condmoden})  is not met these higher order terms must all be retained and the resulting behaviour of the determinant is much more non-trivial to obtain. These arguments can also be applied to the double trumpet with two boundaries and de Sitter case when we calculate the no-boundary wavefunction by analytic continuation from the $(2,0)$ or $(0,2)$ signature metrics as discussed in \ref{dsbasic}.
	
	Let us now show how the conformal anomaly can be used to obtain the coefficient of the first two terms mentioned above, involving the boundary length and the Schwarzian derivative. We can expand  the determinant 
	 \begin{align}
	 \ln\text{det}(-\hat{\nabla}^2)=c_1 \int ds+\epsilon\, c_2 \int du\, \text{Sch}\left(\tan\pqty{\theta(u)\over 2},u\right)+\order(\epsilon^2)\label{adslapdetanz}
	 \end{align}
	 where $u$ is renormalized boundary proper time, eq.\eqref{elements}, and $\theta$ is the coordinate appearing in the line element of the AdS metric in eq.\eqref{metads2}. Our task now simplifies to evaluating the constants $c_1$ and $c_2$. To fix the constants we consider a non-wavy boundary specified by $r=r_B$, and use the conformal transformation property of the determinant to fix its dependence on the value $r_B$. For this it is convenient to work in the coordinate system $\hat{r},\theta$ in which the line element is given by eq.\eqref{edspolar}.
	 	\begin{equation}
	 ds^2=\frac{4}{(1-\hat{r}^2)^2}(d\hat{r}^2+\hat{r}^2d\theta^2),\quad \quad\hat{r}\in [0,1]\label{uniadsm}
	 \end{equation}
	 The boundary $r=r_B$ in the metric eq.\eqref{metads2} is specified in terms of $\hat{r}$ coordinate as 
	 \begin{align}
	 \hat{r}=\hat{r}_B=\sqrt{r_B-1\over r_B+1}\label{eadsrrhbd}
	 \end{align}
	 Defining the coordinate $\bar{\rho}$ as $\hat{r}=\bar{\rho} \hat{r}_B$, we find that the boundary specification now becomes $\bar{\rho}=1$. The line element becomes
	 \begin{align}
	 ds^2=\hat{ g}_{ab}dx^a dx^b=\frac{4\hat{r}_B^2}{(1-\hat{r}^2_B \bar{\rho}^2)^2}(d\bar{\rho}^2+\bar{\rho}^2d\theta^2)\equiv e^{2\sigma}\bar{g}_{ab}dx^a dx^b\label{eadsRmet}
	 \end{align}
	 where $\bar{g}_{ab}$ is defined as 
	 \begin{align}
	 d\bar{s}^2=\bar{ g}_{ab}dx^a dx^b=d\bar{\rho}^2+\bar{\rho}^2d\theta^2, \quad \bar{\rho}\in [0,1],\theta\in [0,2\pi]\label{eadsghdef}
	 \end{align}
	 where as mentioned earlier the boundary is located at $\bar{\rho}=1$. It is easy to see from eq.\eqref{eadsRmet} that $r_B$-dependence is  entirely in the conformal factor with the flat metric $\bar{g}_{ab}$ independent of $r_B$. Now, we note that the conformal transformation property of the determinant of a scalar laplacian with Dirichlet boundary conditions for conformally related metrics 
	 \begin{align}
	 \hat{g}_{ab}=e^{2\sigma}\bar{g}_{ab}\label{adsconfmet}
	 \end{align}
	 is given by 
	 \begin{align}
	 {\text{det}(-\hat{\nabla}^2)_{\hat g}\over \text{det}(-\bar{\nabla}^2)_{\bar{g}}}=&\exp{-S_{\sigma}}\label{adsscallap}
	 \end{align}
	 where $S_\sigma$ is given by 
	 \begin{align}
	 S_\sigma=\frac{1}{6\pi}\left[\half\int d^2x \sqrt{\bar{g}}{(\bar{g}^{ab}\del_a\sigma\del_b\sigma+\bar{R}\sigma)}+\int_\del d\bar{s}\bar{K}\sigma\right]
	 %-\frac{1}{4\pi}\int_\del {K}  ds
	 \label{adssigma}
	 \end{align}
	 From eq.\eqref{eadsRmet} and eq.\eqref{eadsghdef}, we note that 
	 \begin{align}
	 e^{2\sigma}=\frac{4\hat{r}_B^2}{(1-\hat{r}^2_B \bar{\rho}^2)^2}, \quad \bar{R}=0\label{detconffac}
	 \end{align}
	 and normal vector to the boundary normalized with $\bar{ g}_{ab}$  and the corresponding extrinsic trace are
	 \begin{align}
	 \bar{ n}^{\bar{\rho}}=1, \bar{K}=\frac{1}{\bar{\rho}}=1, \quad \hat{K}=e^{-\sigma}(\bar{ K}+\bar{ n}^a\del_a\sigma)=\frac{1}{2\hat{r}_B}+\frac{\hat{r}_B}{2}\label{lunitnok}
	 \end{align}
	 Using these results, we get
	 \begin{align}
	 {\text{det}(-\hat{\nabla}^2)_{\hat g}\over \text{det}(-\bar{\nabla}^2)_{\bar{g}}}=&\exp{-\frac{1}{3}\pqty{\hat{ r}_B^2\over 1-\hat{ r}_B^2}-\frac{1}{3}\ln 2\hat{ r}_B}\label{radetconf}
	 \end{align}
	Now, take the case where  $r_B=\frac{1}{\epsilon}\ll 1 $, so that the boundary length $l \gg 1$. To be more precise, this $\epsilon$ defined by the boundary value of $r_B$ is the same as that in eq.\eqref{elements} to leading order, there will be subleading corrections between the two variables. However, we shall work, for now, consistently with it being defined as defined by the value of $r_B$. But, Using eq.\eqref{eadsrrhbd} we get
	\begin{align}
	 {\text{det}(-\hat{\nabla}^2)_{\hat{g}}\over \text{det}(-\bar{\nabla}^2)_{\bar{g}}}=&\exp(-\frac{1}{6\epsilon}+\frac{1}{6}\ln(4\over e)+\frac{\epsilon}{3})\label{ratdeteps}
	\end{align}
	where we used eq.\eqref{scalelength} to obtain final equality. Note that $\text{det}(-\bar{\nabla}^2)_{\bar{ g}}$ is some constant independent of $r_B$ and hence $\epsilon$. From eq.\eqref{adslapdetanz}, for the boundary at $r_B=\epsilon^{-1}$, we get
	\begin{align}
	\text{det}(-\hat{\nabla}^2)=\exp(\frac{2\pi c_1}{\epsilon}+\pi\epsilon (c_2-c_1))\label{detlapc1c2}
	\end{align}
	So comparing, we get
	\begin{align}
	c_1=-\frac{1}{12\pi},\quad c_2=\frac{1}{4\pi}\label{detc1}
	\end{align}
	and so we have
	\begin{align}
	\ln\text{det}(-\hat{\nabla}^2)=-\frac{1}{12\pi} \int ds+\, \frac{\epsilon}{4\pi} \int du\, \text{Sch}\left(\tan\pqty{\theta(u)\over 2},u\right)+\order(\epsilon^2)\label{detinlesch}
	\end{align}
	
	Note that the dimensional analysis   we had mentioned above  which fixes the powers of $\epsilon$  in each term in eq.(\ref{detlapc1c2}) can be understood as follows. 	The line element eq.(\ref{elements}) is invariant under $(\epsilon, u)\rightarrow (\lambda \epsilon, \lambda u)$. Under this rescaling (with $\tan(\theta/2)$ unchanged ) the 
	Schwarzian term, $\text{Sch}(\tan(\theta/2),u)\rightarrow {1\over \lambda^2} \text{Sch}(\tan(\theta/2),u)$, while the line element $ds$ is invariant. This fixes the powers of $\epsilon$ appearing in the coefficients. Also note that after adding a counter term to cancel the length dependent first term in eq.(\ref{detinlesch})  which goes like $1/\epsilon$, we get are left with the Schwarzian term and additional  subleading corrections which all vanish in the asymptotic AdS limit where  $\epsilon \rightarrow 0$. 
	
	It is also easy to see that with this result for the determinant, the on-shell action, $S_{OS}$ for the $JT$ gravity in the presence of matter fields satisfying vanishing Dirichlet boundary conditions at finite temperature in the semi-classical limit, $G\rightarrow0, N\rightarrow\infty$ with fixed $GN$ is given by 
	\begin{align}
	S_{OS}=S_{JT,\del}+S_{M,qm}\label{sonswprev}
	\end{align}
	where $S_{JT,\del}$ is given by first line in eq.\eqref{acts} and $S_{M,qm}$ is given by 
	\begin{align}
	 S_{M,qm}=\frac{N}{2}\ln\text{det}(-\hat{\nabla}^2).
	\end{align}
	For the finite temperature case taking $\theta(u)={2\pi\over \beta }u, \phi_B=\frac{1}{J\epsilon}+\frac{GN}{3}$, and introducing the  counterterm  mentioned to cancel the first term in  $\text{det}(-\hat{\nabla}^2)$ which is length-dependent,  we see that the value of the on-shell action becomes
	\begin{align}
	S_{OS}=-\frac{\pi}{4GJ\beta}\pqty{1-\frac{2GNJ\epsilon}{3}}\label{onshellact}
	\end{align}
	which indeed matches with the results in \cite{Moitra:2019xoj}.
	
	We should alert the reader to an important issue connected  to the above calculation. The formula relating the scalar laplacian determinant for conformally related metrics given in eq.\eqref{adssigma} is different from the one appearing in \cite{Alvarez:1982zi} by an extra  term 
	\begin{align}
	\Delta S_\sigma= -\frac{1}{4\pi}\int \hat{K} ds\label{lndetdiff}.
	\end{align}
	Indeed, in general,  the bulk conformal anomaly and Wess-Zumino consistency conditions \cite{Polchinski:1998rq} fix the form of the action $S_\sigma$ completely upto the possibility of an additional term of this type.  While \cite{Alvarez:1982zi}  do report that such a term arises for  the determinant with Dirichlet boundary conditions we
	 find that its presence leads to  disagreement with the semi-classical results in \cite{Moitra:2019xoj}, and have accordingly not included it here.

	We now extend the above considerations to compute the value of the determinant of the operator $P^\dagger P$. This is in fact straightforward. 
	Once again, we can expand the determinant in powers of $\epsilon$, with the first two terms being, 
	\begin{align}
	\ln\text{det}'(P^\dagger P)=k_1 \int ds +k_2\epsilon\int \text{Sch}\pqty{\tan(\theta(u)\over 2),u}\label{lndetansz}
	\end{align}
	We will now use the same trick of considering a non-wavy boundary and use the conformal transformation property of $\det'P^\dagger P$ to compute its $r_B$ dependence and then match the coefficients by expanding in $\epsilon$ where $r_B=\frac{1}{\epsilon}$. For conformally related metrics in eq.\eqref{adsconfmet} the determinants of the operator $P^\dagger P$ are related as \cite{Alvarez:1982zi} 
	\begin{align}
		{(\text{det}' P^\dagger P)_{\hat g}\over {(\text{det}' P^\dagger P)_{\bar{ g}}}}=\exp{-26 S_\sigma}\label{detppconf}
	\end{align}
	We note that only the prefactor in the exponent in eq.(\ref{detppconf})  is different from the scalar case due to the difference in the central charges\footnote{We must  mention though that we have not been too careful about the possible presence of a counter term of the form eq.(\ref{lndetdiff}) in eq.(\ref{detppconf}). Also there could be some subtleties due to zero modes.}. So  again by comparing eq.(\ref{lndetansz}) and eq.(\ref{detppconf}) for the geometry, eq.(\ref{metads2}) with boundary at $r_B=\frac{1}{\epsilon}$, we get, upto an $\epsilon$ independent prefactor which we are not retaining, 
	\begin{align}
	\text{det}'(P^\dagger P)=\exp(-\frac{26}{12\pi}\int ds +\frac{26}{4\pi}\epsilon\int \text{Sch}\pqty{\tan(\theta(u)\over 2),u})\label{pdpfinres}
	\end{align}
	This is of the same form as in the scalar case	and once the first term is removed by a suitable counter term again vanishes in the $\epsilon\rightarrow 0$ limit.

	The   determinant  $\text{det}(-\hat{\nabla}^2+2)$ is more complicated. Since it arises  after doing the path integral for a massive scalar of mass $2$, we cannot use the conformal anomaly to obtain useful  information about it . However,  we can still argue from the requirement that the determinant is valued in ${\rm Diff} (S^1)/SL(2,R)$ that it can be expanded as  
	\begin{align}
	\ln\text{det}(-\hat{\nabla}^2+2)=q_1 \int ds +{q_2\epsilon\over 8\pi }\int du\, \text{Sch}\pqty{\tan(\theta(u)\over 2),u}\label{delsqm2anz}
	\end{align}
	where $q_1,q_2\sim\order{(1)}$ constants. Once again it then follows that the dependence on the large diffeomorphisms vanishes in the asymptotic AdS limit. 

{To reiterate a point made earlier, note that the length-dependent terms in various determinants eq.(\ref{detinlesch}), eq.(\ref{pdpfinres}) and eq.\eqref{delsqm2anz} grow like  ${1\over \epsilon}$, i.e. linearly in  the length of the boundary,  and  thus 
	diverge in the asymptotic AdS limit, $\epsilon\rightarrow 0$. To obtain a finite result  in this limit we need to add a boundary term to the JT action, eq.\eqref{bjtact}, 
	\be
	\label{btJT}
	\delta S_{JT,\partial}=A \int_{\partial} ds
	\ee
	and fix  the constant $A$ appropriately so as to cancel this divergence. }
	
	\subsection{Asymptotic AdS limit case}
	\label{asyadsdet}
	We now come to a subtlety having to do with the order in which  various limits are being  taken while evaluating the  determinants. 
	The computations above of the determinants used the Weyl anomaly  and are  valid for an arbitrary boundary. 
	We see above  that the result at leading order in the length $l$ for the determinants go like  
	\be
	\label{genb}
	\text{det} \,{\hat O}\sim e^{(C   l)}
	\ee
	 where $C$ is a constant which depends on the operator ${\hat O}$. 
	 More precisely the determinants above are  
	obtained  for the case of a given geometry with a finite  length boundary  by introducing a cut-off (for large eigenvalues ), regulating the product of eigenvalues of the relevant operators and then taking the cut-off to infinity in a manner which is consistent with the Weyl anomaly. In contrast, in the asymptotic AdS limit, as was mentioned above, we are interested in first taking the limit when the boundary length $l \rightarrow \infty$, keeping the cut-off on 	 eigenvalues  fixed, and thereafter taking the limit where this cut-off goes to infinity \footnote{As we will see below after expanding  modes along the $\theta$ direction in their Fourier components $e^{i m \theta}$,  the cut-off on the eigenvalues of relevance will be  on the mode number $m$ .}. 
	
Here we will show that this second order of limits can  give a different result, and in particular the leading term, eq.(\ref{genb}), which is exponential in $l$ can be absent
in the asymptotic AdS limit. 	To carry out the calculations in this limit we will use a method first discussed by Coleman \cite{Coleman:1985rnk}. 

Let us illustrate this method for the Simple Harmonic Oscillator (SHO). 
	Consider two simple harmonic oscillators with frequencies $w_1$ and $w_2$ constrained to move between $x=0$ and $x=L$. Let $\psi_\lambda^{(1)}$ be the solution to the equation
	\begin{align}
	(-\del_x^2 +w_1^2-\lambda) \psi_\lambda^{(1)}(x)=0\label{shoeveq}
	\end{align}
	with the appropriate boundary conditions. $\psi_\lambda^{(2)}$ satisfies a similar equation with the frequency $w_2$. Suppose we impose Dirichlet boundary conditions at both ends. Coleman's formula then states that 
	\begin{align}
	\label{colfor}
	\frac{\text{det} (-\partial_x^2+w_1^2-\lambda)}{\text{det} (-\partial_x^2+w_2^2-\lambda)}=\frac{\psi_\lambda^{(1)}(L)}{\psi_\lambda^{(2)}(L)}
	\end{align}
	In the above formula the left and right hand sides are to be regarded as a function of the complex variable $\lambda$. The formula  follows from noting that the zeros and poles of 
	the left and right hand slides are the same, and that both sides go to unity as $\lambda\rightarrow \infty$ in any direction except the real axis. 
	It then also follows that upto a constant independent of $\omega$ 
	\be
	\label{conc}
	{\rm det} (-\partial_x^2+w_1^2) = \psi_{(\lambda=0)}(L)
	\ee
	where the RHS is the value of the wavefunction obtained at $x=L$ for the operator with frequency $\omega$  by starting at $x=0$ with the correct boundary conditions. 
	One important point to note here is that the normalization of the solution $\psi_{ \lambda }(x)$ should be  fixed in such a way that when viewed as a function of $\lambda$, any spurious zeros or poles, other than those corresponding to the actual eigenvalues of the operator $(-\del_x^2+\omega_1^2)$ in the solution $\psi_{ \lambda }(x)$ are cancelled and that the ratio of two solutions with different frequencies goes to unity as $\lambda \rightarrow \infty$ in any direction other than along the real axis \footnote{In the present case we achieve this by taking $\partial_x\psi(x=0)=1$.}. The solution $\psi_{ \lambda =0}$ is then obtained by taking the $\lambda=0$ limit of this appropriately normalized $\psi_{ \lambda }$ solution. 
	
	A similar formula also follows for Neumann boundary conditions or mixed boundary conditions where we set 
	$a \psi + b \psi'=0$ (at say both $x=0,L$) with  eq.(\ref{conc}) being  replaced by 
	\be
	\label{cons}
	{\rm det} (-\partial_x^2+w_1^2)= a \psi_0(L)  + b \psi_0'(L).  
	\ee
	where the subscript refers to taking the $\lambda =0$ solution as in eq.\eqref{conc}. The normalization of the solution $\psi_{0}$ is fixed as explained before.

	We will now adopt the same strategy to calculate the determinants $\text{det}(-\hat{\nabla}^2)$, $\text{det}(-\hat{\nabla}^2+2)$ and  $\text{det}'(P^\dagger P)$, in the case of the asymptotic AdS spacetime. 
	In applying this method to the AdS case we  expand the modes for the  operator  in the basis   of modes in the $\theta$ direction, $e^{i m \theta}$ , and then working for any fixed  value of $m$
	obtain a one-dimensional problem in the radial direction. In this one -dimensional problem    we take the boundary to go to infinity $l \rightarrow \infty$ and then use the Coleman method to obtain the determinant of the radial operator. 
	The full determinant is then be obtained by taking the product of contributions  over all values of the mode number $m$, and then taking in this product  $|m|\rightarrow \infty$. We see therefore that in the calculations below while working in the asymptotic AdS limit   with the order of limits mentioned above, first taking $l \rightarrow \infty$ and then taking mode number $m \rightarrow \infty$.

	 Let us first compute $\text{det}(-\hat{\nabla}^2)$ in this manner. The boundary condition at $x=0$ in the SHO is now replaced by the requirement of the regularity of the solution in the interior. The solution to the eigenvalue equation $\hat{\nabla}^2\psi_\lambda=-\lambda\psi_\lambda$ for fixed mode number $m$ which is  regular everywhere in the interior, chosen such that the ratio of two solutions with different mode numbers goes to unity as $\abs{ \lambda}\rightarrow\infty$ other than along the real axis, and without  have any spurious zeros, poles or branch points in the variable $\lambda$, is given by 
	\begin{align}
	&\psi_{\lambda,m}=
	\begin{cases}
	&P_{ v -\half}^{-\abs{m}}(r),\quad m\neq 0\\
	&P_{ v -\half}^{0}(r),\quad\quad\quad m=0
	\end{cases}	\label{nablaevec}\\
 	&\qquad v =\half \sqrt{1-4 \lambda }\nonumber
	\end{align}
		and $P_\alpha^\beta$ is the associated Legendre function of the first kind. In fact, using the asymptotic form of the $P^{-\abs{ m}}_{ v -\half}(r)$ \footnote{We have followed the conventions of \cite{Abramstegun} in obtaining the asymptotic forms here and elsewhere in this manuscript.},
		\begin{align}
		P_{ v -\half}^{-\abs{ m}}(r)=& \left(\frac{(2r)^{- v -\half } \Gamma (- v  ) }{\sqrt{\pi } \Gamma \left(- v  +\left| m\right| +\frac{1}{2}\right)}\right)+O\left({r^{- v -3/2}}\right)+\left(\frac{(2r)^{ v -\half } \Gamma ( v  ) }{\sqrt{\pi } \Gamma \left( v  +\left| m\right| +\frac{1}{2}\right)}\right)+O\left({r^{ v -3/2}}\right)\label{asymPdisk}
		\end{align}
		we see that the asymptotic form has $ v \rightarrow - v $ symmetry.
	For the case of Dirichlet boundary condition at $r=r_B$, the eigenvalues are obtained by solving 
	\begin{align}
	\label{evaldirads}
	\psi_{\lambda,m}(r_B)=0
	\end{align}
	Let the eigenvalue be labelled by $\lambda_{m,n}$. The subscript $m$ in $\lambda_{m,n}$ denotes mode number $m$ and the $n$ labels the various eigenvalues for this particular mode number.
	To compute the determinant in the asymptotic AdS limit we take the asymptotic form of the solution eq.\eqref{adsresol}, with $\lambda=0$, which is given by  
	\begin{align}
	\label{asymscalads}
	\psi_{0,m}=
	\begin{cases}
	\frac{1}{\Gamma(\left| m\right|+1) }-\frac{1}{\Gamma(\abs{m})r}+\order{(r^{-2})}\quad m\neq 0\\
	1\qquad m=0
	\end{cases}
	\end{align}
	Using this asymptotic form, the determinant computed using eq.\eqref{colfor} reads
	\begin{align}
	\ln \text{det}(-\hat{\nabla}^2)=-\sum_{m=-\infty,m\neq 0}^{\infty}\ln({\Gamma(\left| m\right|+1) })\label{nabla2det}
	\end{align}
	which is manifestly independent of $r_B$.
	
	Next consider the determinant for the operator $(-\hat{\nabla}^2+2)$. 
	The eigenvalue equation for the operator reads 
	\begin{align}
	(\hat{\nabla}^2-2)\psi_\lambda=-\lambda\psi_\lambda  \label{evaleqinads} 
	\end{align} 
	which for the mode number $m$ now has the regular solution
	\begin{align}
	\label{adsresol}
	&\psi_{\lambda,m}=
	\begin{cases}
	&P_{ v -\half}^{-\abs{m}}(r),\quad m\neq 0\\
	&P_{v-\half}^{0}(r),\quad\quad\qquad\quad\quad m=0
	\end{cases}	\\
	&\qquad v=\half \sqrt{9-4 \lambda }\nonumber
	\end{align}
	To compute the determinant in the asymptotic AdS limit we take the asymptotic form of the solution eq.\eqref{adsresol}, with $\lambda=0$, which is given by  
	\begin{align}
	\psi_{0,m}=\begin{cases}
	\frac{1}{\Gamma(\abs{ m}+2)}r-\frac{3 \abs{m}^2 }{4r \Gamma(\abs{2+ m})}+O\left(r^{-2}\right)\qquad m\neq 0\\
	r\qquad \qquad \qquad m=0
	\end{cases}\label{asymscalads}
	\end{align}
	We now use this to compute $\det (-\hat{\nabla}^2+2)$ with the Dirichlet boundary condition eq.\eqref{evaldirads} using eq.\eqref{colfor}. This gives
	\begin{align}
	\ln \text{det}(-\hat{\nabla}^2+2)=-2\sum_{m=1}^{\infty}\ln({\Gamma(2+\abs{ m})})\label{scaldetcl} + \sum_{m=-\infty}^{m=\infty} \ln r_B
	\end{align}
	which is independent of $r_B$ since $\sum_{m=-\infty}^{m=\infty}\ln r_B=0$.  
	
	Thus we see that for both the operators considered above, we get no dependence growing exponentially as in eq.(\ref{genb}), in the asymptotic AdS limit. For $\text{det}(- \hat{\nabla}^2)$  we saw in the previous subsections that such a dependence does arise when we consider a different order of limits. 
	
	The computation of the determinant $\det ' P^\dagger P$ is very similar. Let us first study the case when the index $v$ in eq.\eqref{adsresol} is imaginary. We shall later see that there exist one eigenvalue when $v$ is real. For $v$ imaginary, the main difference in this computation would be the boundary conditions on the scalar field. For a general vector field decomposed as in eq.\eqref{veca} the boundary conditions eq.\eqref{veclapdetbc}, in the asymptotic AdS limit gives two possible conditions on the scalar fields.These are
	\begin{align}
	\label{pdpcond}
	\del_r^2\psi_{\lambda,m}=0,\quad \xi_{\lambda,m}\sim {\psi_{\lambda,m}\over r_B}\nonumber\\
	\del_r\xi_{\lambda,m}=0,\quad \psi_{\lambda,m}\sim{\xi_{\lambda,m}\over r_B}
	\end{align}
	The determinant of the eigenvalues determined by the second of the  condition above,$\del_r\xi_{\lambda,m}=0$, is straightforward to compute. The appropriately normalized solution is again given by eq.\eqref{asymscalads}  and so, the product of these eigenvalues is given by taking the derivative of eq.\eqref{asymscalads} which still gives eq.\eqref{scaldetcl}.
	
	 The contribution from the other set of eigenvalues $\del_r^2\psi_{ \lambda ,m}=0$ are more complicated. First, let us note that even though the boundary condition is a second order equation, it can be understood as a mixed boundary condition of the form eq.\eqref{cons}  upon using the eigenvalue equation eq.\eqref{evaleqinads} for $\psi_{ \lambda ,m}$.  The contribution from the modes $m\neq 0$ is straightforward to obtain by taking the second derivative of the corresponding asymptotic expression in eq.\eqref{asymscalads}. However, for $m=0$ mode, since the solution is just $\psi_{0,0}=r$, taking the double derivative gives zero, which using the analog of eq.\eqref{conc} then shows that the determinant for the $m=0$ sector is zero. The zero mode is in fact $\psi_{0,0}$ itself and it corresponds to the $U(1)$ isometry of $AdS_2$ under which 
	 $\theta\rightarrow \theta+c$. 	 We are actually interested in evaluating $\text{det}'(P^\dagger P)$ and would therefore need to evaluate the determinant without the zero eigenmode. 
	 
	 We have not been able to find a fully satisfactory way of dealing with this complication. One might hope to proceed  as follows. We consider in the $m=0$ sector the operator 
	 $\text{det}(-\hat{\nabla}^2+2-\lambda)$ for non-zero $\lambda$ and then take the $\lambda\rightarrow 0$ suitably, removing the extra zero mode and thereby obtaining the determinant for non-zero modes.  For non-zero $\lambda$ the arguments above lead to the conclusion that 
	 \begin{align}
	 {\text{det}(-\hat{\nabla}^2+2-\lambda)\over \lambda}=\frac{\del_r^2\psi_{\lambda,0}}{\lambda}\label{zeromodecol}
	 \end{align}
	 and so we get
	 \begin{align}
	 \text{det}{(-\hat{\nabla}^2+2)}=\lim_{\lambda\rightarrow 0}\frac{\del_r^2\psi_{\lambda,0}}{\lambda}\label{zermcoldel2}
	 \end{align}
	 Computing the solution $\psi_{ \lambda ,0}$ in a perturbation series in $\lambda$ near $\lambda=0$ by imposing regularity near the origin, we get the solution to $\order{(\lambda)}$ to be
	\begin{align}
	\psi_{\lambda,0}=r+\frac{\lambda}{3}(1-r\ln(1+r))\label{lpermzsol}
	\end{align}
	and hence
	\begin{align}
	\lim_{\lambda\rightarrow 0}\frac{\del_r^2\psi_{\lambda,0}}{\lambda}\simeq -{\frac{1}{3r_B}}\label{pdpmzdet}
	\end{align}
	So, the net value of the determinant $\text{det}'(P^\dagger P)$ becomes
	\begin{align}
	\ln\text{det}'(P^\dagger P)=-2\sum_{m=1}^{\infty}\ln({\Gamma(2+\abs{ m})})+2\sum_{m=1}^{\infty}\ln(-\frac{3\abs{ m}^2 }{2\Gamma(2+\abs{ m})r_B^3})+\ln({-1\over 3r_B})\label{pdpzmre}
	\end{align}
	The resulting $r_B$ dependence, after doing the sum by zeta function regularization,  is then given by 
	\begin{align}
	\ln\text{det}'(P^\dagger P)=2\ln r_B\label{pdprbde}
	\end{align}
		This is not a very satisfactory result  though since in the $r_B\rightarrow \infty$ limit  the resulting divergence in the determinant  cannot be removed by a local counter-term (unlike for a term which is growing linearly with $r_B$). We leave a proper resolution of this puzzle for the future. 
		
		Let us end with some  comments. It is easy to see that there exists one additional discrete eigenvalue when $v=\half$ corresponding to $\lambda=2$. Consider the scalar fields $\psi_{ \lambda,m },\xi_{\lambda,m}$ that satisfy the equation $\hat{\nabla}^2\psi_{ \lambda,m }=0=\hat{\nabla}^2\xi_{\lambda,m}$ with mode number $m$, the regular solutions for which are taken to be 
	\begin{align}
	\psi_{ \lambda,m }=\alpha_me^{im\theta}\pqty{r-1\over r+1}^{\abs{m}\over 2},\quad 	\xi_{ \lambda,m }=\beta_m e^{im\theta}\pqty{r-1\over r+1}^{\abs{m}\over 2}\label{lamda2sol}
	\end{align}
	It is then easy to see that near the boundary $r\gg 1$, with the choice of constants $\beta_m=i\, \text{sgn}(m)\alpha_m$, the boundary conditions \eqref{veclapdetbc} are satisfied thus showing that $\lambda=2$ is a genuine eigenvalue, which does not belong to the either of the sets of eigenvalues in eq.\eqref{pdpcond}. Also, note that this discrete eigenvalue exists only for $m\neq 0$, since when $m=0$ the vector field constructed out of these scalar fields vanishes identically everywhere. Further, there are no other eigenvalues apart from the ones we have obtained so far. Including this eigenvalue of course does not change the $r_B$ dependence obtained above. 
	
	The calculation of determinants discussed in this subsection   can be easily extended to the general case when the boundary is located at large but finite value of $r_B$,
	and  also to  de Sitter spacetime.

	\section{Matter coupling to the time reparametrization modes in AdS}
	\label{mcttrp}
	In this appendix we will describe in more detail the coupling of the matter to the time reparametrization modes in the classical action eq.\eqref{class}. The result can be obtained for a general boundary of length $l$ but for simplicity we will work out the case $l\gg 1$ below. We begin with the metric eq.(\ref{metads2}) in which the boundary is at $r_B$ given by eq.\eqref{caleb} in terms of $l$, so that $r_B \gg 1$. 
	We next turn on a large diffeomorphism. Under such a  diffeomorphism the new coordinates asymptotically close to the boundary are given by 
	\be
	\label{newcoor}
	{\tilde \theta} =f(\theta), {\tilde r} =r/f'(\theta)
	\ee
	with the boundary lying at 
	${\tilde r} =r_B$. $f'(\theta)$ denotes a derivative of $f(\theta)$ with respect to $\theta$.
	It is easy to see from our definition of the rescaled proper time $u$, eq.(\ref{elements}),  and eq.(\ref{relive}) that 
	\be
	\label{deaf}
	u={\tilde \theta}=f(\theta)
	\ee
	The infinitesimal version of these transformations follows from eq.(\ref{vcuasy}) and is discussed in eq.(\ref{largedi}) eq.(\ref{default}). 
	Note that the coordinate  $r$   varies along the resulting wavy boundary as
	\be
	\label{wave}
	r(\theta)=r_B f'(\theta)
	\ee
		
	We will consider one massless scalar $\varphi$ here. 
	A general solution to the massless scalar equation $\nabla^2\varphi=0$ 	is given by
	\begin{align}
	\varphi^{(0)}(r,\theta)=\sum_m p_me^{im\theta}\pqty{r-1\over r+1}^{\abs{m}\over 2}\label{fulscsol}
	\end{align}
	where $p_m$ are coefficients which are   fixed by the form of $\varphi$ at the boundary. 
	Near the  boundary at large $r$ we get 
	\begin{align}
	\varphi^{(0)}=\sum_m p_me^{im\theta}\pqty{1-{\abs{m}\over r}}\equiv\varphi_-(\theta)-\frac{1}{r}\varphi_+(\theta)\label{asyscsol}
	\end{align}
	where
	\begin{align}
	\varphi_-(\theta)=\sum_m p_m e^{im\theta},\quad \varphi_+=\sum_m \abs{ m} p_m e^{im\theta}=\int d\theta' F(\theta,\theta')\varphi_-(\theta')\label{phmphpdef}
	\end{align}
	and 
	\begin{align}
	F(\theta,\theta')=\sum_{m=-\infty}^{\infty} \abs{m}e^{im(\theta-\theta')}\label{Fdef}
	\end{align}
	
	If   $\varphi$ is  given by the function  ${\hat \varphi} (u)$ along the boundary, with $u$ being the rescaled proper length as above, then we get (to leading order ) 
	\be
	\label{ass2}
	\varphi_{-}(\theta) = \sum_m p_m e^{i m \theta} = {\hat \varphi} (f(\theta))
	\ee
	which  determines the Fourier coefficients $p_m$ in terms of the functions ${\hat \varphi } $ and $f$. 
	It is easy to see that the classical action for the scalar 
	\be
	\label{classica}
	S= {1\over 2} \int \sqrt{g} d^2x (\partial \varphi)^2
	\ee
	reduces on shell to a boundary term, 
	\be
	\label{bterm}
	S= {1\over 2} \int ds\, \varphi n^\mu \partial_\mu \varphi
	\ee
	where $ds$ is the line element along the boundary and $n^\mu$ the unit normal. 
	This gives 
	\be
	\label{ass3}
	S={1\over 2} \int d\theta_1d\theta_2 \varphi_{-}(\theta_1) \varphi_{-}(\theta_2) F(\theta_1,\theta_2)
	\ee
	where $\varphi_{-}(\theta)$ is given in terms of the boundary function ${\hat \varphi}(u)$ and $f(u)$ by eq.(\ref{ass2}). 	
	Inverting eq.(\ref{deaf}) we can express $\theta$ as a function of $u$
	\be
	\label{deft}
	\theta (u) = f^{-1}(u)
	\ee
	which allows us to also express eq.(\ref{ass3}) as 
	\be
	\label{ass4}
	S= {1\over 2} \int du_1 du_2 \theta'(u_1) \theta'(u_2) {\hat \varphi}(u_1) {\hat \varphi}(u_2) F(\theta(u_1), \theta(u_2) )
	\ee
	This gives the classical action in terms of the boundary time reparametrizations  specified by $\theta(u)$ and the boundary value of the scalar ${\hat \varphi(u)}$. 
	
	At linear order in the diffeomorphisms we have ${\tilde \theta}$ given in terms of $\theta$ in eq.(\ref{largedi}), (\ref{default}),  inserting this  in eq.(\ref{ass4}) gives 
	\be
	\label{linorass}
	S=  \int_\del du_1 du_2 \hat{\varphi}(u_1)\hat{\varphi}(u_2)(\delta\theta'(u_1)F(u_1,u_2)+\delta\theta(u_1)\del_{u_1}F(u_1,u_2))
	\ee
	This result agrees with \eqref{links} in subsection \ref{jtwithmatterads2}, after being generalised to $N$ scalar fields. 
	
	{Also, for the discussion in subsection \eqref{eadsmfr} these formulas need to be extended to $O(\epsilon)$ if we are to include the dependence on the large diffeomorphisms coming from the quantum part, i.e. the scalar laplacian determinant in  \eqref{pisb}}. This can be done in a straightforward fashion along the lines above, but we spare the reader the details. 
	
	Now, in the case of de Sitter spacetime, the matter coupling to the time reparametrization modes can be obtained in an almost similar fashion. The massless scalar field equation in the coordinate system eq.\eqref{minds} now has the solution
	\begin{align}
	\varphi(r,\theta)=\sum_m p_m \pqty{\frac{r-i}{r+i}}^{\abs{m}\over 2}e^{im\theta}\label{dsscesol}
	\end{align}
	This in the limit $r\gg 1$ has the asymptotic form
	\begin{align}
		\varphi(r,\theta)=\sum_m p_m e^{im\theta} \pqty{1-\frac{i\abs{ m}}{r}}=\varphi_{-}(\theta)-\frac{i}{r}\varphi_+(\theta)\label{dsmasye}
	\end{align}
	where $\varphi_{-},\varphi_+$ are as before in eq.\eqref{asyscsol}. The matter coupling can then be obtained in a manner analogous to that in the AdS case above. Doing so, we get
	\begin{align}
	S_{M,cl}={i\over 2} \int d\theta_1d\theta_2 \varphi_{-}(\theta_1) \varphi_{-}(\theta_2) F(\theta_1,\theta_2)\label{dsmcoup}
	\end{align}
	where $F$ is as defined in eq.\eqref{Fdef}. So, comparing eq.\eqref{ass3} and \eqref{dsmcoup}, we see that the expressions upto the factor of $i$ and so the linearized version in eq.\eqref{linorass} will also have an additional factor of $i$.

	\section{de Sitter wavefunction using Euclidean AdS contour}

	\label{nadsds}
	In this appendix we calculate the wavefunction for the de Sitter spacetime in the non-asymptotic limit, by considering modes which have $m>l$ where $l$ is the length of the boundary. Although in such a case we need to carefully calculate various quantities such as determinants, measure for large and small diffeomorphisms which does not decouple, we ignore all such subtleties and evaluate the measure for large diffeomorphisms and try to do the path integral. To evaluate the wavefunction, we follow the Maldacena contour, described in subsection \ref{dsbasic} and so we first do the computation in the negative AdS metric of signature (0,2) and then analytically continue to the (1,1) de Sitter spacetime.
	Consider the metric given by 
	\begin{equation}
	ds^2=-(d\tau^2+\sinh^2\tau d\theta^2)\label{nadsmet}
	\end{equation}
	It is easy to compute the Ricci scalar for this metric which turns out to have the value $R=2$. This metric is the negative of the AdS metric written in global coordinates.  As before, we find that the zero modes of the operator $P^\dagger P$ are given by the vector field as either the gradient or the curl of a scalar which satisfies the scalar Laplacian equation in the background eq.\eqref{nadsmet}. So, we get the vector field as
	\begin{equation}
	V^a=\epsilon^{ab}\nabla_b \psi,\quad \psi=-e^{im\theta}\hat{c}_m
	(\abs{m} +r) \left(\frac{r-1}{r+1}\right)^{\frac{\abs{m} }{2}},\quad  r=\cosh\tau\label{nadscvec}
	\end{equation}
	where $\epsilon^{12}=-\frac{1}{\abs{\sqrt{g}}}$ and $\hat{c}_{-m}=\hat{c}_m^*$ so that the scalar field and the vector field constructed out of it is real. 
 The components of the vector field  written explicitly are
	\begin{equation}
	V^a=\pqty{\frac{i\hat{c}_{m} m e^{i m \theta  }  (\abs{m} +\cosh \tau) \tanh ^{\abs{m} }\left(\frac{\tau }{2}\right)}{\sinh\tau},-\frac{ \hat{c}_{m} e^{i m \theta  }  \left(\abs{m}\cosh \tau   +m^2+\sinh^2\tau\right) \tanh ^{\abs{m} }\left(\frac{\tau }{2}\right)}{\sinh^2\tau}}\label{nadscvcom}
	\end{equation}
	The coefficients
	The corresponding metric perturbations are given by 
	\begin{align}
	\delta g_{\tau\tau}=&{-2 i\hat{c}_{m} m {(m^2-1)}  \text{csch}^2\tau \tanh ^{\abs{m} }\left(\frac{\tau }{2}\right)}e^{i m \theta  }\nonumber\\
	\delta g_{\tau\theta}=&2  \hat{c}_{m} {(m^2-1)} {\abs{m} }  \text{csch}\tau \tanh ^{\abs{m} }\left(\frac{\tau }{2}\right)e^{i m \theta  }\nonumber\\
	\delta  g_{\theta\theta}=&{2 i \hat{c}_{m} m {(m^2-1)}  \tanh ^{\abs{m} }\left(\frac{\tau }{2}\right)}e^{i m \theta  }\label{nadseucmetp}
	\end{align}
	For an arbitrary $\tau=\tau_0$, we have the boundary term for the inner product of two metric perturbations analogous  to that of the eq.\eqref{bdtermpd} to be
	\begin{align}
	\langle PV_{L, m}, PV_{L,-m}\rangle =&2\int_{\tau=\tau_0} d\theta\sqrt{g}g^{\tau\tau}(V^{\tau}\delta g_{\tau\tau}+V^{\theta}\delta g_{\tau \theta})\nonumber\\
	%=&-2\int_{\tau=\tau_0} d\theta\sinh \tau\,(V^{\tau}\delta g_{\tau \tau}+V^{\theta}\delta g_{\tau \theta})\nonumber\\
	=&\sum_{\abs{m}>1}{8 \pi  \hat{c}_{m} \hat{c}_{-m}\abs{m}(m^2-1) \tanh ^{2 \abs{m} }\left(\frac{\tau_0 }{2}\right) \left(2 \abs{m} ^2 \text{csch}^2(\tau_0 )+1 +2 \abs{m} \coth (\tau_0 ) \text{csch}(\tau_0 )\right)}\label{nadseucbter}
	\end{align}
	So the measure for the path integral over $\hat{c}_m$ is given by 
	\begin{equation}
	\hat{M}=\sum_{{ m}>1}16 \pi  \hat{c}_{m} \hat{c}_{-m}\abs{m}(m^2-1) \tanh ^{2 \abs{m} }\left(\frac{\tau_0 }{2}\right) \left(2 \abs{m} ^2 \text{csch}^2(\tau_0 )+1 +2 \abs{m} \coth (\tau_0 ) \text{csch}(\tau_0 )\right)\label{nadseucmes}
	\end{equation}
	The extrinsic for the fluctuations around the saddle $\tau=\tau_0$ is obtained in eq.\eqref{nadsexk}, which expressed in terms of an expansion
	\begin{equation}
	\delta\theta(u)=\sum_m\tilde{c}_m e^{imu}\label{nadseuctex}
	\end{equation}
	is given by 
	\begin{equation}
	\delta K^{(2)}=-\sum_{{m} \geq 2} \tilde{c}_{-m} \tilde{c}_{m} m^2 \left(m^2-1\right) \tanh \tau_0 \text{sech}^2\tau_0\label{nadseucqa}
	\end{equation}
	where the superscript on $\delta K$ is to indicate that this is the quadratic term in time reparametrization modes $\tilde{c}_m$.
	The relation between $\hat{c}_m$ and $\hat{c}_{m}$ is obtained by noting that $\delta\theta=V^\theta$ and is given by 
	\begin{equation}
	\tilde{c}_m=-{\hat{c}_{m} \text{csch}^2\tau \left(\abs{m}\cosh \tau  +\sinh^2 \tau+m^2\right) \tanh ^{\abs{m} }\left(\frac{\tau }{2}\right)}\label{nadseucchcr}
	\end{equation}
	Using this to find the quadratic action in terms of the variables $\hat{c}_{m}$, we obtain
	\begin{align}
	\delta K^{(2)}=
	-&\sum_{m\geq 1}  \hat{c}_{m} \hat{c}_{-m} \abs{m}^2(m^2-1)  \text{csch}^3\tau_0 \text{sech}^3\tau_0 \left(\abs{m}\cosh \tau_0  +\sinh^2 \tau_0+m^2\right)^2 \tanh ^{2 \abs{m} }\left(\frac{\tau_{0}}{2}\right)\label{nadseucqdact}
	\end{align}
The path integral over the modes $\hat{c}_{m}$ is given by,$\Psi_{nAdS}$,
	\begin{align}
	\Psi_{nAdS}=\exp[\frac{\phi_B\cosh\tau_0}{8\pi G}]&\int \hat{M} d\hat{c}_{m}d\hat{c}_{m}^* \exp{\tilde{\gamma}\int_0^{2\pi} du \,\,\phi_B\sinh\tau_0\delta K^{(2)}}\nonumber\\
	=\exp[\frac{\phi_B\cosh\tau_0}{8\pi G}]&\sum_{m>1}\frac{32 \pi  \cosh ^3\tau_0 \left( \half \sinh ^2\tau_0+m^2+ \abs{m} \cosh \tau_0\right)}{\tilde{\gamma}  m \phi_B\left(\abs{m}\cosh \tau_0  +m^2+\sinh^2\tau_0\right)^2}\label{nadseucpian}
	\end{align}
	where $u$ is related to the proper time on the boundary and is defined through the relation eq.\eqref{nadsbdele}. Note that we have ignored the contribution from the topological term eq.\eqref{Stop} and the exponential prefactor above is the classical contribution coming from the leading term in the  extrinsic trace eq.\eqref{nadsexk}. The sum in eq.\eqref{nadseucpian} can be regulated using zeta-function regularization. Defining the variables $\alpha,m_{1}, m_{2},m_{3},m_{4}$, as
	\begin{align}	\alpha=&\frac{32 \pi \cosh ^3\tau_0}{\tilde{\gamma}  \phi_B}, \quad \tilde{\gamma}=\frac{1}{8\pi G}\nonumber\\
	m_{1}=&\frac{1}{4} \left(-2 \cosh \tau_0-\sqrt{2} \sqrt{3-\cosh (2 \tau_{0})}\right),\quad 
	{m_{2}}=&\frac{1}{4} \left(\sqrt{2} \sqrt{3-\cosh (2 \tau_{0})}-2 \cosh \tau_0\right)\nonumber\\
	m_{3} =&\frac{1}{4} \left(-2 \cosh \tau_0-\sqrt{10-6 \cosh (2 \tau_{0})}\right),\quad
	{m_{4}}=&\frac{1}{4} \left(\sqrt{10-6 \cosh (2 \tau_{0})}-2 \cosh \tau_0\right)\label{nadseucmsd}
	\end{align}
	we get the regularized value to be
	\begin{equation}
	\Psi_{nAdS}={\exp[{\tilde{\gamma}\phi_B\cosh\tau_0}]\over\sqrt{(2\pi\alpha)^3}}{\Gamma(2-m_{3})^2\Gamma(2-m_{4})^2 \over  \Gamma(2-m_{1})\Gamma(2-m_{2}) }\label{nadseucpire}
	\end{equation}
	For evaluating large $\tau_0\gg 1$ behaviour, we note that $\pi e^{\tau_0}=l$, and so we get, using the Stirling approximation for the Gamma functions, 
	\begin{align}
	\Psi_{nAdS}\xrightarrow{\tau_0\gg 1}	&\frac{l^3}{32 \sqrt{\pi} }\sqrt{\frac{(\tilde{\gamma}\phi_B)^3}{l^9}}\exp(\frac{\tilde{\gamma}\phi_B l}{2\pi}+\frac{l}{2\pi}\ln({l\over \pi\sqrt{2}e})+l\left(\frac{1}{8}-\frac{1}{\sqrt{3}}\right))\label{nadslartau}
	\end{align}
	Now doing a continuation to the Lorentzian de Sitter by taking 
	\begin{equation}
	\tau_0\rightarrow \tau_0\pm\frac{ i\pi}{2}\Rightarrow l\rightarrow \pm i l\label{nadstodsan}
	\end{equation}
	we get 
	\begin{align}
	\Psi_{dS}
	&=\pm\frac{l^3}{32 \pi i  }\sqrt{ \frac{(\tilde{\gamma}\phi_B)^3}{ (\pm il^9)}}\exp(-\frac{l}{4}\pm\frac{\tilde{\gamma}\phi_B l}{2\pi}\pm \frac{ il}{2\pi}\ln({l\over \pi\sqrt{2}e})\pm il\left(\frac{1}{8}-\frac{1}{\sqrt{3}}\right))\label{nadslartau}
	\end{align}
	We see that there is an exponential damping term for large $l$. The $\pm$ signs in the above expression correspond to the $\pm$ signs in eq.\eqref{nadstodsan} for the different ways of analytic continuation. As can be seen from eq.\eqref{nadslartau}, we find that the exponential damping is independent of the choice of analytic continuation. Moreover this exponential damping cannot be removed by adding a length-dependent counterterm with a real coefficient as that would have an explicit factor of $i$ as in the action eq.\eqref{actsds}.
	
	\section{More on AdS double trumpet calculations}
	\label{dbapdx}
	\subsection{Measure for large diffeomorphisms and Schwarzian action}
	\label{dbldfiscact}
	In this appendix we will elaborate more on the calculation of the measure for the large diffeomorphisms in the double trumpet topology and also show the calculation of the Schwarzian action in explicit detail. The line element is given by eq.\eqref{dbmet} 
	The solutions for the scalar field $\psi$ satisfying eq.\eqref{laps} is given in eq.\eqref{dblapsolrw}. We will use the form in eq.\eqref{dblapsol2} to calculate the measure and the form in eq.\eqref{dblapsolrw} to evaluate the Schwarzian action and finally relate them using eq.\eqref{dbabgdre}.
	We can now construct the vector field corresponding to the large diffeomorphisms. In the disk topology, the modes $m=\pm 1, 0$ for the vector field taken as the curl of the scalar field turned out to be isometries for the spacetime. However, we now have only one isometry corresponding to the $m=0$ mode. So, the large diffeomorphisms correspond to modes with $\abs{m}\geq 1$. 
	The components of the vector field eq.\eqref{steel} computed in terms of the solution eq.\eqref{dblapsol2} is given by
	\begin{align}
	V^r_L&=\del_\theta\psi=i \tilde{m} e^{i \tilde{m} \theta } \left(\frac{r-i}{r+i}\right)^{-\frac{1}{2} (i \tilde{m})} \left(A_m (\tilde{m}+r)+B_m (r-\tilde{m}) \left(\frac{r-i}{r+i}\right)^{i \tilde{m}}\right) \nonumber\\
	V^\theta_L&=-\del_r\psi=-\frac{e^{i \tilde{m} \theta }}{r^2+1} \left(\frac{r-i}{r+i}\right)^{-\frac{1}{2} (i \tilde{m})} \left(A_m \left(\tilde{m}^2+\tilde{m} r+r^2+1\right)+B_m \left(\tilde{m}^2-\tilde{m} r+r^2+1\right) \left(\frac{r-i}{r+i}\right)^{i \tilde{m}}\right)\label{dbcucom}
	\end{align}
	The metric perturbations obtained by $\delta g_{ab}=(PV)_{ab}$ is given by 
	\begin{align}
	(PV_L)_{rr}=&\frac{2 i \left(\tilde{m}^3+\tilde{m}\right)}{\left(r^2+1\right)^2} e^{i \tilde{m}\theta  } \left(\frac{r-i}{r+i}\right)^{-\frac{1}{2} (i \tilde{m})} \left(A_m+B_m \left(\frac{r-i}{r+i}\right)^{i \tilde{m}}\right)\nonumber\\
	(PV_L)_{r\theta}=&-\frac{2 \left(\tilde{m}^3+\tilde{m}\right) }{r^2+1} e^{i \tilde{m}\theta  } \left(\frac{r-i}{r+i}\right)^{-\frac{1}{2} (i \tilde{m})} \left(A_m-B_m \left(\frac{r-i}{r+i}\right)^{i \tilde{m}}\right)\nonumber\\
	(PV_L)_{\theta\theta}=&-2 i \left(\tilde{m}^3+\tilde{m}\right) e^{i \tilde{m}\theta  } \left(\frac{r-i}{r+i}\right)^{-\frac{1}{2} (i \tilde{m})} \left(A_m+B_m \left(\frac{r-i}{r+i}\right)^{i \tilde{m}}\right)
	\label{dbpvlcom}
	\end{align}
	It is now straightforward to compute the measure  for the modes corresponding to the large diffeomorphisms. The measure is obtained by taking the inner product of two metric perturbations $PV_L^{(1)}$ and $PV_L^{(2)}$ using eq.\eqref{metnrm}, which just becomes the boundary term given in eq.\eqref{bdtermpd}. We now have two boundary terms due to the two boundaries as $r\rightarrow\pm \infty$. The value of this boundary term at a single boundary is given by 
	\begin{align}
	BT=&\pm \sum_{\abs{m}>0}2b\tilde{m}(\tilde{m}^2+1)\left(A_m A_{-m} -B_m B_{-m}+\frac{A_m B_{-m} \left(2 \tilde{m}^2+2 \tilde{m} r+r^2+1\right) \left(\frac{r-i}{r+i}\right)^{-i \tilde{m}}}{r^2+1}\right)\nonumber\\
	&\mp 2b\tilde{m}(\tilde{m}^2+1)\left(\frac{A_{-m} B_m \left(2 \tilde{m}^2-2 \tilde{m} r+r^2+1\right) \left(\frac{r-i}{r+i}\right)^{i \tilde{m}}}{r^2+1}\right)\label{dbmes}
	\end{align}
	where the upper sign is to be used at the right boundary ($r\rightarrow\infty$) and the lower sign at the left boundary ($r\rightarrow-\infty$). The relative sign between the two boundaries arises due to the change in the sign of the outward normal used to compute this boundary term in eq.\eqref{bdtermpd}. In the asymptotic limit, eq.\eqref{dbmes} becomes,
	\begin{align}
	BT=\pm 2b\tilde{m}(\tilde{m}^2+1)\left(-B_m B_{-m}+A_m A_{-m} +A_m B_{-m}  \left(\frac{r-i}{r+i}\right)^{-i \tilde{m}}-A_{-m} B_m  \left(\frac{r-i}{r+i}\right)^{i \tilde{m}}\right)\label{dbasymes}
	\end{align}
	We will now evaluate each of the boundary terms separately. The contribution to the measure coming from the boundary term at $r\rightarrow \infty$, denoted $M_2$, can be evaluated by following the conventions in eq.\eqref{dnlnz},\eqref{defat}. We see that as $r\rightarrow\infty$
	\begin{align}
	\pqty{r-i\over r+i}^{\pm {i\tilde{m}\over 2}}=\exp({\pm i\tilde{m}\over 2}\ln(r-i\over r+i))\simeq 1\label{dbrpili}
	\end{align}
	  and so from eq.\eqref{dbrpili}, we find
	\begin{align}
	M_2=\sum_{\abs{m}\geq 1}4b \tilde{m}(\tilde{m}^2+1)A_m B_{-m}\label{dbmpmes}
	\end{align}
	Noting that as $r\rightarrow -\infty$,
	\begin{align}
	\pqty{r-i\over r+i}^{\pm {i\tilde{m}\over 2}}=\exp({\pm i\tilde{m}\over 2}\ln(r-i\over r+i))\simeq \exp(\pm \tilde{m}\pi)\label{dbrmili}
	\end{align}
	the contribution to the measure from the boundary term at $r\rightarrow-\infty$, denoted $M_1$, is given by 
	\begin{align}
	M_1=-\sum_{\abs{m}\geq 1}4b \tilde{m}(\tilde{m}^2+1)A_m B_{-m}e^{-2 \tilde{m}\pi}\label{dbmmmes}
	\end{align}
	So the total measure, $M$,  becomes
	\begin{align}
	M=M_1 +M_2= \sum_{\abs{m}\geq 1}4b \tilde{m}(\tilde{m}^2+1)A_m B_{-m}(1-e^{-2 \tilde{m}\pi`})\label{dbtomes}
	\end{align}
	Noting that 
	\begin{align}
	\label{dbrealgd}
	\delta_m=-\delta_{-m}^*, \gamma_m=-\gamma_{-m}^*
	\end{align} which follow from the requirement that the solution eq.\eqref{dblapsolrw} be real, we can rewrite this in term of the modes $\gamma_m,\delta_m$ using eq.\eqref{dbabgdre} as 
	\begin{align}
	M=\sum_{\abs{m}\geq 1}4 b \tilde{m}(\tilde{m}^2+1)\pqty{(\gamma_m\gamma_m^*+\delta_m\delta_m^*)\sinh(2 \tilde{m} \pi)+2(\gamma_m \delta_m^*+\delta_m \gamma_m^*)\sinh(\tilde{m}\pi)}
	\label{dbtotm}
	\end{align}
	Further expressing the complex variables $\gamma_m,\delta_m$  in terms of the real variables $p_m, q_m, r_m,s_m$ as in eq.\eqref{dbgdpqrs}	we find that the measure becomes
	\begin{align}
	M=\sum_{m\geq 1}16 b \tilde{m}(\tilde{m}^2+1)\sinh(\tilde{m}\pi)\pqty{(p_m^2+q_m^2+r_m^2+s_m^2)\cosh(\tilde{m} \pi)+2(p_m r_m+q_m s_m)}\label{dbgusmes}
	\end{align}
	Reading off the measure for the large diffeomorphism modes $p_m, q_m, r_m, s_m,$ from the above, we have
	\begin{align}
	\int D[PV_L]=\int \prod_{m\geq 1}\left[(16 b \tilde{m} \left(\tilde{m}^2+1\right) \sinh (\pi  \tilde{m}))^2dp_m\,dq_m\,dr_m\,ds_m\right]\label{dbldfmes2}
	\end{align}
	We will now evaluate the action for the large diffeomorphisms. The action is given by the boundary term in the JT action eq.\eqref{bjtact}. 	Using eq.\eqref{dbrpili},\eqref{dbrmili}, we see that the scalar field solution eq.\eqref{dblapsolrw} as $r\rightarrow \infty$ becomes
	\begin{align}
	\psi\big\vert_{r\rightarrow\infty}\simeq 2\sum_{m}e^{i\tilde{m}\theta}\delta_m r\sinh(\tilde{m}\pi)\label{dbprp}
	\end{align}
	and at $r\rightarrow-\infty$, we get
	\begin{align}
	\psi\big\vert_{r\rightarrow -\infty}\simeq -2\sum_{m}e^{i\tilde{m}\theta}\gamma_m r\sinh(\tilde{m}\pi)\label{dbprm}
	\end{align}
	which shows that the the large diffeomorphism at the left and right boundaries are independent and so we can independently compute the action at each of the boundaries. In the asymptotic AdS limit, for the parametrization of $\theta(u)$ in terms of the diffeomorphism as
	\begin{align}
	\theta(u)=\frac{b}{2\pi}u+V_L^\theta(u)\label{dbthurel}
	\end{align}
	the boundary term in the JT action, eq.\eqref{dbbjtact} to the quadratic order in large diffeomorphisms, becomes
	\begin{align}
	S_{JT,\del}=\frac{b^2}{16\pi G J \beta}+\frac{1}{8 GJ\beta}\int_0^{2\pi} du \left[(\del_u V_L^\theta)^2+\left(\frac{2\pi}{b}\right)^2(\del_u^2 V_L^\theta)^2\right]\label{dbjtbac}
	\end{align}
	where we used the fact the dilaton is of the form eq.\eqref{scaled} at the boundary in the asymptotic AdS limit.
	Noting that near the boundary at $r\rightarrow\infty$, the large diffeomorphism is given by 
	\begin{align}
	V_L^\theta(u)\simeq 2\sum_{m}e^{i{\tilde{m}b\over 2\pi}u}\delta_m \sinh(\tilde{m}\pi)\label{dbrpvec}
	\end{align}
	the term quadratic in the large diffeomorphism in the action is given by
	\begin{align}
	S_{JT,\del_2}=\frac{b^2}{16\pi G J \beta_1}+\frac{1}{2\pi G}\frac{b^2}{J\beta_1} \sum_{m\geq 1} \tilde{m}^2 (\tilde{m}^2+1)\sinh^2(\tilde{m}\pi)\delta_m\delta_m^*\label{dbactrp}
	\end{align}
	where we have used eq.\eqref{dbrealgd} and the first line in eq.\eqref{limeades2} to obtain the above equation.
	Similarly the action for the boundary term near $r\rightarrow -\infty$ can be obtained by noting the vector field corresponding to the large diffeomorphisms is given by 
	\begin{align}
	V_L^\theta(u)\simeq -2\sum_{m}e^{i\tilde{m}u}\gamma_m \sinh(\tilde{m}\pi)\label{dbrmvec}
	\end{align}
	and the term quadratic in the large diffeomorphism in the action is given by 
	\begin{align}
	S_{JT,\del_1}=\frac{b^2}{16\pi G J \beta_2}+\frac{1}{2\pi G}\frac{b^2}{J\beta_2} \sum_{m\geq 1}\tilde{m}^2 (\tilde{m}^2+1)\sinh^2(\tilde{m}\pi) \gamma_m \gamma_m^*\label{dbactrm}
	\end{align}
	where again we have used eq.\eqref{dbrealgd} and the second line in eq.\eqref{limeades2} to obtain the above result.   Combining eq.\eqref{dbactrm} and eq.\eqref{dbactrp} and expressing in terms of $p_m, q_m, r_m,s_m$ using eq.\eqref{dbgdpqrs}, the net action then becomes
	\begin{align}
	S_{JT,\del}=\frac{b^2}{16\pi G J}\pqty{\frac{1}{\beta_1}+\frac{1}{\beta_2}}+\sum_{m\geq 1}\frac{1}{2\pi G}\frac{b^2}{J} \tilde{m}^2 (\tilde{m}^2+1)\sinh^2(\tilde{m}\pi) \pqty{\frac{p_m^2+q_m^2}{\beta_1}+\frac{r_m^2+s_m^2}{\beta_2}}\label{dbldtact}
	\end{align}
	
	We will now elaborate on the orthogonality of different classes of metric perturbations namely, those corresponding to twist, $b$-modulus, small and large diffeomorphisms. The discussion regarding the inner product of small and large diffeomorphisms is presented towards the end of \ref{offdiagme} which in the asymptotic AdS limit satisfies the inequality
	\begin{align}
	{\langle PV_s,PV_L\rangle\over \sqrt{\langle PV_s,PV_s\rangle\langle PV_L, PV_L\rangle}}\leq \order(r^{-\frac{3}{2}})\label{dtvsvlinp}
	\end{align}
	with the notation being the same as in appendix \ref{offdiagme}. Let us first consider the inner product of the metric perturbation corresponding to $b$-modulus with others. From eq.\eqref{dbmodper}, eq.\eqref{dbtwper}, noting the metric eq.\eqref{dbmet} and using the definition eq.\eqref{metnrm}, it is straightforward to see that
	\begin{align}
	\langle PV_{mod}, PV_{tw}\rangle=0\label{dttwmod}
	\end{align}
	Further since there is not $\theta$ dependence in $PV_{b}$ where as the large diffeomorphisms have the dependence on $\theta$ as $e^{im\theta},m\geq 1$, we have
	\begin{align}
	\langle PV_{b}, PV_{L,m}\rangle =0\label{dtmodlar}
	\end{align}
	Considering the inner product of a metric perturbation for small diffeomorphism with $PV_{mod}$ we get
	\begin{align}
	\langle PV_{b}, PV_{s,\lambda,m}\rangle =&\langle P^\dagger PV_{b}, V_{s,\lambda,m}\rangle+\int_{\del}V_{s,\lambda,m}^a n^b (PV_{b})_{ab}\nonumber\\
	=&\int_{\del}V_{s,\lambda,m}^\theta n^r (PV_{b})_{\theta r} =0
	\label{dtmodlarb}
	\end{align}
	where in obtaining the second line we used eq.\eqref{pdpvbtw} and for a small diffeomorphism that eq.\eqref{rsqmval} is satisfied at the boundary. The vanishing of the second line then follows by noting that $PV_{b}$ does not have off-diagonal components, see eq.\eqref{dbtwper}. Now, we shall consider the inner product of $V_{tw}$ with other metric perturbations. The inner product with $PV_b$ is already obtained in eq.\eqref{dttwmod}. The inner product of $PV_{tw}$ with large diffeomorphisms also vanishes since as before the large diffeomorphisms have a non-trivial $\theta$ dependence whereas the twist perturbation has no $\theta$ dependence as it corresponds to $m=0$ sector, see eq.\eqref{dbtwper}. The inner product of $PV_{tw}$ and $PV_{s,\lambda,m}$, can be simplified following the steps in eq.\eqref{dtmodlarb} to obtain
	\begin{align}
	\langle PV_{tw},PV_{s,\lambda,m}\rangle =&\int_{\del}V_{s,\lambda,m}^\theta n^r (PV_{tw})_{\theta r} \label{dtvtvsinp}
	\end{align}
	which is non-zero for $m=0$ mode of $V_{s,\lambda,m}$ after noting  eq.\eqref{dbtwper}. So, to estimate it, we consider the quantity, 
	\begin{align}
	\langle PV_{tw},PV_{s,\lambda,m}\rangle\over \sqrt{\langle PV_{tw},PV_{tw}\rangle \langle PV_{s,\lambda,-m},PV_{s,\lambda,m}\rangle}\label{dtvtvsnoin}
	\end{align}
	We shall show that in the asymptotic AdS limit, the above quantity goes as $\order{(r^{-3/2})}$. The quantity $\langle PV_{tw},PV_{tw}\rangle$ has already been computed in eq.\eqref{moduliinp}. For $m=0$, using eq.\eqref{smalldifexp}, we have that 
	\begin{align}
	V^\theta_{s,\lambda,0}=\del_r\psi_{ \lambda ,0}\sim \order{(r^{-3/2})}\label{dtvsor}
	\end{align}
	It is also easy to see from line element eq.\eqref{dbmet} that the normal vector at either of the boundaries $r=-r_{B1}$ or $r=r_{B2}$, has the behaviour
	\begin{align}
	n^r\sim \order{(r)}\label{dtnoror}
	\end{align}
	The analog of the calculation leading to eq.\eqref{vsvsinp} for the double trumpet shows that 
	\begin{align}
	\abs{\langle PV_{s,\lambda,-m},PV_{s,\lambda,m} \rangle} \geq \abs{(\lambda+2)\pi r\psi_{\lambda ,m}\psi_{\lambda ,-m}\bigg\vert_{\del}}\sim \order{(r^0)}\label{dtvsvsinp}
	\end{align}
	where $\del$ stands for both the boundaries. From eq.\eqref{dbtwper}, we have
	\begin{align}
	(PV_{tw})_{r,\theta}=-\frac{2t}{\pi (r^2+1)}\sim\order{(r^{-2})}\label{dbpvtwrt}
	\end{align}
	Further noting that $\langle PV_{tw}, PV_{tw} \rangle\sim \order{(r^0)} $ from eq.\eqref{moduliinp}, we have, combining eq.\eqref{dtvsor}, \eqref{dtnoror}, \eqref{dtvsvsinp} that 
	\begin{align}
	{\langle PV_{tw},PV_{s,\lambda,m}\rangle\over \sqrt{\langle PV_{tw},PV_{tw}\rangle \langle PV_{s,\lambda,m},PV_{s,\lambda,-m}\rangle}}\sim \order{(r^{-3/2})}\label{dtvtwvs}
	\end{align}
	which vanishes in the asymptotic AdS limit. So, in total, noting all the above results, eq.\eqref{measures} then follows immediately.

	\subsection{Matter in double trumpet calculations}
	\label{dbmatapx}
	
	In this appendix we shall elaborate on the details used in the discussion in subsection \ref{dbtwm}. We shall carefully evaluate the determinant of scalar laplacian, $\text{det}(-\hat{\nabla}^2)$. We will consider massless scalar in the background of the double trumpet topology with the metric written in conformally flat coordinate system, eq.\eqref{edletrs}, as
	\begin{align}
	ds^2={dr_*^2+d\theta^2\over \cos^2{r_*}},\quad \quad r_*\in\left[-\frac{\pi}{2},\frac{\pi}{2}\right],\quad \theta\in [0,b]\label{dbconfmet}
	\end{align}
	We can compute the dependence on $b$ by noting that the metric above is conformally flat and so we can use the conformal anomaly to evaluate the contribution due to the conformal factor and then compute the contribution from the flat metric separately.
	The $b$ dependence coming from the conformal anomaly can be evaluated using the conformal anomaly since the theory of a massless scalar field is a conformal field theory. The relation between determinants of conformally related metrics $\hat{g}_{ab}=e^{2\sigma} \bar{ g}_{ab}$ is given by 
	\begin{align}
	{\text{det}(-\hat{\nabla}^2)\over \text{det}(-\bar{\nabla}^2)}=&\exp{-\frac{1}{6\pi}\left[\half\int d^2x \sqrt{\bar{g}}{(\bar{g}^{ab}\del_a\sigma\del_b\sigma+\bar{R}\sigma)}+\int_\del d\bar{s}\bar{K}\sigma\right]
	}\label{detndco}
	\end{align}
	where quantities denoted by hats are calculated with respect to the metric $\hat{ g}$. In the case at hand
	\begin{align}
	\sigma=-\ln \cos r_*\label{dbcfcof}
	\end{align}
	and $\bar{R}=0$.
	For a non-wavy boundary, the boundary can be specified by 
	\begin{align}
	r_*=-r_{*1},r_*=r_{*2}, \quad r_{*1,2}>0\label{dbconbd}
	\end{align}
	where the subscripts $1$ and $2$ correspond to the left and right boundaries respectively. The normal vector normalized with respect to the flat metric, $\bar{ g}$, at the left and the right boundaries is given by 
	\begin{align}
	\bar{n}^\mu_2=(1,0), \bar{n}^\mu_1=(-1,0)\label{dbconnor}
	\end{align}
	and the extrinsic curvature $\bar{K}=0$. So, we have
	\begin{align}
	{\text{det}(-\hat{\nabla}^2)\over \text{det}(-\bar{\nabla}^2)}=& \exp(-\frac{b}{12\pi}\int_{-r_{*1}}^{r_{*2}}(\del_{r_*}\sigma)^2	)	\nonumber\\
	=&\exp[-\frac{b}{12\pi}\left(\tan r_{*1}+\tan r_{*2}-{r_{*1}}-{r_{*2}}\right)]\label{dbdetconf}
	\end{align}
	We shall now show the computation of the $b$ dependence in the determinant coming from the flat metric given by 
	\begin{align}
	ds^2=dr_*^2+d\theta^2 \quad \quad r_*\in[0,\pi],\quad \theta\in [0,b]\label{dbcff}
	\end{align}
	where we have shifted the range of $r_*$ by $\pi\over 2$ so as to simplify the computations.  Taking $\theta$ circle as the time direction in the Euclidean scalar field theory, the periodicity of the $\theta$ circle determines the temperature at which we need to calculate the thermal partition function. So, we have	
	\begin{align}
	\int {D}\varphi e^{-\half\int({\hat\nabla}\varphi)^2 }=Z_{M,flat}[b]=\text{Tr}\left(e^{-bH}\right)\label{dbcfpf}
	\end{align}
	the subscript $f$ in $Z_f$ to denote that the computation is being done for the flat metric,  $H$ is the Hamiltonian given by
	\begin{align}
	H=\int dr_* (\pi_\varphi^2 + (\partial_{r_*} \varphi)^2))\label{dbschamil}
	\end{align}
	where 
	\begin{align}
	\pi_\varphi=\del_t\varphi, \quad t=\theta\label{dbhadef}
	\end{align}
	The solution for the matter equation
	\begin{align}
	\bar{\nabla}^2\varphi=0\label{dbscel}
	\end{align}
	with the boundary conditions 
	\begin{align}
	\varphi(0,t)=0=\varphi(\pi,t)\label{dbscbdeq}
	\end{align}
	is given by 
	\begin{align}
	\varphi=\sum_{n=0}^\infty \frac{\sin(nr_*)}{n\sqrt{\pi}}\pqty{\alpha_n e^{-i\omega_n t}+\alpha_{-n}e^{i w_n t}},\quad \omega_n=n\label{dbscexp}
	\end{align}
	The Hamiltonian in terms of the modes is then given by 
	\begin{align}
	H=\half\sum_{n=1}^{\infty} (\alpha_n\alpha_n^\dagger+\alpha_n^\dagger \alpha_n)=\sum_{n=1}^{\infty} \alpha_n^\dagger \alpha_n-\frac{1}{24}\label{dbhammode}
	\end{align}
	The partition function then becomes
	\begin{align}
	{ Z}_{M,flat}[b]&=\text{Tr}(e^{-bH})=\prod_{n=1}^{\infty}\frac{e^{b\over 24}}{1-e^{-bn}}=\frac{1}{\eta(\tau)}\label{dbpareta}
	\end{align}
	where $\eta(\tau)$ is the Dedekind eta function and $\tau$ is related to $b$ by
	\begin{align}
	\tau=\frac{ib}{2\pi}\label{dbtaudef}
	\end{align}
	Using the modular transformation property of $\eta(\tau)$, we can study the behaviour of the partition function near $b=0$. So, we have
	\begin{align}
	{ Z}_{M,flat}[b]=\frac{1}{\eta(\tau)}=\frac{\sqrt{-i \tau}}{\eta(-\tau^{-1})}=\sqrt{\frac{b}{2\pi}}	{ Z}_{M,flat}\left[\frac{4\pi^2}{b}\right]\nonumber\\
	\Rightarrow \lim_{b\rightarrow 0}	{ Z}_{M,flat}[b]=\sqrt{\frac{b}{2\pi}}\lim_{b\rightarrow 0}	{ Z}_{M,flat}\left[\frac{4\pi^2}{b}\right]=\sqrt{\frac{b}{2\pi}}e^{\frac{\pi^2}{6b}}\label{dbflpart}
	\end{align}
	Putting together eq.\eqref{dbdetconf} and eq.\eqref{dbpareta}, we get the full dependence of $\det(-\hat{\nabla}^2)$ as
	\begin{align}
	\text{det}(-\hat{\nabla}^2)=\left(\eta\left({ib\over 2\pi}\right)\right)^2\exp[-\frac{b}{12\pi}\left(\tan r_{*1}+\tan r_{*2}-{r_{*1}}-{r_{*2}}\right)]\label{dbfuldet}
	\end{align}
	The terms $\tan r_{*1}, \tan r_{*2}$ in the exponent above diverges when $r_{*1}, r_{*2}\rightarrow{\pi \over 2}$. This  can be avoided by adding a length-dependent term with the appropriate coefficient, following which we have, in the limit $r_{*1}, r_{*2}\rightarrow{\pi \over 2}$, 
	\begin{align}
	\text{det}(-\hat{\nabla}^2)&=\left(\eta\left({ib\over 2\pi}\right)\right)^2\exp[-\frac{b}{12\pi}\left(\tan r_{*1}-\sec r_{*1}+\tan r_{*2}-\sec r_{*2}-\pi \right)]\nonumber\\
	&\simeq\left(\eta\left({ib\over 2\pi}\right)\right)^2e^{\frac{b}{12}}
	\label{dbfuldetrs}
	\end{align}
	
	{One can also compute the contribution to the determinant from the flat metric eq.\eqref{dbcff} directly by noting that the eigenvalues in the geometry eq.\eqref{dbcff} with Dirichlet boundary conditions along the $\theta$ and $r_*$ directions are given by 
	\begin{align}
	\lambda= n^2+\tilde{m}^2\label{dtadsevals}
	\end{align}
	where $n$ is the mode number along the $r_*$ direction and $\tilde{m}$ is related to the mode number $m$ in the $\theta$ direction by eq.\eqref{tmmrel}.}

	\subsection{Coleman method for computing determinants in AdS double trumpet}
	\label{adsdtcoldets}
	In this section, we shall show in detail how to evaluate the various determinants in the Euclidean AdS double trumpet geometry. The metric for the double trumpet geometry is given by 
	\begin{align}
	ds^2=\frac{dr^2}{r^2+1}+(r^2+1)d\theta^2\label{dbtnet}
	\end{align}
	The left and right boundaries are taken to be located at $r=-r_{B1}$ and $r=r_{B2}$ respectively where $r_{B1}, r_{B2}>0$.
	The solution for the wave equation 
	\begin{align}
	\hat{\nabla}^2\psi=-\lambda \psi\label{delevaleq}
	\end{align}
	for the modes of the form $e^{i \tilde{m} \theta}$ for the $\theta$ dependence is given by 
	\begin{align}
	\psi_{\lambda,{m}}=k_1 P_{v-\half}^{i\tilde{m}}(ir)+k_2 Q_{v-\half}^{i\tilde{m}}(ir)\label{esol}
	\end{align}
	where 
	\begin{align}
	v=\sqrt{
		\frac{1}{4}-\lambda},\quad \tilde{m}=\frac{2\pi m}{b}\label{nuvalde}
	\end{align}
	For $m\neq 0$ an equivalently good independent basis of solutions is 
	\begin{align}
	\psi_{\lambda,{m}}=k_1 P_{v-\half}^{i\tilde{m}}(ir)+k_2 P_{v-\half}^{-i\tilde{m}}(ir)\label{delaltbs}
	\end{align}
	We shall use this form of the general solution to compute the determinants of $(-\hat{\nabla}^2)$ and $(-\hat{\nabla}^2+2)$ with Dirichlet boundary conditions. To begin with, let us compute the value of the determinant of $(-\hat{\nabla}^2)$ using Coleman Trick. To do this, we need to impose the required boundary conditions and find the appropriately normalized solution to get the correct $\tilde{m}$ dependence. We note however, that the overall $r_B$ normalization of the solution, where $r_B>0$ referring to either $r_{B1}$ or $r_{B2}$ is the location of the boundary, cannot be uniquely fixed. We will discuss more about this point later. Consider first the case of $\tilde{m}\neq0$ modes. For this, imposing the Dirichlet boundary conditions at the left boundary, say specified as $r=-r_{B1}$ we get the relation between the constant $k_1$ and $k_2$ as
	\begin{align}
	k_2=-k_1\frac{P^{i\tilde{m}}_{v-\half}(-ir_{B1})}{P^{-i\tilde{m}}_{v-\half}(-ir_{B1})}\label{dircdelk1k2}
	\end{align}
	and so the solution becomes
	\begin{align}
	\psi_{\lambda,{m}}=\frac{k_1}{P_{v-\half}^{-i\tilde{m}}(-ir_{B1})} \pqty{P_{v-\half}^{i\tilde{m}}(ir)P_{v-\half}^{-i\tilde{m}}(-ir_{B1})- P_{v-\half}^{-i\tilde{m}}(ir)P_{v-\half}^{i\tilde{m}}(-ir_{B1})}\label{k2rek1}
	\end{align}
	Choose $k_2=k P_{v-\half}^{-i\tilde{m}}(-ir_{B1})$ and so the solution becomes
	\begin{align}
	\psi_{\lambda,{m}}=k \pqty{P_{v-\half}^{i\tilde{m}}(ir)P_{v-\half}^{-i\tilde{m}}(-ir_{B1})- P_{v-\half}^{-i\tilde{m}}(ir)P_{v-\half}^{i\tilde{m}}(-ir_{B1})}\label{delintrk}
	\end{align}
	Now, we need to fix the constant $k$ based on the following requirements. Viewed as a complex function of $\lambda$, the ratio of two solutions with different mode numbers $m_1$ and $m_2$, should have no extra poles of zeroes other those corresponding to the eigenvalues of the operator $(-\hat{\nabla}^2)$. Also, the ratio of two solutions should approach unity as $\abs{\lambda}$ goes to infinity in any direction other than the positive real axis. This completely fixed the $\tilde{m}$ dependence of the constant $k$ although the $r_{B1},r_{B2}$ dependence is ambiguous. This can be fixed by the requirement that the final answer should be independent of $r_B$ since otherwise, in the asymptotic AdS limit that would mean that we would get a $\ln r_B$ divergence in the net action, which cannot be cancelled away by a local counterterm. First, it is useful to note the asymptotic expansions of the associated Legendre functions $P_{a}^b(\pm i r_B)$. From eq.8.1.5 of \cite{Abramstegun}, we see that for at the left boundary, 
	\begin{align}
	P_{\nu}^{\mu}(z)=&\frac{{2^{-\nu-1}}\Gamma (-\half-\nu )z^{\mu-1-\nu}}{\sqrt{\pi}(z^2-1)^{\mu\over 2}\Gamma(-\nu-\mu)}F\pqty{\frac{1+\nu-\mu}{2},\frac{2+\nu-\mu}{2},\nu+\frac{3}{2},z^{-2}}\nonumber\\
	&\frac{{2^{\nu}}\Gamma (\half+\nu )z^{\mu+\nu}}{\sqrt{\pi}(z^2-1)^{\mu\over 2}\Gamma(1+\nu-\mu)}F\pqty{\frac{-\nu-\mu}{2},\frac{1-\nu-\mu}{2},-\nu+\frac{1}{2},z^{-2}}\label{abmsnaiexp}
	\end{align}
	Using this and the fact that $F(a,b,c,z)\simeq 1+\order(z)$ for small $z$, we find that the asymptotic forms at $-ir_{B1}$ and $ir_{B2}$ are
	\begin{align}
	P_{v-\half}^{i\tilde{m}}(\pm ir_B)=e^{\mp i\pi \over 4}\pqty{e^{\mp i\pi v \over 2}{(2r_B)^{- v -\half}\over \sqrt{\pi}}\frac{\Gamma(- v )}{\Gamma(\half-i \tilde{m}- v )}+e^{\pm i\pi v \over 2}{(2r_B)^{ v -\half}\over \sqrt{\pi}}\frac{\Gamma( v )}{\Gamma(\half-i \tilde{m}+ v )}}\label{abmsasasp}
	\end{align}
	However, there is a small subtlety which is as follows. After imposing the Dirichlet boundary conditions at the left boundary and relating the coefficients, the solution has to be continued through $r=0$ to the right boundary. In doing so, one has to use eq.8.1.4 of \cite{Abramstegun} and so our asymptotic expansions will have a relative factor of $e^{\tilde{m}\pi}$ between $z=-ir_B$ and $z=ir_B$ due to the presence of the branch cut between $(1,-\infty)$. The physics problem of continuing the solution through $r=0$ is unambiguous as it is perfectly well-defined point in the double trumpet, where as the expansions in \cite{Abramstegun} are defined with a different choice of branch which can be seen from the factor $(z^2-1)^{-\frac{\mu}{2}}$ in eq 8.1.4 and 8.1.5 . Thus, the solution is not continuous across $r=0$. So, to get rid of this problem, we multiply the asymptotic form at $ir_B$ by an extra factor of $e^{\pi \tilde{m}}$. Doing so, we have
	\begin{align}
	&P_{v-\half}^{i\tilde{m}}(- ir_B)=e^{ i\pi \over 4}\pqty{e^{ i\pi v \over 2}{(2r_B)^{- v -\half}\over \sqrt{\pi}}\frac{\Gamma(- v )}{\Gamma(\half-i \tilde{m}- v )}+e^{- i\pi v \over 2}{(2r_B)^{ v -\half}\over \sqrt{\pi}}\frac{\Gamma( v )}{\Gamma(\half-i \tilde{m}+ v )}}\nonumber\\
	&P_{v-\half}^{i\tilde{m}}( ir_B)=e^{\pi \tilde{m}}e^{- i\pi \over 4}\pqty{e^{- i\pi v \over 2}{(2r_B)^{- v -\half}\over \sqrt{\pi}}\frac{\Gamma(- v )}{\Gamma(\half-i \tilde{m}- v )}+e^{ i\pi v \over 2}{(2r_B)^{ v -\half}\over \sqrt{\pi}}\frac{\Gamma( v )}{\Gamma(\half-i \tilde{m}+ v )}}\label{pcorasyexp}
	\end{align}
	Using these asymptotic forms, we can immediately calculate the solution at $r=ir_{B2}$ to be
	\begin{align}
	\psi_{\lambda,{m}}=&k \pqty{P_{v-\half}^{i\tilde{m}}(ir_{B2})P_{v-\half}^{-i\tilde{m}}(-ir_{B1})- P_{v-\half}^{-i\tilde{m}}(ir_{B2})P_{v-\half}^{i\tilde{m}}(-ir_{B1})}\nonumber\\
	=&\frac{k}{\pi}(4 r_{B2} r_{B1})^{-\half}\pqty{\frac{2\sinh(\tilde{m}\pi)(4 r_{B2} r_{B1})^{ v }\Gamma( v )^2}{\Gamma(\half-i\tilde{m}+ v )\Gamma(\half-i\tilde{m}+ v )}-e^{2\pi i  v }\frac{\sinh(2\tilde{m}\pi)}{ v \sin(\pi  v )}\pqty{\frac{r_{B2}}{r_{B1}}}^v +( v \rightarrow- v )}\label{evaleqinw}
	\end{align}
	From the above we see that  as $\abs{ v }\rightarrow\infty$, the second term and the corresponding $ v \rightarrow  - v $ are subleading and so we need to choose $k$ such that $\tilde{m}$ dependence in the remaining terms cancels in this limit and that there are no extra poles or zeroes. So, in the limit $\abs{\lambda}\rightarrow\infty$, we get
	\begin{align}
	\psi_{\lambda,{m}}
	=&\frac{2\sinh(\tilde{m}\pi)k}{\pi\sqrt{4 r_{B2} r_{B1}}}\pqty{(4r_{B2} r_{B2})^v +(4r_{B2} r_{B2})^{- v }}\label{nuinf}
	\end{align}
	From the above expression, we see that there will be no spurious poles or zeroes in the ratio of two solutions. So we choose k as
	\begin{align}
	k=\frac{1}{2\sinh(\tilde{m}\pi)}\label{valk}
	\end{align}
	The solution then becomes
	\begin{align}
	\psi_{\lambda,{m}}=&\frac{(4 r_{B2} r_{B1})^{-\half}}{2\pi\sinh(\tilde{m}\pi)}\pqty{\frac{2\sinh(\tilde{m}\pi)(4 r_{B2} r_{B1})^{ v }\Gamma( v )^2}{\Gamma(\half-i\tilde{m}+ v )\Gamma(\half+i\tilde{m}+ v )}-e^{2\pi i  v }\frac{\sinh(2\tilde{m}\pi)}{ v \sin(\pi  v )}\pqty{\frac{r_{B2}}{r_{B1}}}^v +( v \rightarrow- v )}\label{delcolsol}
	\end{align}
	Now, the final step in the Coleman method is to evaluate the solution at $\lambda=0$ to get the  value of the determinant. Doing so, we have $ v =\half$, the expression in eq.\eqref{evaleqinw} becomes to leading order
	\begin{align}
	\psi_{0, \tilde{m}}=\frac{1}{\Gamma(1-i\tilde{m})\Gamma(1+i \tilde{m})}=\frac{\sinh(\tilde{m}\pi)}{\tilde{m}\pi}\label{coltricans}
	\end{align}
	Now, it remains to evaluate the contribution from the $\tilde{m}=0$ sector. The contribution from this sector can only be dependent on $r_{B2},r_{B1}$. We shall choose the normalization so as to cancel this dependence. Thus, we can ignore the contribution from the $\tilde{m}=0$ sector. So, the value of the determinant for the operator $(-\hat{\nabla}^2)$ is given by 
	\begin{align}
	\ln \text{det}(-\hat{\nabla}^2)=\sum_{m=-\infty,m\neq 0}^{m=\infty}\ln\frac{\sinh(\tilde{m}\pi)}{\tilde{m}\pi}=2\ln\eta\pqty{i b\over 2\pi}-\ln {2\pi^{3/2}}\label{lndetdel}
	\end{align}
	which in the limit $b\rightarrow 0$ gives
	\begin{align}
	\text{det}(-\hat{\nabla}^2)\simeq\frac{1}{b}e^{-\frac{\pi^2}{3b}}\label{detb0}
	\end{align}
	{Comparing this with the answer obtained earlier for the determinant in eq.\eqref{dbfuldet}, we see that there is a mismatch of the exponential term coming from the conformal part of the metric in the earlier calculation. We could not satisfactorily establish the reason for this but it could be due to the different order of limits that are being implemented in the Coleman method used above in calculating the determinant, as was also mentioned in eq.\ref{asyadsdet}. }
	Now, we will evaluate  $\text{det}(-\hat{\nabla}^2+2)$ in the same manner for Dirichlet boundary conditions. The solution is the same as in eq.\eqref{delaltbs} except that the value of $ v $ is now given by 
	\begin{align}
	 v =\sqrt{\frac{9}{4}-\lambda}\label{numdelval}
	\end{align}
	The arguments leading to eq.\eqref{delcolsol} remain the same and hence we get the same expression as in eq.\eqref{delcolsol}. In the final step of computing the determinant when we set $\lambda=0$, we get $ v =\frac{3}{2}$ and so the leading term in the solution becomes
	\begin{align}
	\psi_{0, \tilde{m}}=\frac{ r_{B2} r_{B1}}{\Gamma(2-i\tilde{m})\Gamma(2+i\tilde{m})}=\frac{\sinh(\tilde{m}\pi)}{\pi \tilde{m} (1+\tilde{m}^2)}r_{B2} r_{B1}\label{mdelcolval}
	\end{align}
	To get rid of the $r_{B1},r_{B2}$ dependence, we can further normalize the solution by a factor of $\frac{1}{r_{B1}r_{B2}}$. The reasoning for this  is  same as in the calculation of $(-\hat{\nabla}^2)$, namely that the absence of local counterterms to cancel $\ln r_B$ divergent term in the determinant. 
	We would also get dependence on $r_{B2}, r_{B1}$ from the $\tilde{m}=0$ sector which can again be normalized to unity and so we will ignore the contribution from the $\tilde{m}=0$ sector. 
	The value of the determinant of $(-\hat{\nabla}^2+2)$ is then given by 
	\begin{align}
	\ln \text{det}(-\hat{\nabla}^2+2)=\sum_{m=-\infty,m\neq 0}^{m=\infty}\ln\frac{\sinh(\tilde{m}\pi)}{\tilde{m}\pi(1+\tilde{m}^2)}=2\ln\eta\pqty{i b\over 2\pi}-\ln( {2\sqrt{\pi}b\sinh(b/2)})\label{lnmdelval}
	\end{align}
	which in the limit $b\rightarrow 0$ gives
	\begin{align}
	\text{det}(-\hat{\nabla}^2+2)\simeq\frac{1}{b^3}e^{-\frac{\pi^2}{3b}}\label{mdelb0}
	\end{align}
	Now, we shall repeat the steps for the operator $P^\dagger P$. As in the case of the disk, we have two sets of eigenvalues corresponding to the boundary conditions $\del_r\xi=0$ and $\del_r^2\psi=0$, see eq.\eqref{evalset1},\eqref{evalset2}. Let us first compute the contribution to the eigenvalues from $\xi$ with the boundary condition
	\begin{align}
	\del_r\xi=0.\label{1stsetpdf}
	\end{align}
	The solution for $\xi$ is given
	\begin{align}
	\xi_{\lambda,{m}}=k_1 P_{v-\half}^{i\tilde{m}}(ir)+k_2 P_{v-\half}^{-i\tilde{m}}(ir)\label{esolxipdp}
	\end{align}
	Now, imposing the boundary condition eq.\eqref{1stsetpdf}, the constants $k_1,k_2$ are related by 
	\begin{align}
	k_2=-k_1\frac{\partial_{r_{B1}}P^{i\tilde{m}}_{v-\half}(-ir_{B1})}{\partial_{r_{B1}}P^{-i\tilde{m}}_{v-\half}(-ir_{B1})}\label{pdp1k1k2}
	\end{align}
	Choosing $k_1=k\partial_{r_{B1}}P^{-i\tilde{m}}_{v-\half}(-ir_{B1}) $, we get the solution to be 
	\begin{align}
	\xi_{\lambda,{m}}=k \pqty{P_{v-\half}^{i\tilde{m}}(ir)\partial_{r_{B1}}P_{v-\half}^{-i\tilde{m}}(-ir_{B1})- P_{v-\half}^{-i\tilde{m}}(ir)\partial_{r_{B1}}P_{v-\half}^{i\tilde{m}}(-ir_{B1})}\label{pdp1ksol}
	\end{align}
	So, we have
	\begin{align}
	\del_r \xi_{\lambda,{m}}(r)\big\vert_{r=r_{B2}}=k \pqty{\partial_{r_{B2}}P_{v-\half}^{i\tilde{m}}(ir_{B2})\partial_{r_{B1}}P_{v-\half}^{-i\tilde{m}}(-ir_{B1})- \partial_{r_{B2}}P_{v-\half}^{-i\tilde{m}}(ir_{B2})\partial_{r_{B1}}P_{v-\half}^{i\tilde{m}}(-ir_{B1})}\label{delrxipdp1}
	\end{align}
	which using eq.\eqref{pcorasyexp} becomes
	\begin{align}
	\del_r \xi_{\lambda,{m}}(r)\big\vert_{r=r_{B2}}=&{k\over(4 r_{B1} r_{B2})^{\frac{3}{2}}}\pqty{\frac{2\sinh(\tilde{m}\pi)(2v-1)^2\Gamma(v)^2(4r_{B1} r_{B2})^v}{\Gamma(\half -i \tilde{m} +v)\Gamma(\half +i \tilde{m} +v)}-\frac{\pi \sinh(2 \tilde{m} \pi)(1-4v^2)}{v e^{-2\pi i v}\sin(\pi v)}\pqty{r_{B2}\over r_{B1}}^v}\nonumber\\
	&+(v\rightarrow -v)\label{pdp1normali}
	\end{align}
	From the above expression it is easy to see that for large $\abs{\lambda}$, we need to choose $k$ as in eq.\eqref{valk}. Doing so and setting $\lambda=0\Rightarrow v=\frac{3}{2}$ in eq.\eqref{pdp1normali}, we get the leading term to be
	\begin{align}
	\partial_{r_{B2}}\xi(r_{B2})=\frac{\sinh(\tilde{m}\pi)}{\pi \tilde{m} (1+\tilde{m}^2)}\label{pdpmnoval}
	\end{align}
	We see that there is no $r_{B1},r_{B2}$ dependence in the above expression. Again, we ignore the $\tilde{m}=0$ sector with the understanding that it only gives an $ r_B$ dependence which can be cancelled away by an appropriate normalization. Let us now evaluate the value of the determinant coming from the eigenvalues corresponding to the boundary condition
	\begin{align}
	\del_r^2\psi=0\label{pdp2bdcond}
	\end{align}
	at both the ends. The solution to begin with is given in eq.\eqref{delaltbs}. Imposing the boundary condition eq.\eqref{pdp2bdcond} at the left end $r=-r_{B1}$ gives the relation
	\begin{align}
	k_2=-k_1\frac{\partial_{r_{B1}}^2P^{i\tilde{m}}_{v-\half}(-ir_{B1})}{\partial_{r_{B1}}^2P^{-i\tilde{m}}_{v-\half}(-ir_{B1})}\label{pdp2k1k2}
	\end{align}
	Choosing $k_1=k\partial_{r_{B1}}^2P^{-i\tilde{m}}_{v-\half}(-ir_{B1}) $, we get the solution to be 
	\begin{align}
	\psi_{\lambda,{m}}=k \pqty{P_{v-\half}^{i\tilde{m}}(ir)\partial_{r_{B1}}^2P_{v-\half}^{-i\tilde{m}}(-ir_{B1})- P_{v-\half}^{-i\tilde{m}}(ir)\partial_{r_{B1}}^2P_{v-\half}^{i\tilde{m}}(-ir_{B1})}\label{pdp2ksol}
	\end{align}
	So, we have
	\begin{align}
	\del_r^2 \psi_{\lambda,{m}}(r)\big\vert_{r=r_{B2}}=k \pqty{\partial_{r_{B2}}^2P_{v-\half}^{i\tilde{m}}(ir_{B2})\partial_{r_{B1}}^2P_{v-\half}^{-i\tilde{m}}(-ir_{B1})- \partial_{r_{B2}}^2P_{v-\half}^{-i\tilde{m}}(ir_{B2})\partial_{r_{B1}}^2P_{v-\half}^{i\tilde{m}}(-ir_{B1})}\label{delrxipdp2}
	\end{align}
	which using eq.\eqref{pcorasyexp} becomes
	\begin{align}
	\del_r^2 \psi_{\lambda,{m}}(r)\big\vert_{r=r_{B2}}=k(4 r_{B1} r_{B2})^{-\frac{5}{2}}\pqty{\frac{2\sinh(\tilde{m}\pi)(2v-1)^2(2v-3)^2\Gamma(v)^2(4r_{B1} r_{B2})^v}{\Gamma(\half -i \tilde{m} +v)\Gamma(\half +i \tilde{m} +v)}}\nonumber\\
	+k(4 r_{B1} r_{B2})^{-\frac{5}{2}}\pqty{-\frac{\pi e^{2\pi i v}\sinh(2 \tilde{m} \pi)(1-4v^2)(9-4v^2)}{v \sin(\pi v)}\pqty{r_{B2}\over r_{B1}}^v}+(v\rightarrow -v)
	\label{pdp2normali}
	\end{align}
	Again noting that in the limit of $\abs{\lambda}\gg 1$, the first terms dominates and so we choose $k$ as in eq.\eqref{valk}. With this choice of normalization in eq.\eqref{pdp2ksol}, setting $\lambda=0$, we get the leading term as
	\begin{align}
	\del_r^2 \psi_{\lambda,{m}}(r)\big\vert_{r=r_{B2}}=\frac{\tilde{m}\sinh(\tilde{m}\pi)}{\pi r_{B1}^3 r_{B2}^3}\label{pdp2mneq0}
	\end{align}
	In getting the above result, one has to keep the subleading terms arising from expanding the hypergeometric functions appearing in eq.\eqref{abmsnaiexp} when evaluated for $\nu=v-\half =1$.
	Again ignoring the $r_{B1},r_{B2}$ dependence with the understanding that they can be cancelled by an appropriate normalization and also the $\tilde{m}=0$ sector, we get the total value of the determinant of $P^\dagger P$ as
	\begin{align}
	\ln \text{det}'P^\dagger P=\sum_{m=-\infty,m\neq 0}^{m=\infty}\pqty{\ln(\frac{\sinh(\tilde{m}\pi)}{\tilde{m}\pi(1+\tilde{m}^2)})+\ln(\frac{\tilde{m}\sinh(\tilde{m}\pi)}{\pi })}\label{lnpdpval}
	\end{align}
	The prime in the determinant is to indicate the exclusion of zero modes that appear in the $m=0$ sector. 
	Using eq.\eqref{lnpdpval} and eq.\eqref{lnmdelval}, we get
	\begin{align}
	\ln {\sqrt{\text{det}'P^\dagger P}\over\text{det}(-\hat{\nabla}^2+2) }=\sum_{m=-\infty,m\neq 0}^{m=\infty}\half \ln(\tilde{m}^2(1+\tilde{m}^2))\xrightarrow{b\rightarrow 0}2\ln\frac{b}{4\pi^2}\label{pdpovmdel}
	\end{align}
	So, we find that there is a non-trivial $b$-dependence in the ratio $\sqrt{\text{det}'P^\dagger P}\over\text{det}(-\hat{\nabla}^2+2) $. However, using results in string theory for the partition function of the ghost fields, it it straightforward to compute the determinant of the operator $P^\dagger P$ directly. To do so we use the conformally flat form of the double trumpet geometry as in eq.\eqref{dbconfmet}. The contribution due to the conformal factor is same as in eq.\eqref{dbdetconf} with a factor of $-26$ multiplied in the exponent. The contribution from the flat part is then obtained by reading off the result from eq.7.4.1 of  \cite{Polchinski:1998rq}, the string theory vacuum amplitude for open strings on a cylinder which gives
	\begin{align}
		\sqrt{\text{det}'(P^\dagger P)}\simeq\pqty{\frac{\eta({ib\over 2\pi})}{b}}^2\label{pdpstrings}
	\end{align}
	which then has the correct $b\rightarrow 0$ limit  as in eq.\eqref{mdelb0}.

	\section{dS double trumpet determinants}
	\label{dsdbdets}
	
	\subsection{Coleman method computation}
	\label{ctcdsdt}
	In this section, we shall compute the determinants in the de Sitter double trumpet topology. We will find that the computation of the determinants is very similar to that in the case of AdS double trumpet and so we shall only work out in detail the case of scalar laplacian determinant. To do so, we shall view the metric of the de Sitter double trumpet as arising from the analytic continuation of the $-AdS$ double trumpet which is given by 
	\begin{align}
	ds^2=-\frac{dr^2}{1+r^2}-(1+r^2)d\theta^2\label{dsdbsmet}
	\end{align}
	 and continuing by 
	\begin{align}
	r\rightarrow\pm ir\label{dbadsdscont}
	\end{align}
	at both the ends of the AdS double trumpet, we get the metric of $dS$ double trumpet as
	\begin{align}
	ds^2=-\frac{dr^2}{r^2-1}+(r^2-1)d\theta^2,  \theta\sim\theta+b\label{dbmadsmet}
	\end{align}
	The solutions for an eigenvector of the scalar laplacian in the AdS double trumpet, with mode number $m$, satisfying the equation
	\begin{align}
	\hat{\nabla}^2\psi=-\lambda\psi\label{dsadseeq}
	\end{align}
	whose solution is given in eq.\eqref{esol}.
	Continuing this solution by analytic continuation we get the eigenvector for the $dS$ double trumpet topology as
	\begin{align}
	\psi_{\lambda, m}=k_1 P_{v-\half}^{i \tilde{m}}(r)+k_2Q_{v-\half}^{i \tilde{m}}(r)\label{dsbdslapesol}
	\end{align}
	where 
	\begin{align}
	v=\sqrt{\frac{1}{4}+\lambda},\quad \tilde{m}=\frac{2\pi m}{b}\label{dsdbvmt}
	\end{align}
	An equivalently good basis of solution for $\tilde{m}\neq0$ is $P_{v-\half}^{\pm \tilde{m}}$ in which the solution above becomes
	\begin{align}
	\psi_{\lambda, m}=k_1 P_{v-\half}^{i \tilde{m}}(r)+k_2P_{v-\half}^{-i \tilde{m}}(r)\label{dsdbdslapesol}
	\end{align}
	Now, to have the correct asymptotic expansions at the left and the right boundaries, we need to look at the contour a bit more carefully. Due to the presence of the branch cut from $(-1,\infty)$ in the  complex $r$ plane, we take the contour for the de Sitter double trumpet to start at $r=-\infty$ just below the real axis  and first rotate counterclockwise to $r=-ir_{B1},r_{B1}\rightarrow\infty$. We then end up in the AdS double trumpet geometry in which we go from the left end to the right end along the imaginary axis from which we pick a relative factor of $e^{\tilde{m}\pi}$, see discussion after  eq.\eqref{abmsasasp}. From the point $r=i r_{B2}, r_{B2}\rightarrow\infty$, we rotate clockwise to end up at $r=\infty$. However, note that there are other equivalently good choices of contour, say rotating counterclockwise in the last step or beginning from above the cut at $r=-\infty$. In total, we have four possible choices of contours, which are depicted in Fig.\ref{eefig3}.  Our choice corresponds to the path $EDCBF$. Other choices will only change the phase factor $e^{\pm i\pi v}$ and will end up giving the same final answer. 
	Now, following the same steps as in the AdS double trumpet, imposing the Dirichlet boundary conditions at one end of the boundary $r=- r_{B1}$, we get the form of the solution in eq.\eqref{delintrk} as
	\begin{align}
	\psi_{\lambda,{m}}=k \pqty{P_{v-\half}^{i\tilde{m}}(r)P_{v-\half}^{-i\tilde{m}}(-r_{B1})- P_{v-\half}^{-i\tilde{m}}(r)P_{v-\half}^{i\tilde{m}}(-r_{B1})}\label{dbelintrk}
	\end{align}
	With the choice of contour we made,  the asymptotic forms of the associated Legendre functions now read
	\begin{align}
	\label{ddsaspexp}
	&P_{v-\half}^{i\tilde{m}}(- r_B)=e^{i\pi \over 2}\pqty{{(2r_B)^{-v-\half}\over \sqrt{\pi}}\frac{\Gamma(-v)e^{ i\pi v}}{\Gamma(\half-i \tilde{m}- v )}+{(2r_B)^{ v -\half}\over \sqrt{\pi}}\frac{\Gamma( v )e^{ -i\pi v }}{\Gamma(\half-i \tilde{m}+ v )}}\nonumber\\
	&P_{ v -\half}^{i\tilde{m}}( r_B)=e^{\pi \tilde{m}}\pqty{{(2r_B)^{- v -\half}\over \sqrt{\pi}}\frac{\Gamma(- v )}{\Gamma(\half-i \tilde{m}- v )}+{(2r_B)^{ v -\half}\over \sqrt{\pi}}\frac{\Gamma( v )}{\Gamma(\half-i \tilde{m}+ v )}}
	\end{align}
	and hence we get, at $r= r_{B2}\gg 1$, the value of the scalar field solution as
	\begin{align}
	\psi_{\lambda,\tilde{m}}=&k \pqty{P_{v-\half}^{i\tilde{m}}(r_{B2})P_{v-\half}^{-i\tilde{m}}(-r_{B1})- P_{v-\half}^{-i\tilde{m}}(r_{B2})P_{v-\half}^{i\tilde{m}}(-r_{B1})}\nonumber\\
	=&\frac{k}{\pi}(4 r_{B2} r_{B1})^{-\half}e^{i\pi \over 2}\pqty{\frac{2\sinh(\tilde{m}\pi)(4 r_{B2} r_{B1})^{ v }\Gamma( v )^2}{\Gamma(\half-i\tilde{m}+ v )\Gamma(\half+i\tilde{m}+ v )}e^{-i\pi   v }-e^{2i\pi v}\frac{\sinh(2\tilde{m}\pi)}{ v \sin(\pi  v )}\pqty{\frac{r_{B2}}{r_{B1}}}^v +( v \rightarrow- v )}\label{ddseaveqinw}
	\end{align}
	The same reasoning as in the case of AdS double trumpet, around eq.\eqref{nuinf} shows that $k$ should be chosen as in eq.\eqref{valk} and so the wavefunction and consequently, the laplacian determinant are as given by eq.\eqref{lndetdel}. It should be noted that the only difference in the associated Legendre function expansions in the two cases as seen from eq.\eqref{pcorasyexp} and eq.\eqref{ddsaspexp} is only in the overall constant phases and factors of $v$-dependent exponentials, due to the fact that argument of the associated Legendre function is imaginary and real in the AdS and $dS$ double trumpets respectively. However, this doesn't make a difference in the value of $k$ and hence the value of the solution $\psi_{\lambda,m}$ for $v=\half$ becomes
	\begin{align}
	\psi_{0,\tilde{m}}=&\frac{\sinh \tilde{m}\pi}{\tilde{m}\pi}\label{ddspsi}
	\end{align}
	which is the same as in eq.\eqref{coltricans}. It then immediately follows that the value of the scalar laplacian determinant is same as before eq.\eqref{lndetdel} upto an irrelevant numerical constant\footnote{It can be argued easily that the other choices of contours in Fig.\ref{eefig3} will only result in an overall unimportant constant in the final value of the determinant.}. The limit $b\rightarrow 0$ of this determinant is therefore same as before as in eq.\eqref{lndetdel}. To calculate the determinant of the operator $(-\hat{\nabla}^2+2)$, we can use the same result in eq.\eqref{ddseaveqinw} except that the  definition of $ v $ becomes
	\begin{align}
	v=\sqrt{\frac{9}{4}+\lambda}\label{ddsdel2nu}
	\end{align}
	and hence we get the same result again for $(-\hat{\nabla}^2+2)$ as in eq.\eqref{lnmdelval} upto irrelevant numerical constant. The same arguments apply for the operator $P^\dagger  P$ and so we get the result eq.\eqref{lnpdpval} for this  operator.

	\subsection{Direct calculation of scalar laplacian}
	\label{dcdtscl}
	
	We shall now show an alternative but direct way of evaluating the scalar determinant with Dirichlet boundary conditions at both ends. This calculations is analogous to the one in AdS double trumpet in appendix \ref{dbmatapx} by writing the metric in conformally flat form and then separately evaluating the conformal factor and flat metric contributions. We consider the analogue of the Maldacena contour here where we consider the $-AdS$ metric for the double trumpet and analytically continue it to the $dS$ double trumpet. The $-AdS$ double trumpet meric is given by 
	\begin{align}
	ds^2=-\frac{1}{\sin^2r_{*, AdS}}(dr_{*, AdS}^2+d\theta^2),\quad r_{*, AdS}\in [0,\pi],\theta\in [0,b]\label{adsdtconfmet}
	\end{align}. Now we first do the analytic continuation at the left end of the above $-AdS$ line element by doing 
	\begin{align}
	r_{*, AdS}= \pm i r_{*,dS}\label{leenddsdtcon}
	\end{align}
	The double trumpet metric in the conformally flat form is given by 
	\begin{align}
	ds^2=\frac{1}{\sinh^2 r_{*, dS}}(-dr_{*,dS}^2+d\theta^2)\label{dsdtconfmet}
	\end{align}
	The contribution from the conformal factor can be obtained by using eq.\eqref{adsscallap} with the conformal factor given by 
	\begin{align}
	e^{2\sigma}=-\frac{1}{\sin^2 r_{*, AdS}}\label{dsdtconf}
	\end{align}
	for the part of the contour through AdS double trumpet. 
	We are then left to evaluate the contribution from flat part of the line element.  Taking the solution for the eigenvalue equation of $-\hat{\nabla}^2$ with the mode number $m$, the equation for which reads
	\begin{align}
	\hat{\nabla}^2\psi=-\lambda\psi\label{scaleq}
	\end{align}
	the solution satisfying the Dirichlet boundary conditions is given by 
	\begin{align}
	\psi\sim e^{i\tilde{m}\theta}\sin(\sqrt{\lambda-\tilde{m}^2}r_{*, AdS})\label{adsdtsol}
	\end{align}
	where $\tilde{m}$ is related to $m$ by eq.\eqref{dsdbvmt}. Continuing this solution to the $dS$ using eq.\eqref{leenddsdtcon} gives
	\begin{align}
	\psi\sim \pm e^{i\tilde{m}\theta}\sinh(\sqrt{\lambda-\tilde{m}^2}r_{*, AdS})\label{dsdtscsol}
	\end{align}
	which correctly satisfies the Dirichlet boundary conditions as $r_{*, AdS}\rightarrow 0$. Now to obtain the continuation at the right end where $r_{*, AdS}\rightarrow \pi$, we do the continuation
	\begin{align}
	r_{*, AdS}=\pi\pm i r_{*,dS}\label{adsdsrecon}
	\end{align}
	as a result of which the solution eq.\eqref{adsdtsol} then becomes
	\begin{align}
	\psi\simeq e^{i\tilde{m}\theta}\sin(\sqrt{\lambda-\tilde{m}^2}(\pi\pm ir_{*,dS}))\label{dsdtresol}
	\end{align}
	Requiring this to vanish at $r_{*,dS}=0$ then gives
	\begin{align}
	\lambda=\tilde{m}^2+n^2	\label{dsdteval}
	\end{align}
	which are the same as one would get in the case of AdS double trumpet and so we get
	\begin{align}
	\ln \text{det}(-\hat{\nabla}^2)=\sum_{m,n}\ln (\tilde{m}^2+n^2)\label{ddtdeld2dire}
	\end{align}
	which combined with the contribution from the conformal factor gives the full dependence. Note the flat part contribution is the same as in the AdS case since the eigenvalues are the same and hence we will get the same $b\rightarrow 0$ dependence as in eq.\eqref{dbflpart} for the scalar partition function.

	%-----------------END OF APPENDICES-----------------------

	\bibliographystyle{JHEP}
	\bibliography{refs}
\end{document}